\documentclass[12pt,english]{article}
\usepackage{times}
\usepackage[T1]{fontenc}
\usepackage{geometry}
\usepackage{rotating}
\usepackage{graphicx}
\usepackage{setspace}
\usepackage{amssymb}

\makeatletter

\providecommand{\tabularnewline}{\\}

\usepackage{bm}

\usepackage{babel}
\makeatother
\begin{document}

\section*{Learned Global Optimization for Inverse Scattering Problems - Matching
Global Search with Computational Efficiency}

\noindent \vfill

\noindent M. Salucci,$^{(1)}$ \emph{Member, IEEE}, L. Poli,$^{(1)}$
\emph{Member, IEEE}, P. Rocca, $^{(1)(2)}$ \emph{Senior Member, IEEE},
and A. Massa,$^{(1)(3)(4)}$ \emph{Fellow, IEEE}

\noindent \vfill

\noindent {\footnotesize $^{(1)}$} \emph{\footnotesize CNIT} {\footnotesize -
\char`\"{}University of Trento\char`\"{} Research Unit}{\footnotesize \par}

\noindent {\footnotesize Via Sommarive 9, 38123 Trento - Italy}{\footnotesize \par}

\noindent \textit{\emph{\footnotesize E-mail:}} {\footnotesize \{}\emph{\footnotesize marco.salucci}{\footnotesize ,}
\emph{\footnotesize lorenzo.poli, paolo.rocca}{\footnotesize ,} \emph{\footnotesize andrea.massa}{\footnotesize \}@}\emph{\footnotesize unitn.it}{\footnotesize \par}

\noindent {\footnotesize Website:} \emph{\footnotesize www.eledia.org/eledia-unitn}{\footnotesize \par}

\noindent {\footnotesize ~}{\footnotesize \par}

\noindent {\footnotesize $^{(2)}$} \emph{\footnotesize ELEDIA Research
Center} {\footnotesize (}\emph{\footnotesize ELEDIA}{\footnotesize @}\emph{\footnotesize XIDIAN}
{\footnotesize - Xidian University)}{\footnotesize \par}

\noindent {\footnotesize P.O. Box 191, No.2 South Tabai Road, 710071
Xi'an, Shaanxi Province - China }{\footnotesize \par}

\noindent {\footnotesize E-mail:} \emph{\footnotesize paolo.rocca@xidian.edu.cn }{\footnotesize \par}

\noindent {\footnotesize Website:} \emph{\footnotesize www.eledia.org/eledia-xidian}{\footnotesize \par}

\noindent {\footnotesize ~}{\footnotesize \par}

\noindent {\footnotesize $^{(3)}$} \emph{\footnotesize ELEDIA Research
Center} {\footnotesize (}\emph{\footnotesize ELEDIA}{\footnotesize @}\emph{\footnotesize UESTC}
{\footnotesize - UESTC)}{\footnotesize \par}

\noindent {\footnotesize School of Electronic Engineering, Chengdu
611731 - China}{\footnotesize \par}

\noindent \textit{\emph{\footnotesize E-mail:}} \emph{\footnotesize andrea.massa@uestc.edu.cn}{\footnotesize \par}

\noindent {\footnotesize Website:} \emph{\footnotesize www.eledia.org/eledia}{\footnotesize -}\emph{\footnotesize uestc}{\footnotesize \par}

\noindent {\footnotesize ~}{\footnotesize \par}

\noindent {\footnotesize $^{(4)}$} \emph{\footnotesize ELEDIA Research
Center} {\footnotesize (}\emph{\footnotesize ELEDIA@TSINGHUA} {\footnotesize -
Tsinghua University)}{\footnotesize \par}

\noindent {\footnotesize 30 Shuangqing Rd, 100084 Haidian, Beijing
- China}{\footnotesize \par}

\noindent {\footnotesize E-mail:} \emph{\footnotesize andrea.massa@tsinghua.edu.cn}{\footnotesize \par}

\noindent {\footnotesize Website:} \emph{\footnotesize www.eledia.org/eledia-tsinghua}{\footnotesize \par}

\noindent \vfill

\noindent \vfill

\newpage
\section*{Learned Global Optimization for Inverse Scattering Problems - Matching
Global Search with Computational Efficiency}

~

\noindent ~

\noindent ~

\begin{flushleft}M. Salucci, L. Poli, P. Rocca, and A. Massa\end{flushleft}

\noindent \vfill

\begin{abstract}
\noindent The computationally-efficient solution of fully non-linear
microwave inverse scattering problems (\emph{ISP}s) is addressed.
An innovative System-by-Design (\emph{SbD}) based method is proposed
to enable, for the first time to the best of the authors' knowledge,
an effective, robust, and \emph{time-efficient} exploitation of an
evolutionary algorithm (\emph{EA}) to perform the global minimization
of the data-mismatch cost function. According to the \emph{SbD} paradigm
as suitably applied to \emph{ISP}s, the proposed approach founds on
(\emph{i}) a \emph{smart} re-formulation of the \emph{ISP} based on
the definition of a minimum-dimensionality and representative set
of degrees-of-freedom (\emph{DoF}s) and on (\emph{ii}) the artificial-intelligence
(\emph{AI})-driven integration of a customized global search technique
with a \emph{digital twin} (\emph{DT}) predictor based on the Gaussian
Process (\emph{GP}) theory. Representative numerical and experimental
results are provided to assess the effectiveness and the efficiency
of the proposed approach also in comparison with competitive state-of-the-art
inversion techniques.
\end{abstract}
\noindent \vfill

\noindent \textbf{Key words}: Inverse Scattering (\emph{IS}), Evolutionary
Algorithms (\emph{EAs}), System-by-Design (\emph{SbD}), Digital Twin
(\emph{DT}), Artificial Intelligence (\emph{AI}), Learning-by-Examples
(\emph{LBE}), Gaussian Processes (\emph{GP}s)

\newpage
\section{Introduction }

\noindent In microwave imaging, an electromagnetic (\emph{EM}) source
illuminates an inaccessible investigation domain to be non-invasively
reconstructed by inverting the scattered field data collected in an
external observation domain \cite{Chen 2018}. Depending on the application
at hand, both qualitative (i.e., detection, localization, and shaping)
and quantitative (i.e., \emph{EM} properties characterization) reconstructions
can be yielded by solving an inverse scattering problem (\emph{ISP}).
\emph{ISP}s arise in free-space imaging, biomedical diagnostics \cite{Abubakar 2002}-\cite{Song 2019},
subsurface and ground penetrating radar (\emph{GPR}) investigations
\cite{Cui 2001}-\cite{Salucci 2017b}, non-destructive testing and
evaluation (\emph{NDT/NDE}) \cite{Liu 2018}-\cite{Caorsi 2001},
and through-the-wall imaging (\emph{TWI}) \cite{Xu 2018b}-\cite{Fallahpour 2015}.
Recently, microwave imaging techniques, based on inverse scattering
(\emph{IS}) formulations, have been also successfully applied to innovative
contexts such as, for instance, food quality assessment \cite{LoVetri 2020}-\cite{Occhiuzzi 2020}.
However, solving an \emph{ISP} is not a trivial task and it poses
several challenges due to the intrinsic complexity of the scattering
phenomena in the microwave regime described by the Maxwell's equations.
First, the non-uniqueness of the solution, caused by the presence
of non-radiating currents induced in the investigation domain, that
do not contribute to the scattered data. Second, the non-linearity
related to the multiple scattering effects \cite{Chen 2018}. To properly
address such issues for yielding robust/reliable data-inversions,
many effective strategies have appeared in the state-of-the-art literature.
For instance, Born-based \cite{Chew 1990} and Rytov-based \cite{Zhang 2009}
approximations simplify the \emph{IS} equations as linearly depending
on the unknown contrast distribution. However, they have limited applications
to weak scatterers. Otherwise, innovative reformulations of the scattering
equations as, for instance, the contraction integral equation (\emph{CIE})
method, have been introduced to deal with the non-linearity by properly
redefining the contrast function \cite{Xu 2018b}\cite{Zhong 2016}.
Differently, contrast source inversion (\emph{CSI}) techniques proved
to be an effective alternative to the linearization of the data equation
\cite{Kleinman 1997}, even though they are subject to the non-uniqueness
of the arising inverse source problem so that multiplicative regularizations
have been investigated \cite{vandenBerg 2003}\cite{Xu 2015}. 

\noindent Regardless of the formulation and unless closed-form solutions,
\emph{ISP}s are generally solved with deterministic (\emph{DO}) or
global (\emph{GO}) optimization techniques. Strategies belonging to
the former class include the subspace optimization method (\emph{SOM})
\cite{Chen 2010}-\cite{Zhong 2011}, the conjugate gradient (\emph{CG})
\cite{Harada 1995}, and the inexact Newton method (\emph{INM}) \cite{Salucci 2017c}.
To deterministically explore the solution space, these methods typically
require the analytic/numerical differentiation of the cost function
to be minimized. Consequently, they exhibit a high computational efficiency,
but they can be trapped into local-minima/false-solutions, unless
properly initialized within the so-called {}``attraction basin''
of the global optimum. 

\noindent As for \emph{GO} methods, nature-inspired strategies (i.e.,
evolutionary algorithms \emph{}(\emph{EA}s) \cite{Rocca 2009}-\cite{Goudos 2021})
such as genetic algorithms (\emph{GA}s) \cite{Caorsi 2001}, particle
swarm optimization (\emph{PSO}) \cite{Salucci 2017b}, and differential
evolution (\emph{DE}) \cite{Rocca 2011} have been successfully applied
to solve \emph{ISP}s. Thanks to the {}``hill-climbing'' features,
they perform an effective global exploration of the solution space
by evolving a population of trial solutions with stochastic operators
\cite{Rocca 2009} to {}``escape'' from local minima, while converging
towards the global optimum. Although successful in several \emph{ISP}
applications and more effective than \emph{DO}s in sampling nonlinear
cost functions, \emph{EA-GO}s are inherently limited by the computational
burden. Indeed, the \emph{CPU} cost of a stochastic \emph{GO} is directly
linked to the number of agents that evolve throughout the optimization
process, which is in turn proportional to the number of degrees-of-freedom
(\emph{DoF}s) that define the dimensionality of the solution space.
To partially counteract such a limitation, one practical and effective
solution is the integration of \emph{EA-GO}s with multi-resolution
(\emph{MR}) strategies such as the iterative multi-scaling approach
(\emph{IMSA}) \cite{Caorsi 2003}. By adaptively refining the spatial
resolution of the reconstruction only within the so-called regions-of-interest
(\emph{RoI}), where the unknown scatterer has been detected, the number
of unknowns is strongly reduced at each \emph{MR} step \cite{Salucci 2017b}\cite{Donelli 2009}
by making computationally-feasible an \emph{EA-GO}-based optimization.

\noindent On the other hand, artificial intelligence (\emph{AI})-based
techniques, belonging to the so-called deep learning (\emph{DL}) framework
\cite{Massa 2019}-\cite{Zhou 2021}, have shown an unprecedented
computational efficiency in addressing the pixel-wise inversion of
scattered data. However, they still present some unsolved challenges
such as the need of huge amounts of training datasets to calibrate
thousands of hyper-parameters that define the underlying complex neural
network (\emph{NN}) architecture composed by several hidden layers
\cite{Massa 2019}. Within the \emph{AI} context, the System-by-Design
(\emph{SbD}) has rapidly emerged as an innovative paradigm for the
optimization-driven solution of complex \emph{EM} problems \cite{Massa 2021}.
The problem at hand is first decomposed into a set of sub-tasks implemented
into suitably-defined functional blocks jointly designed with the
shared goal of an effective, reliable, and computationally-efficient
exploitation of \emph{GO}s. Such a goal is attained by (\emph{i})
re-formulating the problem at hand as a \emph{GO} one described by
a minimum-dimensionality set of \emph{DoF}s and (\emph{ii}) integrating
\emph{EA}-based strategies with fast analysis tools or digital twins
(\emph{DT}s), generated with learning-by-examples (\emph{LBE}s) techniques
\cite{Massa 2017}, to speed up the evaluation (i.e., the cost function
computation) of each trial solution. Thanks to its effectiveness and
efficiency, the \emph{SbD} has been already successfully applied to
many \emph{EM} design problems including the synthesis of single radiators
\cite{Salucci 2019b}, wide angle impedance matching layers \cite{Oliveri 2017},
reflectarrays \cite{Oliveri 2020}, and meta-material devices \cite{Salucci 2019},
but not to \emph{ISP}s. This paper is then aimed at assessing the
\emph{SbD} in reliably solving fully non-linear \emph{ISPs} with a
computational efficiency, comparable to that of \emph{DO}s, towards
the {}``holy-grail'' of a global real-time optimization. 

\noindent The paper is organized as follows. The \emph{ISP} is described
and mathematically formulated in Sect. II. Section III details the
customization of the \emph{SbD} paradigm to \emph{ISP}s and its implementation.
Numerical and experimental results are shown in Sect. IV to prove
the effectiveness and the efficiency of the proposed method in different
operative conditions. Eventually, some conclusions and final remarks
are drawn (Sect. V).

\section{Mathematical Formulation}

Without loss of generality, let us consider a two-dimensional (\emph{2D})
scenario comprising a square investigation domain $D$ located within
a homogeneous, lossless (i.e., conductivity $\sigma=\sigma_{0}=0$
{[}S/m{]}), and non-magnetic (i.e., permeability $\mu=\mu_{0}$) background
medium of permittivity $\varepsilon_{0}$. By assuming a time-harmonic
dependence $\exp\left(-j2\pi ft\right)$, $f$ being the working frequency,
and a transverse magnetic (\emph{TM}) (i.e., $z$-oriented) polarization
of the \emph{EM} field, the scattering phenomena excited by a set
of $V$ monochromatic \emph{}incident fields, \{$\mathcal{I}^{\left(v\right)}\left(x,\, y\right)$;
$v=1,...,V$\}, which illuminate the investigation domain $D$, in
any $\left(x,\, y\right)\in D$ are modeled by the following \emph{State}
\emph{Equation} \cite{Chen 2018}\begin{equation}
\mathcal{I}^{\left(v\right)}\left(x,\, y\right)=\mathcal{T}^{\left(v\right)}\left(x,\, y\right)-\int_{D}\mathcal{G}\left(x,\, y,\, x',\, y'\right)\mathcal{J}^{\left(v\right)}\left(x',\, y'\right)dx'dy'\label{eq:state-equation}\end{equation}
where

\begin{equation}
\mathcal{J}^{\left(v\right)}\left(x,\, y\right)=\tau\left(x,\, y\right)\mathcal{T}^{\left(v\right)}\left(x,\, y\right)\label{eq: W Rossiglione}\end{equation}
is the $v$-th ($v=1,...,V$) equivalent current induced within $D$,
$\mathcal{T}^{\left(v\right)}\left(x,\, y\right)$ is the total field,
and\begin{equation}
\tau\left(x,\, y\right)=\left[\varepsilon_{r}\left(x,\, y\right)-1\right]+j\frac{\sigma\left(x,\, y\right)}{2\pi f\varepsilon_{0}}\label{eq:}\end{equation}
is the contrast function that mathematically models the presence,
within $D$, of an unknown scatterer with support $\Omega$ (i.e.,
$\tau\left(x,\, y\right)\neq0$ when $\left(x,\, y\right)\in\Omega$)
whose relative permittivity and conductivity distributions are equal
to $\varepsilon_{r}\left(x,\, y\right)$ {[}$\varepsilon_{r}\left(x,\, y\right)\triangleq\frac{\varepsilon\left(x,\, y\right)}{\varepsilon_{0}}${]}
and $\sigma\left(x,\, y\right)$, respectively. Moreover,\begin{equation}
\mathcal{G}\left(x,\, y,\, x',\, y'\right)=j\frac{k_{0}^{2}}{4}\mathcal{H}_{0}^{\left(1\right)}\left(k_{0}\sqrt{\left(x-x'\right)^{2}+\left(y-y'\right)^{2}}\right)\label{eq:}\end{equation}
is the \emph{2D} Green's function of the background medium, $\mathcal{H}_{0}^{\left(1\right)}$
being the zero-th order Hankel's function of the first kind, and $k_{0}$
is the wavenumber ($k_{0}\triangleq2\pi f\sqrt{\varepsilon_{0}\mu_{0}}$).

\noindent Otherwise, the \emph{EM} interactions in the external observation
domain $O\notin D$ ($O\cap D=\left\{ 0\right\} $) \cite{Chen 2018}
are described by the \emph{Data Equation}\begin{equation}
\mathcal{S}^{\left(v\right)}\left(x,\, y\right)=\int_{D}\mathcal{G}\left(x,\, y,\, x',\, y'\right)\mathcal{J}^{\left(v\right)}\left(x',\, y'\right)dx'dy',\label{eq:data-equation}\end{equation}
where $\mathcal{S}^{\left(v\right)}\left(x,\, y\right)$ {[}$\mathcal{S}^{\left(v\right)}\left(x,\, y\right)\triangleq\mathcal{T}^{\left(v\right)}\left(x,\, y\right)-\mathcal{I}^{\left(v\right)}\left(x,\, y\right)${]}
is the scattered field radiated in free-space by the $v$-th ($v=1,\,...,\, V$)
equivalent source, $\mathcal{J}^{\left(v\right)}\left(x,\, y\right)$,
and embedding the information on the unknown scatterer distribution
in $D$.

\noindent To numerically deal with (\ref{eq:data-equation}), the
method-of-moments (\emph{MoM}) is applied by partitioning $D$ into
$N$ square sub-domains, $D_{n}$ being the $n$-th ($n=1,...,N$)
discretization domain ($D=\sum_{n=1}^{N}D_{n}$) centered at $\left(x_{n},\, y_{n}\right)$
and using $M$ Dirac's test functions to sample the scattered field
at $M$ locations in $O$, $\underline{\mathcal{S}}^{\left(v\right)}=\left\{ \mathcal{S}^{\left(v\right)}\left(x_{m},\, y_{m}\right);\, m=1,\,...,\, M\right\} $.
The discrete form of (\ref{eq:data-equation}) is then derived\begin{equation}
\underline{\mathcal{\mathcal{S}}}^{\left(v\right)}=\underline{\underline{G}}_{O}\underline{\mathcal{J}}^{\left(v\right)}\label{eq:scattered-field-computation}\end{equation}
where $\underline{\mathcal{J}}^{\left(v\right)}=\left\{ \mathcal{J}^{\left(v\right)}\left(x_{n},\, y_{n}\right);\, n=1,...,N\right\} $
and $\underline{\underline{G}}_{O}$ is the $\left(M\times N\right)$
external Green's matrix whose $\left(m,n\right)$-th ($m=1,...,M$;
$n=1,...,N$) entry is given by $\left.\underline{\underline{G}}_{O}\right\rfloor _{mn}$
$=$ $j\frac{k_{0}^{2}}{4}$ $\int_{D_{n}}$ $\mathcal{H}_{0}^{\left(1\right)}\left(k_{0}\rho_{m}\right)$
$dx'dy'$ being $\rho_{m}\triangleq\sqrt{\left(x_{m}-x'\right)^{2}+\left(y_{m}-y'\right)^{2}}$.

\noindent Accordingly, the inverse problem at hand can be stated as
follows

\begin{quotation}
\noindent \textbf{\emph{ISP}} - Starting from the knowledge of the
incident, \{$\mathcal{I}^{\left(v\right)}\left(x_{n},\, y_{n}\right)$;
$n=1,....,N$\}, and the scattered, \{$\mathcal{S}^{\left(v\right)}\left(x_{m},\, y_{m}\right)$;
$m=1,...,M$\}, data samples, determine the contrast function distribution,
\{$\tau\left(x_{n},\, y_{n}\right)$; $n=1,...,N$\}, by solving (\ref{eq:scattered-field-computation}).
\end{quotation}
In order to solve this full non-linear \emph{ISP}, an innovative \emph{SbD}-based
is adopted according to the implementation detailed in Sect. \ref{sec:SbD-Based-Inversion-Method}.

\section{\emph{SbD}-Based \emph{}Inversion Method \label{sec:SbD-Based-Inversion-Method}}

According to the \emph{SbD} paradigm, the solution of the \emph{ISP}
relies on the exploitation of four interconnected \emph{functional
blocks,} each performing a specific sub-task (Fig. 1). The design
and implementation of each block is strongly correlated to the other
ones and it is driven by the following shared goals \cite{Massa 2021}:
(\emph{i}) to yield an effective and reliable solution of the fully
non-linear \emph{ISP}. From an optimization viewpoint, it means to
guarantee the convergence towards the global optimum; (\emph{ii})
to reduce the computational burden required by a standard non-deterministic
exploration of the solution space. In other words, the proposed \emph{SbD}
approach is aimed at overcoming the limitation of \emph{DO}s, which
cannot avoid being trapped into local minima unless properly initialized
in the {}``attraction basin'' of the actual-solution/global-optimum,
while yielding competitive computational performance in solving the
\emph{ISP} so that the following condition on the required \emph{CPU}-time
holds true\begin{equation}
\Delta t_{SbD}\simeq\Delta t_{DO}\ll\Delta t_{GO}.\label{eq:time-goal}\end{equation}
More specifically, the \emph{SbD} as applied to \emph{ISP}s is implemented
by defining the following blocks (Fig. 1):

\begin{enumerate}
\item \emph{Problem Formulation} (\emph{PF}) - This block reformulates the
\emph{ISP} to enable an effective, reliable, and computationally-efficient
exploitation of \emph{GO}s by coding the \emph{ISP} unknowns into
a minimum-dimension (yet highly-flexible) set of $K$ degrees-of-freedom
(\emph{DoF}s), $\underline{\xi}=\left\{ \xi_{k};\, k=1,...,K\right\} $,
to give a {}``smart'' representation of the solution space. Moreover,
it defines a suitable cost function, $\Phi\left(\underline{\xi}\right)$,
which quantifies the quality of the solution in terms of data mismatch
and it represents the unique link between the computational world
and the physical one;
\item \emph{Data Computation} (\emph{DC}) - In this block, the set of \emph{SbD-DoF}s,
$\underline{\xi}$, is mapped into a pixel-based representation of
the equivalent currents induced within $D$, $\left\{ \underline{\mathcal{J}}^{\left(v\right)};\, v=1,...,V\right\} $,
by means of (\ref{eq: W Rossiglione}) and (\ref{eq:state-equation})
to compute, through (\ref{eq:scattered-field-computation}), the scattered
field distribution in $O$;
\item \emph{Cost Function Evaluation} (\emph{CFE}) - This block efficiently
evaluates the cost function with a computationally-fast digital twin
(\emph{DT}) \cite{Massa 2017} of the accurate, but time-consuming,
full-wave solver. It is the {}``engine'' of the \emph{SbD}-based
inversion and it exploits the \emph{DC} block for the computation
of the scattered data, $\underline{\mathcal{\widetilde{\mathcal{S}}}}^{\left(v\right)}$,
in correspondence with each coded trial solution, $\underline{\xi}$;
\item \emph{Solution Space Exploration} (\emph{SSE}) - This block performs
an effective sampling of the \emph{ISP} solution space by leveraging
on (\emph{a}) the {}``hill-climbing'' features of a properly customized
\emph{EA} strategy and on (\emph{b}) the smart interaction with the
\emph{DT} to yield a fast and reliable convergence towards the global
optimum. The \emph{SSE} block receives as external inputs the samples
of the incident, \{$\mathcal{I}^{\left(v\right)}\left(x_{n},\, y_{n}\right)$,
$\left(x_{n},\, y_{n}\right)\in D$; $n=1,....,N$\}, and the scattered,
\{$\mathcal{S}^{\left(v\right)}\left(x_{m},\, y_{m}\right)$, $\left(x_{m},\, y_{m}\right)\in O$;
$m=1,...,M$\}, fields, while it uses the unknowns coding, $\underline{\xi}$,
and the cost function definition, $\Phi$, from the \emph{PF} block.
The \emph{SSE} output is the \emph{SbD} solution, $\underline{\xi}^{\left(SbD\right)}$,
and its mapping in a contrast distribution, $\underline{\tau}^{\left(SbD\right)}$.
\end{enumerate}
Each \emph{SbD} block is detailed in the following by pointing out
the key-item for its integrated implementation.

\subsection{Problem Formulation (\emph{PF}) \label{sub:Problem-Formulation}}

Concerning the identification of a suitable parametric model of the
\emph{ISP} solution in terms of a limited set of $K$ descriptors,
$\underline{\xi}=\left\{ \xi_{k};\, k=1,...,K\right\} $, it is worth
noticing that the number of \emph{DoF}s $K$ is directly proportional
to the size of the population of trial-solutions, $P$, used in the
multiple-agent minimization of $\Phi\left(\underline{\xi}\right)$,
and it determines the overall computational cost of the inversion
process. Therefore, it is paramount to seek for the smartest coding
of the solution that minimizes the computational burden of the optimization,
while enabling a careful exploration of the solution space towards
the global optimum $\underline{\xi}^{\left(opt\right)}$ ($\Phi\left(\underline{\xi}^{\left(opt\right)}\right)\triangleq0$).
Moreover, one should consider that the definition of a minimum-dimensionality
representation of the \emph{ISP} solution facilitates the generation
of an accurate surrogate model (i.e., the \emph{DT}) able to predict
$\Phi\left(\underline{\xi}\right)$ from a reduced set of training
observations (see Sect. \ref{sub:Cost-Function-Evaluation}). Following
this line of reasoning, a standard pixel-based representation of the
unknown distribution of the \emph{EM} profile of $D$, $\underline{\tau}$
$=$ \{$\Re\left(\tau_{n}\right)$, $\Im\left(\tau_{n}\right)$; $n=1,...,N$\},
$\Re\left(\,.\,\right)$/$\Im\left(\,.\,\right)$ being the real/imaginary
part and $\tau_{n}=\tau\left(x_{n},\, y_{n}\right)$ ($n=1,\,...,\, N$),
is sub-optimal because of the huge dimension of the corresponding
solution space (i.e., $K=2\times N$) \cite{Salucci 2017b}. To reduce
the cardinality of the problem at hand, spline basis functions \cite{Salucci 2019}
are exploited here to model the external contour $\partial\Omega\left(x,\, y\right)$
of the homogeneous%
\footnote{\noindent The extension of the spline representation to doubly-connected
contours (e.g., inhomogeneous concentric contrast distributions) as
well as to multiple disconnected objects is straightforward as discussed
and proved in Sect. \ref{sec:Performance-Assessment}.%
} scatterer (i.e., $\tau\left(x,\, y\right)=\tau_{\Omega}$, $\left(x,\, y\right)\in\Omega$)
of extension/support $\Omega$ (Fig. 2). More in detail, the 2\emph{-D}
profile $\partial\Omega\left(x,\, y\right)$ is expanded into $Q$
quadratic Bezier spline functions\begin{equation}
\partial\Omega\left(x,\, y\right)=\sum_{q=1}^{Q}\mathcal{B}^{\left(q\right)}\left(\alpha\right),\label{eq:}\end{equation}
the $q$-th basis function ($q=1,\,...,\, Q$) being given by\begin{equation}
\mathcal{B}^{\left(q\right)}\left(\alpha\right)=\left(1-\alpha\right)^{2}\left[\begin{array}{c}
\mathcal{V}_{x}^{\left(q\right)}\\
\mathcal{V}_{y}^{\left(q\right)}\end{array}\right]+2\alpha\left(1-\alpha\right)\left[\begin{array}{c}
\mathcal{C}_{x}^{\left(q\right)}\\
\mathcal{C}_{y}^{\left(q\right)}\end{array}\right]+\alpha^{2}\left[\begin{array}{c}
\mathcal{V}_{x}^{\left(q+1\right)}\\
\mathcal{V}_{y}^{\left(q+1\right)}\end{array}\right]\label{eq:}\end{equation}
where $\alpha\in\left[0,\,1\right]$ and $\mathcal{C}^{\left(q\right)}=\left(\mathcal{C}_{x}^{\left(q\right)},\,\mathcal{C}_{y}^{\left(q\right)}\right)$
is the $q$-th ($q=1,...,Q$) \emph{control point} of the spline profile
whose coordinates are (Fig. 2)

\begin{equation}
\left\{ \begin{array}{l}
\mathcal{C}_{x}^{\left(q\right)}=x_{\Omega}+\rho^{\left(q\right)}\times\cos\left(\left(q-1\right)\frac{2\pi}{Q}\right)\\
\mathcal{C}_{y}^{\left(q\right)}=y_{\Omega}+\rho^{\left(q\right)}\times\sin\left(\left(q-1\right)\frac{2\pi}{Q}\right)\end{array}\right.,\label{eq:}\end{equation}
while $\rho^{\left(q\right)}$ ($\rho^{\left(q\right)}>0$) is the
radial distance of the $q$-th control point from the barycenter of
$\Omega$, $\left(x_{\Omega},\, y_{\Omega}\right)$, (Fig. 2)\begin{equation}
\rho^{\left(q\right)}=\sqrt{\left(x_{\Omega}-\mathcal{C}_{x}^{\left(q\right)}\right)^{2}+\left(y_{\Omega}-\mathcal{C}_{y}^{\left(q\right)}\right)^{2}}.\label{eq:}\end{equation}
Moreover, $\mathcal{V}^{\left(q\right)}=\left(\mathcal{V}_{x}^{\left(q\right)},\,\mathcal{V}_{y}^{\left(q\right)}\right)$
is the $q$-th ($q=1,...,Q$) spline \emph{virtual point}\begin{equation}
\left\{ \begin{array}{l}
\mathcal{V}_{x}^{\left(q\right)}=\frac{\mathcal{C}_{x}^{\left(q\right)}+\mathcal{C}_{x}^{\left(q+1\right)}}{2}\\
\mathcal{V}_{y}^{\left(q\right)}=\frac{\mathcal{C}_{y}^{\left(q\right)}+\mathcal{C}_{y}^{\left(q+1\right)}}{2}\end{array}\right.\label{eq:}\end{equation}
and the condition $\mathcal{C}_{x}^{\left(Q+1\right)}=\mathcal{C}_{x}^{\left(1\right)}$
and $\mathcal{C}_{y}^{\left(Q+1\right)}=\mathcal{C}_{y}^{\left(1\right)}$
holds true so that $\partial\Omega\left(x,\, y\right)$ is a simply-connected
curve (Fig. 2).

\noindent Owing to such a parametric description of the scatterer
support $\Omega$, the \emph{ISP} solution is coded into the following
$K=\left(4+Q\right)$ \emph{SbD}-\emph{DoF}s\begin{equation}
\underline{\xi}=\left\{ x_{\Omega},\, y_{\Omega},\,\Re\left(\tau_{\Omega}\right),\,\Im\left(\tau_{\Omega}\right),\,\underline{\rho}\right\} \label{eq:SbD-DoFs}\end{equation}
where $\underline{\rho}=\left\{ \rho^{\left(q\right)};\, q=1,\,...,\, Q\right\} $.
It is worth highlighting that such a parametric modeling yields also,
as a by-product, a profitable regularization of the \emph{ISP} by
enforcing a physical \emph{a-priori} knowledge on the unknown target.

\noindent As for the second task of the \emph{PF} block, the \emph{ISP}
is re-formulated into an optimization/minimization one\begin{equation}
\underline{\xi}^{\left(opt\right)}=\arg\left\{ \min_{\underline{\xi}}\left[\Phi\left(\underline{\xi}\right)\right]\right\} \label{eq:SbD-Solution}\end{equation}
whose solution is the global minimum of the cost function $\Phi\left(\underline{\xi}\right)$
set here to the normalized mismatch between measured, \{$\mathcal{S}^{\left(v\right)}\left(x_{m},\, y_{m}\right)$;
$m=1,...,M$\}, and estimated, \{$\widetilde{\mathcal{S}}^{\left(v\right)}\left(\left.x_{m},\, y_{m}\right|\underline{\xi}\right)$;
$m=1,...,M$\}, scattered data\begin{equation}
\Phi\left(\underline{\xi}\right)=\frac{\sum_{v=1}^{V}\sum_{m=1}^{M}\left|\mathcal{S}^{\left(v\right)}\left(x_{m},\, y_{m}\right)-\widetilde{\mathcal{S}}^{\left(v\right)}\left(\left.x_{m},\, y_{m}\right|\underline{\xi}\right)\right|^{2}}{\sum_{v=1}^{V}\sum_{m=1}^{M}\left|\mathcal{S}^{\left(v\right)}\left(x_{m},\, y_{m}\right)\right|^{2}}.\label{eq:cost-function}\end{equation}
In (\ref{eq:cost-function}), $\underline{\mathcal{\widetilde{\mathcal{S}}}}^{\left(v\right)}\left(\underline{\xi}\right)=\left\{ \widetilde{\mathcal{S}}^{\left(v\right)}\left(\left.x_{m},\, y_{m}\right|\underline{\xi}\right);\, m=1,\,...,\, M\right\} $
is the set of field data scattered in the observation domain $O$
from the scatterer, coded by $\underline{\xi}$, when illuminated
by the $v$-th ($v=1,\,...,\, V$) incident field, $\mathcal{I}^{\left(v\right)}$.

\subsection{Data Computation (\emph{DC}) \label{sub:DC}}

In order to compute $\underline{\mathcal{\widetilde{\mathcal{S}}}}^{\left(v\right)}\left(\underline{\xi}\right)$
($v=1,...,V$), let us remember that it is the scattered data vector
radiated by the $v$-th ($v=1,...,V$) equivalent current distribution
$\underline{\mathcal{J}}^{\left(v\right)}\left(\underline{\xi}\right)$
according to (\ref{eq:scattered-field-computation}). Thus, $\underline{\xi}$
is first mapped into the corresponding $v$-th ($v=1,...,V$) equivalent
current vector $\underline{\mathcal{J}}^{\left(v\right)}\left(\underline{\xi}\right)$
whose generic $n$-th ($n=1,...,N$) entry is defined as\begin{equation}
\mathcal{J}_{n}^{\left(v\right)}\left(\underline{\xi}\right)\triangleq\mathcal{T}^{\left(v\right)}\left(\left.x_{n},\, y_{n}\right|\underline{\xi}\right)\tau\left(\left.x_{n},\, y_{n}\right|\underline{\xi}\right).\label{ccu}\end{equation}
Because of the spline-based representation of the unknown scattering
profile of support $\Omega$, the relation between the $n$-th ($n=1,...,N$)
contrast value $\tau\left(\left.x_{n},\, y_{n}\right|\underline{\xi}\right)$
and $\underline{\xi}$ is based on the Jordan curve theorem \cite{Shimrat 1962}
that allows one to state whether a point $\left(x_{n},\, y_{n}\right)$
belongs or not to the scatterer region $\Omega$ enclosed by the spline
contour $\partial\Omega\left(\left.x,\, y\right|\underline{\xi}\right)$\begin{equation}
\tau\left(\left.x_{n},\, y_{n}\right|\underline{\xi}\right)=\left\{ \begin{array}{cc}
\tau_{\Omega} & \mathrm{if}\,\,\left(x_{n},\, y_{n}\right)\in\partial\Omega\left(\left.x,\, y\right|\underline{\xi}\right)\\
0 & otherwise\end{array}\right..\label{mapping spline-coded sol TAU}\end{equation}
On the other hand, the $n$-th ($n=1,...,N$) sample of the $v$-th
($v=1,...,V$) total field $\mathcal{T}^{\left(v\right)}\left(\left.x_{n},\, y_{n}\right|\underline{\xi}\right)$
is numerically derived from the \emph{MoM}-discretized version of
(\ref{eq:state-equation})\begin{equation}
\underline{\mathcal{T}}^{\left(v\right)}\left(\underline{\xi}\right)=\left[\underline{\underline{I}}-\underline{\underline{G}}_{D}\,\underline{\underline{\tau}}\left(\underline{\xi}\right)\right]^{-1}\underline{\mathcal{I}}^{\left(v\right)}\label{eq: etot vector (state inverse)}\end{equation}
where $\underline{\mathcal{I}}^{\left(v\right)}=\left\{ \mathcal{I}^{\left(v\right)}\left(x_{n},\, y_{n}\right);\, n=1,\,...,\, N\right\} $,
$\underline{\underline{\tau}}\left(\underline{\xi}\right)=\mathrm{diag}\left\{ \tau\left(\left.x_{n},\, y_{n}\right|\underline{\xi}\right);\, n=1,...,N\right\} $,
$\underline{\underline{I}}$ is the identity matrix, and $\underline{\underline{G}}_{D}$
is the $\left(N\times N\right)$ internal Green's operator whose $\left(n,\, p\right)$-th
($n$, $p=1,...,N$) entry is equal to $\left.\underline{\underline{G}}_{D}\right\rfloor _{np}=j\frac{k_{0}^{2}}{4}\int_{D_{p}}\mathcal{H}_{0}^{\left(1\right)}\left(k_{0}\rho_{n}\right)dx'dy'$.

\noindent Once $\underline{\mathcal{J}}^{\left(v\right)}\left(\underline{\xi}\right)$
($v=1,...,V$) has been obtained by substituting (\ref{eq: etot vector (state inverse)})
and (\ref{mapping spline-coded sol TAU}) in (\ref{ccu}), the corresponding
scattered field vector $\underline{\mathcal{\widetilde{\mathcal{S}}}}^{\left(v\right)}\left(\underline{\xi}\right)$
($v=1,...,V$) is then computed through (\ref{eq:scattered-field-computation}).

\subsection{Cost Function Evaluation (\emph{CFE}) \label{sub:Cost-Function-Evaluation}}

To efficiently compute the data mismatch cost function (\ref{eq:cost-function}),
by avoiding the time-consuming call to the forward (\emph{FW}) solver
in (\ref{eq:data-equation}), the \emph{LBE} paradigm \cite{Massa 2017}
is exploited to build a fast yet accurate surrogate of $\Phi\left(\underline{\xi}\right)$,
$\widehat{\Phi}\left(\underline{\xi}\right)$, which is adaptively
{}``reinforced'' at each $i$-th ($i=1,...,I_{SbD}$) iteration
of the optimization process performed in the \emph{SSE} block (Sect.
\ref{sub:Solution-Space-Exploration}). More specifically, a Gaussian
Process (\emph{GP})-based \emph{DT} \cite{Forrester 2008}\cite{Jones 1998}
of $\Phi\left(\underline{\xi}\right)$ is built at each $i$-th ($i=1,...,I_{SbD}$)
iteration of the optimization, $\widehat{\Phi}_{i}\left(\underline{\xi}\right)$,
from a \emph{training set} of $S_{i}$ known input/output (\emph{I/O})
pairs according to the following {}``three-step'' strategy leveraging
on the interconnections among all \emph{SbD} functional blocks (Fig.
1):

\begin{itemize}
\item \emph{Input-Space Reduction} - Input the minimum set of $K$ highly-informative
\emph{SbD-DoF}s (\ref{eq:SbD-DoFs}), which univocally describe the
\emph{ISP} solution $\underline{\xi}$, from the \emph{PF} block (Sect.
\ref{sub:Problem-Formulation});
\item \emph{Input-Space Representative Sampling} - Build the smallest size
$i$-th ($i=1,...,I_{SbD}$) training set \begin{equation}
\Lambda_{i}=\left\{ \left[\underline{\xi}^{\left(s\right)},\,\Phi\left(\underline{\xi}^{\left(s\right)}\right)\right];\, s=1,...,S_{i}\right\} \label{eq:}\end{equation}
 of $S_{i}$ \emph{I/O} pairs to suitably represent the $K$-dimensional
input space. It means that for each $s$-th ($s=1,\,...,\, S_{i}$)
sample, $\underline{\xi}^{\left(s\right)}$, decoded with the \emph{DC}
block, $\Phi\left(\underline{\xi}^{\left(s\right)}\right)$ is computed
with a \emph{FW} solver. At the initialization ($i=0$), the $\left.S_{i}\right\rfloor _{i=0}$
samples are selected according to the \emph{Latin Hypercube Sampling}
(\emph{LHS}) strategy (see \emph{Appendix I}) to uniformly explore
the \emph{SbD-DoF}s thanks to its {}``input space filling'' property
\cite{Garud 2017}, while new \emph{I/O} pairs are adaptively selected
in the \emph{SSE} block and added to the training set of the previous
iteration, $\Lambda_{i-1}$, to build the $i$-th ($i=1,...,I_{SbD}$)
training set, $\Lambda_{i}$, otherwise (i.e., $1\le i\le I_{SbD}$); 
\item \emph{DT Generation} - Starting from the $i$-th ($i=1,...,I_{SbD}$)
training set, $\Lambda_{i}$, define the $i$-th ($i=1,...,I_{SbD}$)
\emph{GP} predictor \cite{Forrester 2008}\cite{Jones 1998} of $\Phi\left(\underline{\xi}\right)$,
$\widehat{\Phi}_{i}\left(\underline{\xi}\right)$, as follows\begin{equation}
\widehat{\Phi}_{i}\left(\underline{\xi}\right)=\chi_{i}+\left[\underline{r}_{i}\left(\underline{\xi}\right)\right]^{T}\underline{\underline{R}}_{i}^{-1}\left[\underline{\Phi}_{i}-\underline{1}_{i}\mathbf{\chi}_{i}\right],\label{eq:GP-prediction}\end{equation}
where $\chi_{i}$ is a scalar term given by\begin{equation}
\chi_{i}\triangleq\frac{\underline{1}_{i}^{T}\underline{\underline{R}}_{i}^{-1}\underline{\Phi}_{i}}{\underline{1}_{i}^{T}\underline{\underline{R}}_{i}^{-1}\underline{1}_{i}},\label{eq:}\end{equation}
$.^{T}$ being the transpose operator, $\underline{r}_{i}\left(\underline{\xi}\right)$
is a $\left(S_{i}\times1\right)$-dimensional vector whose $s$-th
entry is equal to\begin{equation}
r_{i}\left(\underline{\xi}^{\left(s\right)},\underline{\xi}\right)=\prod_{k=1}^{K}\exp\left(-\gamma_{i,k}\left|\xi_{k}^{\left(s\right)}-\xi_{k}\right|^{\beta_{i,k}}\right),\label{eq:correlation-vector}\end{equation}
 $\underline{\underline{R}}_{i}$ is the $\left(S_{i}\times S_{i}\right)$
correlation matrix of $\Lambda$ whose $\left(s,u\right)$-th ($s$,
$u=1,...,S_{i}$) element is $r_{i}\left(\underline{\xi}^{\left(s\right)},\underline{\xi}^{\left(u\right)}\right)$
(\ref{eq:correlation-vector}), $\underline{\Phi}_{i}=\left[\Phi\left(\underline{\xi}^{\left(s\right)}\right);\, s=1,...,S_{i}\right]^{T}$,
and $\underline{1}_{i}$ is a $\left(S_{i}\times1\right)$ unitary
column vector. Moreover, $\gamma_{i,k}$ and $\beta_{i,k}$ are the
$k$-th ($k=1,...,K$) elements of the \emph{GP} hyper-parameter vectors
$\underline{\gamma}_{i}$ and $\underline{\beta}_{i}$, respectively,
which are yielded from the maximization of the \emph{concentrated
log-likelihood} function \cite{Forrester 2008}\begin{equation}
\Gamma\left(\underline{\gamma}_{i},\,\underline{\beta}_{i}\right)=-\frac{1}{2}\left\{ S_{i}\times ln\left(\nu_{i}^{2}\right)+ln\left[\det\left(\underline{\underline{R}}_{i}\right)\right]\right\} \label{eq:}\end{equation}
where\begin{equation}
\nu_{i}\triangleq\frac{1}{S_{i}}\left[\left(\underline{\Phi}_{i}-\underline{1}_{i}\chi_{i}\right)^{T}\underline{\underline{R}}_{i}^{-1}\left(\underline{\Phi}_{i}-\underline{1}_{i}\chi_{i}\right)\right],\label{eq:_GP.6}\end{equation}
$ln\left(\,.\,\right)$ and $\det\left(\,.\,\right)$ being the natural
logarithm and the matrix determinant operators.
\end{itemize}
It is worth noticing that the choice of the \emph{GP} to build the
\emph{DT} of $\Phi\left(\underline{\xi}\right)$, unlike other regression
strategies such as, for instance, the \emph{Support Vector Regression}
(\emph{SVR}) \cite{Massa 2017}, ensures an exact prediction of the
actual value of the cost function when a trial solution, $\underline{\xi}$,
coincides with a training sample, $\underline{\xi}^{\left(s\right)}$
(i.e., $\widehat{\Phi}_{i}\left(\underline{\xi}^{\left(s\right)}\right)=\Phi\left(\underline{\xi}^{\left(s\right)}\right)$;
$s=1,...,S_{i}$). Moreover, it must be pointed out that the definition
of the $i$-th ($i=1,...,I_{SbD}$) \emph{GP} surrogate model, $\widehat{\Phi}_{i}\left(\underline{\xi}\right)$
in (\ref{eq:GP-prediction}) is based on the assumption that the actual
value of the cost function, $\Phi\left(\underline{\xi}\right)$, is
the realization of a normally-distributed random variable with average
value $\widehat{\Phi}_{i}\left(\underline{\xi}\right)$ and variance
\cite{Forrester 2008} equal to\begin{equation}
\delta_{i}^{2}\left(\underline{\xi}\right)=\nu_{i}^{2}\left[1-\left[\underline{r}_{i}\left(\underline{\xi}\right)\right]^{T}\underline{\underline{R}}_{i}^{-1}\underline{r}_{i}\left(\underline{\xi}\right)+\frac{\left[1-\underline{1}_{i}^{T}\underline{\underline{R}}_{i}^{-1}\underline{r}_{i}\left(\underline{\xi}\right)\right]^{2}}{\underline{1}_{i}^{T}\underline{\underline{R}}_{i}^{-1}\underline{1}_{i}}\right].\label{eq:Uncertainty}\end{equation}
This latter quantity provides an estimate of the reliability of the
\emph{GP}-based \emph{DT}, greater values of $\delta_{i}^{2}\left(\underline{\xi}\right)$
corresponding to a lower {}``reliability'' of the associated prediction
$\widehat{\Phi}_{i}\left(\underline{\xi}\right)$. Indeed, the value
of $\delta_{i}^{2}\left(\underline{\xi}\right)$ depends on $\underline{r}_{i}\left(\underline{\xi}\right)$
(\ref{eq:Uncertainty}), which in turn is related to the $\left(\underline{\gamma}_{i},\,\underline{\beta}_{i}\right)$-weighted
distance between $\underline{\xi}$ and the $s$-th ($s=1,...,S_{i}$)
training sample, $\underline{\xi}^{\left(s\right)}$ (\ref{eq:correlation-vector}).
Thus, if $\underline{\xi}$ is very far from all the $S_{i}$ training
samples, $\left\{ \underline{\xi}^{\left(s\right)};\, s=1,...,S_{i}\right\} $,
then $\underline{r}_{i}\left(\underline{\xi}\right)\rightarrow\underline{0}$
and the uncertainty reaches its maximum (i.e., $\delta_{i}^{2}\left(\underline{\xi}\right)\rightarrow\nu_{i}^{2}$).
On the contrary, the uncertainty is minimal in correspondence of the
training samples {[}i.e., $\delta_{i}^{2}\left(\underline{\xi}^{\left(s\right)}\right)=0$
($s=1,...,S_{i}$){]} since $\left[\underline{r}_{i}\left(\underline{\xi}^{\left(s\right)}\right)\right]^{T}\underline{\underline{R}}_{i}^{-1}\underline{r}_{i}\left(\underline{\xi}^{\left(s\right)}\right)=\underline{1}_{i}^{T}\underline{\underline{R}}_{i}^{-1}\underline{r}_{i}\left(\underline{\xi}^{\left(s\right)}\right)=1$.
Finally, let us consider that, according to the \emph{GP} theory \cite{Jones 1998},
the actual value of the cost function $\Phi\left(\underline{\xi}\right)$
fulfils at least to $95\%$ probability \cite{Jones 1998} the following
condition\begin{equation}
\mathcal{L}_{i}\left(\underline{\xi}\right)\leq\Phi\left(\underline{\xi}\right)\leq\mathcal{U}_{i}\left(\underline{\xi}\right),\label{eq:}\end{equation}
$\mathcal{L}_{i}\left(\underline{\xi}\right)$ and $\mathcal{U}_{i}\left(\underline{\xi}\right)$
being the lower and the upper {}``confidence bounds'', respectively,
defined as\begin{equation}
\left\{ \begin{array}{l}
\mathcal{L}_{i}\left(\underline{\xi}\right)=\widehat{\Phi}_{i}\left(\underline{\xi}\right)-2\delta_{i}\left(\underline{\xi}\right)\\
\mathcal{U}_{i}\left(\underline{\xi}\right)=\widehat{\Phi}_{i}\left(\underline{\xi}\right)+2\delta_{i}\left(\underline{\xi}\right)\end{array}\right.\label{eq:LCB-UCB}\end{equation}
so that $\mathcal{L}_{i}\left(\underline{\xi}^{\left(s\right)}\right)=\mathcal{U}_{i}\left(\underline{\xi}^{\left(s\right)}\right)=\widehat{\Phi}_{i}\left(\underline{\xi}^{\left(s\right)}\right)=\Phi\left(\underline{\xi}^{\left(s\right)}\right)$
($s=1,...,S_{i}$).

\subsection{Solution Space Exploration (\emph{SSE}) \label{sub:Solution-Space-Exploration}}

To explore in a smart way the $K$-dimensional \emph{SbD} solution
space for solving the non-linear \emph{ISP}, nature-inspired \emph{EA}s
are the most suitable candidates to effectively implement such a task
without requiring, unlike \emph{DO}s, the differentiation of the data
mismatch cost function (\ref{eq:cost-function}) \cite{Rocca 2009}.
However, a {}``bare'' integration of an \emph{EA}-\emph{GO} with
a forward solver (\emph{FW}) would imply an overall inversion time
equal to\begin{equation}
\Delta t_{GO}=\left(P\times I_{GO}\right)\times\Delta t_{FW},\label{eq: Delta_t_GO}\end{equation}
$P$ and $\Delta t_{FW}$ being the number of trial solutions evolved
through $I_{GO}$ iterations and the time of a single full-wave evaluation
of (\ref{eq:cost-function}), which clearly becomes unpractical in
many applicative scenarios requiring a fast inversion. If a significant
reduction of $P$ can be yielded with a minimum-dimensionality coding
of the unknown scattering profile (e.g., the spline-based strategy
in Sect. \ref{sub:Problem-Formulation}), it is not enough towards
a computationally-competitive global inversion/optimization. In order
to break down the computational burden required by the iterated (multi-agent)
evaluation of (\ref{eq:cost-function}) to comply with (\ref{eq:time-goal})
by reducing $\Delta t_{GO}$ (\ref{eq: Delta_t_GO}), there are two
different strategies. The former is that of minimizing the number
of iterations of the \emph{EA} to reach the global optimum $\underline{\xi}^{\left(opt\right)}$,
$I_{GO}$. Towards this end, it is mandatory to choose an \emph{EA}
that provides a proper balance between \emph{exploration} and \emph{exploitation}
to enable {}``\emph{hill-climbing}'' features for effectively escaping
from local minima/false solutions as well as to guarantee a quick
convergence towards the attraction basin of the global minimum of
the cost function $\Phi\left(\underline{\xi}\right)$. Accordingly,
the Particle Swarm Optimization (\emph{PSO}) algorithm \cite{Rocca 2009}
is chosen as a robust and effective evolutionary strategy particularly
suitable for the exploration of the real-valued solution space of
the \emph{SbD-DoF}s (\ref{eq:SbD-DoFs}). During $I_{SbD}$ iterations,
the \emph{PSO} processes a swarm of $P$ particles/agents, $\mathcal{A}=\left\{ A^{\left(p\right)};\, p=1,...,P\right\} $,
by changing their velocities, $\mathcal{V}=\left\{ \underline{\iota}^{\left(p\right)};\, p=1,...,P\right\} $,
to evolve their positions in the solution space, $\mathcal{P}=\left\{ \underline{\xi}^{\left(p\right)};\, p=1,...,P\right\} $,
until reaching the global optimum (i.e., $\underline{\xi}^{\left(opt\right)}=\arg\left\{ \min_{\underline{\xi}}\left[\Phi\left(\underline{\xi}\right)\right]\right\} $).

\noindent The second method to shorten (\ref{eq: Delta_t_GO}) is
that of building a surrogate model in the \emph{CFE} block (see Sect.
\ref{sub:Cost-Function-Evaluation}) to replace the \emph{FW} solver
during the optimization so that $\Delta t_{DT}^{test}\ll\Delta t_{FW}$.
However, the definition of a globally-accurate predictor would generally
require a huge number of training samples $S$ {[} $S\gg\left(P\times I\right)${]},
which not linearly depends on the number of scatterer descriptors,
$K$, because of the so-called {}``curse-of-dimensionality'' \cite{Massa 2021}.
On the other hand, it is worth to consider that the \emph{DT} is required
to predict the value of the cost function $\Phi\left(\underline{\xi}\right)$
(\ref{eq:cost-function}) for guiding the \emph{GO} search throughout
the solution space with an accuracy adaptively enhanced and very high
only in the attraction basin (i.e., in the proximity) of the global
optimum. Owing to such considerations, a {}``collaborative'' framework
is implemented between the \emph{PSO}, which is responsible of sampling
the solution space with the swarm $\mathcal{A}$ of $P$ trial agents,
and the \emph{DT} model based on the \emph{GP} regression strategy
\cite{Forrester 2008}\cite{Jones 1998} that gives not only a prediction
of the cost function associated to each trial solution, $\widehat{\Phi}\left(\underline{\xi}^{\left(p\right)}\right)$
($p=1,...,P$), but also an estimate of its {}``degree of reliability'',
$\delta\left(\underline{\xi}^{\left(p\right)}\right)$. This latter
is an additional information to be profitably exploited for identifying
{}``promising'' solutions for which the cost function (\ref{eq:cost-function})
is expected to be lower than any previously-explored solution set.
Moreover, the value $\delta$ can be used as a threshold for triggering
adaptive refinements/reinforcements, obtained by simulating selected
particles to enhance the accuracy only {}``where needed'', of the
predictor during the optimization loop. The resulting \emph{SSE} block
then works as follows:

\begin{enumerate}
\item \emph{Initialization} ($i=0$) - With the \emph{CFE} block (Sect.
\ref{sub:Cost-Function-Evaluation}), build the initial training set
of $S_{0}$ \emph{I/O} pairs, $\Lambda_{0}=\left\{ \left[\underline{\xi}^{\left(s\right)},\,\Phi\left(\underline{\xi}^{\left(s\right)}\right)\right];\, s=1,...,S_{0}\right\} $,
to train the initial \emph{GP} predictor $\widehat{\Phi}_{i}\left(\underline{\xi}\right)$.
Randomly initialize the positions of the swarm $\mathcal{A}_{0}$
of $P$ particles, $\mathcal{P}_{0}=\left\{ \underline{\xi}_{0}^{\left(p\right)};\, p=1,...,P\right\} $,
with random velocities, $\mathcal{V}_{0}=\left\{ \underline{\iota}_{0}^{\left(p\right)};\, p=1,...,P\right\} $,
and set the personal best position of each $p$-th ($p=1,...,P$)
particle to the initial one (i.e., $\underline{\zeta}_{0}^{\left(p\right)}=\underline{\xi}_{0}^{\left(p\right)}$);
\item \emph{SbD Optimization Loop} ($i=1,...,I_{SbD}$)

\begin{enumerate}
\item \emph{Cost Function Prediction} - For each $p$-th ($p=1,\,...,\, P$)
particle of the current $i$-th swarm, $\mathcal{A}_{i}$, predict
the values of $\Phi\left(\underline{\xi}_{i}^{\left(p\right)}\right)$,
$\mathcal{L}\left(\underline{\xi}_{i}^{\left(p\right)}\right)$, and
$\mathcal{U}\left(\underline{\xi}_{i}^{\left(p\right)}\right)$, with
the $i$-th \emph{DT} $\widehat{\Phi}_{i}\left(\underline{\xi}\right)$;
\item \emph{Particles Ranking} - Determine the {}``best promising'' (\emph{BP})
position of a particle of $\mathcal{A}_{i}$\begin{equation}
\underline{\xi}_{i}^{\left(BP\right)}=\arg\left\{ \min_{p=1,...,P}\left[\mathcal{L}_{i}\left(\underline{\xi}_{i}^{\left(p\right)}\right)\right]\right\} ;\label{eq:}\end{equation}

\item \emph{DT Adaptive Updating} - If $\mathcal{L}\left(\underline{\xi}_{i}^{\left(BP\right)}\right)<\min_{s=1,...,S_{i}}\left[\Phi\left(\underline{\xi}^{\left(s\right)}\right)\right]$
perform the following operations, otherwise set $S_{i}\leftarrow S_{i-1}$
and $\Lambda_{i}\leftarrow\Lambda_{i-1}$ and jump to Step 2(\emph{d}):

\begin{enumerate}
\item Exploit the \emph{DC} bock (Sect. \ref{sub:DC}) to derive the $v$-th
($v=1,\,...,\, V$) induced equivalent current, $\underline{\mathcal{J}}^{\left(v\right)}\left(\underline{\xi}_{i}^{\left(BP\right)}\right)$
from $\underline{\xi}_{i}^{\left(BP\right)}$ , then compute the corresponding
scattered field, $\underline{\mathcal{\widetilde{\mathcal{S}}}}^{\left(v\right)}\left(\underline{\xi}_{i}^{\left(BP\right)}\right)$;
\item Compute $\Phi\left(\underline{\xi}_{i}^{\left(BP\right)}\right)$
with (\ref{eq:cost-function});
\item Update the training set by adding the \emph{BP} training set sample,
$\Lambda_{i}=\Lambda_{i-1}\cup\left\{ \underline{\xi}_{i}^{\left(BP\right)},\,\Phi\left(\underline{\xi}_{i}^{\left(BP\right)}\right)\right\} $,
and let $S_{i}\leftarrow\left(S_{i-1}+1\right)$ ;
\item Use the \emph{CFE} block (Sect. \ref{sub:Cost-Function-Evaluation})
to re-train the \emph{GP} predictor using the updated/reinforced training
information within $\Lambda_{i}$.
\end{enumerate}
\item \emph{Personal Best Updating} - Update the personal best position
of each $p$-th ($p=1,...,P$) particle, $\underline{\zeta}_{i}^{\left(p\right)}=\left\{ \zeta_{i,k}^{\left(p\right)};\, k=1,...,K\right\} $,
according to the \emph{SbD}-updating rules in Fig. 3(\emph{a});
\item \emph{Global Best Updating} - Update the global best, $\underline{\psi}_{i}=\left\{ \psi_{i,k};\, k=1,...,K\right\} $
according to the work-flow in Fig. 3(\emph{b}); 
\item \emph{Convergence Check} - Stop the optimization if $i=I_{SbD}$ and
output the \emph{SbD} solution, set to the current global best swarm
position, $\underline{\xi}^{\left(SbD\right)}=\underline{\psi}_{i=I_{SbD}}$,
along with its pixel-wise representation $\underline{\tau}^{\left(SbD\right)}=\left\{ \tau^{\left(SbD\right)}\left(x_{n},\, y_{n}\right);\, n=1,\,...,\, N\right\} $
yielded from the \emph{DC} bock (Fig. 1 - Sect. \ref{sub:DC}). Otherwise,
proceed to Step 2(\emph{g});
\item \emph{Velocities Updating} - Update the velocity vector ($\mathcal{V}_{i}\to\mathcal{V}_{i+1}$)
by computing the $k$-th ($k=1,...,K$) component of the velocity
of the $p$-th ($p=1,...,P$) particle of the swarm $\mathcal{A}_{i+1}$
according to the \emph{PSO} mechanism\begin{equation}
\iota_{i+1,k}^{\left(p\right)}=w\iota_{i,k}^{\left(p\right)}+\mathcal{\ell}_{1}\varsigma_{1}\left(\xi_{i,k}^{\left(p\right)}-\zeta_{i,k}^{\left(p\right)}\right)+\ell_{2}\varsigma_{2}\left(\xi_{i,k}^{\left(p\right)}-\psi_{i,k}\right)\label{eq:}\end{equation}
where $\varsigma_{1}$ and $\varsigma_{2}$ are real random values
within the interval $\left[0,\,1\right]$, the acceleration coefficients
$\ell_{1}$ and $\ell_{2}$ are positive user-defined real values,
and $w$ is the constant inertial weight;
\item \emph{Swarm Updating} - Update the position vector ($\mathcal{P}_{i}\to\mathcal{P}_{i+1}$)
by adding to the $k$-th ($k=1,...,K$) component of the current position
of the $p$-th ($p=1,...,P$) particle of the swarm $\mathcal{A}_{i+1}$
the corresponding term of the velocity vector $\mathcal{V}_{i+1}$\begin{equation}
\xi_{i+1,k}^{\left(p\right)}=\xi_{i,k}^{\left(p\right)}+\iota_{i+1,k}^{\left(p\right)},\label{eq:}\end{equation}
then let $i\leftarrow\left(i+1\right)$ and go to Step 2(\emph{a}).
\end{enumerate}
\end{enumerate}
It is worth pointing out that the \emph{SSE} block implements a novel
{}``\emph{time-constrained reinforced} \emph{PSO}'' strategy to
allow the user to \emph{a-priori} fulfil the \emph{CPU}-time target
(\ref{eq:time-goal}) by properly setting the size $S_{0}$ of the
initial training set, $\Lambda_{0}$, and the maximum number of \emph{DT}
{}``reinforcements'', $I_{SbD}$, performed during the global minimization
of (\ref{eq:cost-function}). Indeed, the total number of calls to
the \emph{FW} solver \emph{}during a \emph{SbD} inversion, thus the
\emph{SbD} time cost, as well, is upper-bounded to $S=\left(S_{0}+I_{SbD}\right)$
\footnote{\noindent If $S\ll10^{3}$ (Sect. \ref{sec:Performance-Assessment}),
the overall time required to train ($\Delta t_{DT}^{train}$) and
to test ($\Delta t_{DT}^{test}$) the \emph{DT} model can be neglected
since $I_{SbD}\times\Delta t_{DT}^{train}\ll\Delta t_{FW}$ and $P\times I_{SbD}\times\Delta t_{DT}^{test}\ll\Delta t_{FW}$
\cite{Massa 2021}.%
}so that a \emph{SbD} inversion turns out to be computationally advantageous
with respect to a standard \emph{GO} solution when $S\ll\left(P\times I_{GO}\right)$,
with a time saving equal to

\begin{equation}
\Delta t_{sav}\simeq\left(\frac{\left(P\times I_{GO}\right)-S}{\left(P\times I_{GO}\right)}\right)=\left(\frac{\left(P\times I_{GO}\right)-\left(S_{0}+I_{SbD}\right)}{\left(P\times I_{GO}\right)}\right).\label{eq:time-saving}\end{equation}

\section{Performance Assessment\label{sec:Performance-Assessment}}

\noindent This section is aimed at presenting a set of representative
numerical and experimental results drawn from an extensive validation
of the proposed \emph{SbD}-based inversion method. Unless stated otherwise,
a square investigation domain $D$ of side $L_{D}=2\times\lambda$
has been probed by $V=18$ incident plane waves impinging from the
$V$ angular directions \{$\varphi_{v}\triangleq2\pi\frac{\left(v-1\right)}{V}$;
$v=1,...,V$\}. The scattered field samples have been collected at
$M=18$ probing locations uniformly distributed on a circular observation
domain $O$ of radius $\rho_{O}=3\times\lambda$. As for the generation
of the synthetic scattered field data, the \emph{MoM} solution of
the \emph{FW} problem (\ref{eq:state-equation})(\ref{eq:data-equation})
has been performed by partitioning the investigation domain into $N_{FW}=40\times40$
square sub-domains, while $N=20\times20$ pixel bases have been adopted
in the inversion process to avoid the \emph{inverse crime} (see \cite{Chen 2018}
p. 174). Moreover, an additive Gaussian noise has been added to the
synthetically-generated data samples to test the robustness of the
inversion to different signal-to-noise ratios (\emph{SNR}s). Furthermore,
owing to the stochastic nature of the \emph{SbD}-based approach, a
set of $\Upsilon=50$ random executions has been run for each inversion
dataset to ensure the statistic meaningfulness of the results.

\noindent Concerning the imaging results/performance and besides the
pictorial representation of the reconstruction in terms of color-maps
of the dielectric profile of \emph{D}, the accuracy of the data inversion
is quantified by the error index\begin{equation}
\Xi=\frac{1}{N}\sum_{n=1}^{N}\frac{\left|\tau\left(x_{n},y_{n}\right)-\widetilde{\tau}\left(x_{n},y_{n}\right)\right|}{\tau\left(x_{n},y_{n}\right)+1},\label{eq:error_cal}\end{equation}
where $\tau\left(x_{n},y_{n}\right)$ and $\widetilde{\tau}\left(x_{n},y_{n}\right)$
stand for the actual and the retrieved contrast value of the $n$-th
($n=1,...,N$) pixel $D_{n}$ ($D_{n}\in D$), respectively.

\noindent The first test case deals with the noiseless reconstruction
of the scatting profile in Fig. 4(\emph{a}) having contrast $\tau_{\Omega}=4.0$.
The \emph{SbD}-based inversion has been carried out by considering
a spline description of the scatter with $Q=4$ control points ($\Rightarrow K=8$
- Tab. I) and choosing, according to the guidelines in \cite{Rocca 2009}
a swarm size of $P=10$ particles, a constant inertial weight equal
to $w=0.4$, and acceleration coefficients with values $\ell_{1}=\ell_{2}=2.0$.
To investigate on the dependence of the prediction accuracy of the
\emph{DT} of the \emph{FW} solver on the size of the initial training
set $S_{0}$, a set of experiments has been run by varying the $S_{0}/K$
ratio and the adaptive generation of a fixed amount of $I_{SbD}=100$
additional training samples according to the \emph{SSE} procedure
(Sect. \ref{sub:Solution-Space-Exploration}). For each test, the
normalized prediction error $\eta$\begin{equation}
\eta\left(S_{0}/K\right)\triangleq\frac{1}{P}\sum_{p=1}^{P}\frac{\left|\Phi\left(\underline{\xi}_{i=I_{SbD}}^{\left(p\right)}\right)-\widehat{\Phi}\left(\left.\underline{\xi}_{i=I_{SbD}}^{\left(p\right)}\right|S_{0}\right)\right|}{\Phi\left(\underline{\xi}_{i=I_{SbD}}^{\left(p\right)}\right)}\label{eq:}\end{equation}
of the \emph{GP} surrogate trained with $S=\left(S_{0}+I_{SbD}\right)$
samples has been evaluated along with the time saving (\ref{eq:time-saving})%
\footnote{\noindent For fair comparisons, all inversions have been executed
using non-optimized FORTRAN codes on a standard laptop equipped with
$16$ {[}GB{]} of RAM memory and an Intel(R) Core(TM) i5-8250U CPU
@ 1.60 {[}GHz{]}.%
} with respect to a {}``bare'' \emph{GO} based on the standard \emph{PSO}
(i.e., a \emph{PSO}-based inversion method that exploits the same
spline-based coding, but that computes the cost function of each trial
solution by solving the corresponding \emph{FW} problem with the \emph{MoM})
run with equal swarm size, $P$, for the maximum number of iterations
($I_{GO}=I_{SbD}$). As expected, the plot of $\eta$ and $\Delta t_{sav}$
versus $S_{0}/K$ (Fig. 5) indicates that the prediction accuracy
improves widening the initial training set (e.g., $\left.\eta\right|_{S_{0}/K=1.25}=27$
\% $\to$ $\left.\eta\right|_{S_{0}/K=20}=5$ \%), but the time saving
reduces ($\left.\Delta t_{sav}\right|_{S_{0}/K=1.25}=89$ \% $\to$
$\left.\Delta t_{sav}\right|_{S_{0}/K=20}=74$ \%) albeit in a less
evident way. The ratio $S_{0}/K=5$ ($\to$ $\left.\eta\right|_{S_{0}/K=5}\approx7$
\% and $\left.\Delta t_{sav}\right|_{S_{0}/K=5}=86$ \%) has been
then chosen as the optimal trade-off threshold to fit (\ref{eq:time-goal}).
More specifically, the size of the initial and the final training
datasets have been set here to $S_{0}=5\times K=40$ and $S=140$,
respectively, so that the total execution time of the \emph{SbD is}
equal to that of a \emph{DO} method (i.e., $\Delta t_{SbD}\approx\Delta t_{DO}$),
which is based on a standard implementation of the Conjugate Gradient
(\emph{CG}) technique, running for $I_{DO}=400$ iterations \cite{Kleinman 1997}.
By using such a setup, Figure 6 shows the evolution of the optimal
value of the cost function, $\Phi_{i}\triangleq\Phi\left(\underline{\psi}_{i}\right)$,
during the \emph{SSE} minimization ($i=1,...,I_{SbD}$). As it can
be observed, $I_{SbD}=100$ iterations are enough to decrease the
data mismatch of about two orders of magnitude (i.e., $\frac{\Phi_{I_{SbD}}^{\left(SbD\right)}}{\Phi_{0}^{\left(SbD\right)}}=2.07\times10^{-2}$
- Fig. 6). Moreover, the values of $\Phi_{i}$ are very similar to
those when applying a standard \emph{GO} (e.g., $\frac{\Phi_{I_{SbD}}^{\left(SbD\right)}}{\Phi_{I_{GO}}^{\left(GO\right)}}=1.03$
- Fig. 6). This proves the reliability of the \emph{SbD} algorithm
to faithfully sample/explore the solution space looking for the global
optimum even though guided by a \emph{DT} model of the \emph{FW} solver.
For completeness, the behavior of the \emph{DO} minimization is reported,
as well. To better understand the optimization performance of the
three inversion approaches, Figure 7 shows the \emph{2-D} parametric
representation of the functional described by the following equation\begin{equation}
\Phi\left(a,b\right)=\Phi\left\{ b\times\left[\left(a+1\right)\times\underline{\xi}^{\left(1\right)}-a\times\underline{\xi}^{\left(act\right)}\right]+\left(b-1\right)\times a\times\underline{\xi}^{\left(2\right)}\right\} \label{eq:2D-cost}\end{equation}
in the ranges $-1.5\le a\le0.5$ and $-0.5\le b\le1.5$ when setting
$\underline{\xi}^{\left(1\right)}=\underline{\xi}^{\left(SbD\right)}$
and $\underline{\xi}^{\left(2\right)}=\underline{\xi}^{\left(DO\right)}$
{[}Fig. 7(\emph{a}){]} or $\underline{\xi}^{\left(2\right)}=\underline{\xi}^{\left(GO\right)}$
{[}Fig. 7(\emph{b}){]}, $\underline{\xi}^{\left(act\right)}$ being
the actual solution%
\footnote{\noindent According to (\ref{eq:2D-cost}), it can be easily verified
that $\Phi\left(-1,1\right)=\Phi\left(\underline{\xi}^{\left(act\right)}\right)$,
$\Phi\left(0,1\right)=\Phi\left(\underline{\xi}^{\left(1\right)}\right)$,
and $\Phi\left(-1,\,0\right)=\Phi\left(\underline{\xi}^{\left(2\right)}\right)$.%
}. The landscape in Fig. 7(\emph{a}) proves that the \emph{DO} solution
is trapped into a local minimum of the cost function (i.e., a false
solution for the inversion) without any possibility to escape from
such a {}``wrong'' valley. This is even more evident by looking
at the plot of the cost function along the \emph{1-D} cut of the solution
space passing through $\underline{\xi}^{\left(DO\right)}$ and the
actual solution {[}i.e., $\left.\Phi\left(a,\, b\right)\right|_{a=-1}^{\underline{\xi}^{\left(2\right)}=\underline{\xi}^{\left(DO\right)}}$
- Fig. 7(\emph{c}){]} %
\footnote{\noindent It is worth pointing out that a standard definition of the
\emph{DO-DoF}s, i.e., $\underline{\xi}^{\left(DO\right)}=\left\{ \mathcal{T}^{\left(v\right)}\left(x_{n},\, y_{n}\right);\,\tau\left(x_{n},\, y_{n}\right);\, v=1,...,V;\, n=1,...,N\right\} $,
has been adopted according to the reference literature on gradient-based
local search algorithms \cite{Salucci 2015}.%
}. Otherwise, the \emph{SbD} solution $\underline{\xi}^{\left(SbD\right)}$
belongs to the {}``attraction basin'' of the actual solution $\underline{\xi}^{\left(act\right)}$
analogously to the \emph{GO} solution $\underline{\xi}^{\left(GO\right)}$
{[}Figs. 7(\emph{b})-7(\emph{c}){]}. Such outcomes are confirmed by
the corresponding reconstructions in Figs. 4(\emph{b})-4(\emph{d}).
Indeed, the \emph{DO} inversion is unsatisfactory and remarkably worse
than the \emph{SbD} one, as quantified by the integral errors (i.e.,
$\frac{\left.\Xi\right|_{DO}}{\left.\Xi\right|_{SbD}}=51.3$ - Fig.
8), even though the execution time of the two iterative minimizations
is approximately the same ($\left.\Delta t\right|_{DO}=480$ {[}sec{]}
vs. $\left.\Delta t\right|_{SbD}=490$ {[}sec{]} - Fig. 8). Furthermore,
the computational efficiency of the \emph{SbD} is disruptive when
compared to the standard \emph{GO} since $\frac{\left.\Delta t\right|_{SbD}}{\left.\Delta t\right|_{GO}}=0.14$
($\Rightarrow$ $\Delta t_{sav}=86\%$ - Fig. 8), while yielding the
same accuracy (i.e., $\frac{\left.\Xi\right|_{GO}}{\left.\Xi\right|_{SbD}}=0.99$
- Fig. 8).

\noindent In order to assess the robustness of the reconstruction
process to blurred/corrupted data, Figure 9(\emph{a}) compares the
behavior of the cost function for the \emph{SbD} and the \emph{GO}
optimizations when varying the \emph{SNR} of the scattered field samples.
As expected, the data matching gets worse as the noise increases from
$SNR=20$ {[}dB{]} up to $SNR=5$ {[}dB{]} {[}i.e., $\left.\Phi\left(\underline{\xi}^{\left(SbD\right)}\right)\right|_{SNR=20\,[\mathrm{dB}]}=6.84\times10^{-2}$
vs. $\left.\Phi\left(\underline{\xi}^{\left(SbD\right)}\right)\right|_{SNR=10\,[\mathrm{dB}]}=1.32\times10^{-1}$
vs. $\left.\Phi\left(\underline{\xi}^{\left(SbD\right)}\right)\right|_{SNR=5\,[\mathrm{dB}]}=2.70\times10^{-1}$
- Fig. 9(\emph{a}){]}, but the \emph{SbD} still performs as the \emph{GO},
while reducing the inversion time {[}$\Delta t_{sav}=86\%$ - Fig.
9(\emph{b}){]}, despite the need of predicting the cost function values
starting from non-ideal (blurred) data. The reliability of the \emph{SbD}
in emulating a \emph{GO} when exploring highly-nonlinear solution
spaces is confirmed by the comparison of the corresponding reconstruction
errors {[}Fig. 9(\emph{b}){]}, which are almost identical whatever
the amount of noise and, always, significantly lower than the \emph{DO}
ones. As a matter of fact, the \emph{DO} is unable either to find
a satisfactory reconstruction {[}Figs. 11(\emph{g})-11(\emph{i}){]}
or to localize the attraction basin of the global optimum {[}Fig.
10(\emph{a}), Fig. 10(\emph{c}), and Fig. 10(\emph{e}){]}. It is also
worth noticing that the \emph{SbD} is effective even under very harsh
operative conditions (e.g., $SNR=5$ {[}dB{]}) as confirmed pictorially
in Fig. 11(\emph{c}) and quantitatively by the value of the error
index {[}i.e., $\left.\Xi\right|_{SNR=5\,[\mathrm{dB}]}^{SbD}=5.6\times10^{-2}$
- Fig. 9(\emph{b}){]}.

\noindent The next set of results are concerned with the dependence
of the data inversion on the contrast value of the scatterer, $\tau_{\Omega}$,
still considering the extremely challenging scattering environment
with $SNR=5$ {[}dB{]}. Figure 12(\emph{a}) gives some indications
on the iterative minimization of the cost function. As expected, the
weaker the scatterer more effective is the optimization process as
denoted by the smaller and smaller values of the cost function at
the convergence {[}i.e., $\left.\Phi\left(\underline{\xi}^{\left(SbD\right)}\right)\right|_{\tau_{\Omega}=1}=2.54\times10^{-1}$,
$\left.\Phi\left(\underline{\xi}^{\left(SbD\right)}\right)\right|_{\tau_{\Omega}=2}=2.57\times10^{-1}$,
and $\left.\Phi\left(\underline{\xi}^{\left(SbD\right)}\right)\right|_{\tau_{\Omega}=10}=5.72\times10^{-1}$
being $\left.\Phi\left(\underline{\xi}^{\left(GO\right)}\right)\right|_{\tau_{\Omega}=1}=2.50\times10^{-1}$,
$\left.\Phi\left(\underline{\xi}^{\left(GO\right)}\right)\right|_{\tau_{\Omega}=2}=2.56\times10^{-1}$,
and $\left.\Phi\left(\underline{\xi}^{\left(GO\right)}\right)\right|_{\tau_{\Omega}=10}=5.56\times10^{-1}${]}.
This implies that the reconstruction quality decreases as $\tau_{\Omega}$
increases {[}Fig. 12(\emph{a}) and Fig. 13{]}. However, it has to
be observed that the performance of the \emph{GO}-based methods are
significantly better than those from the \emph{DO}, which results
unable to handle high contrasts {[}e.g., $\tau_{\Omega}=10$ - Fig.
13(\emph{i}){]} that cause high non-linearities.

\noindent The second test case is related to a more complex scatterer
profile {[}Fig. 16(\emph{a}){]} with $\tau_{\Omega}=4.0$ and described
by a larger number of spline control points ($Q=8$), thus a greater
dimensionality (i.e., $K=12$ - Tab. I) of the solution space. Therefore,
a larger initial training set has been chosen to keep the optimal
setup of the $S_{0}/K$ ratio (i.e., $S_{0}=5\times K=60$), while
the number of \emph{SbD} iterations has been reduced (i.e., $I_{SbD}=I_{GO}=80$)
to fit the time constraint (\ref{eq:time-goal}). Despite the smaller
number of optimization iterations, the higher dimensionality, and
the non-negligible noise level of the scattered data (i.e., $SNR=10$
{[}dB{]}), the \emph{SbD} solution is very close to the \emph{GO}
one {[}i.e., $\Phi\left(\underline{\xi}^{\left(SbD\right)}\right)=2.07\times10^{-1}$
vs. $\Phi\left(\underline{\xi}^{\left(GO\right)}\right)=1.79\times10^{-1}$
- Fig. 14(\emph{a}){]} and, unlike the \emph{DO}, it belongs to the
{}``attraction basin'' of the actual solution {[}Figs. 14(\emph{b})-14(\emph{c}){]}.
Consequently, the retrieved contrast distributions {[}Figs. 16(\emph{b})-16(\emph{d}){]}
quite faithfully reproduce the actual one {[}Fig. 16(\emph{a}){]}
with similar values of the reconstruction error and significantly
smaller than those of the \emph{DO} (i.e., $\left.\Xi\right|_{SbD}=5.46\times10^{-2}$
vs. $\left.\Xi\right|_{GO}=4.54\times10^{-2}$ vs. $\left.\Xi\right|_{DO}=4.36\times10^{-1}$
- Fig. 15). On the other hand, the \emph{CPU}-time of the \emph{SbD}
inversion is remarkably lower that of the \emph{GO} (i.e., $\Delta t_{sav}=82.5\%$
- Fig. 15) and very close to the \emph{DO}. 

\noindent The flexibility of the adopted minimum-dimensionality encoding
of the unknown scattering profiles as well as the feasibility of representing
doubly connected (\emph{DC}) contours/inhomogeneous objects is assessed
in the third test case {[}Fig. 18(\emph{a}){]}. More in detail, the
scatterer has been modeled with the following set of $K=11$ ($\Rightarrow$
$S_{0}=5\times K=55$, $I_{SbD}=I_{GO}=85$) descriptors \begin{equation}
\underline{\xi}_{DC}=\left\{ x_{\Omega},\, y_{\Omega},\,\Re\left(\tau_{\Omega}^{\left(out\right)}\right),\,\Im\left(\tau_{\Omega}^{\left(out\right)}\right),\,\Re\left(\tau_{\Omega}^{\left(int\right)}\right),\,\Im\left(\tau_{\Omega}^{\left(int\right)}\right),\,\underline{\rho}^{\left(out\right)},\,\upsilon\right\} \label{eq:DD-DoFs}\end{equation}
where the superscript $\left(out\right)$ {[}$\left(int\right)${]}
refers to the outer {[}internal{]} contour $\partial\Omega^{\left(out\right)}$
{[}$\partial\Omega^{\left(int\right)}${]}, while $0<\upsilon<1$
is the scale factor between the two borders, the $q$-th ($q=1,...,Q$;
$Q=4$) control point of $\partial\Omega^{\left(int\right)}$ {[}i.e.,
$\underline{\rho}^{\left(int\right)}=\left\{ \rho^{\left(q,\, int\right)};\, q=1,...,Q\right\} ${]}
being $\rho^{\left(q,int\right)}=\upsilon\rho^{\left(q,out\right)}$
(Tab. I). The outcomes from such a benchmark are summarized in Fig.
17(\emph{a}) in terms of reconstruction errors and execution time.
Once again, these results confirm the superior trade-off between computational
efficiency and effectiveness of the \emph{SbD} method over the \emph{GO}
and the \emph{DO} ones. As for the retrieved contrast, Figure 18 shows
that the \emph{SbD} reconstruction provides a reliable estimation
of both the object shape and the contrast value ($\tau_{\Omega}^{\left(out\right)}=3$,
$\tau_{\Omega}^{\left(int\right)}=1$, $\upsilon=0.6$ - Tab. I) well
detecting the presence of a {}``hole'' {[}$\left.\Xi\right|_{SbD}=3.30\times10^{-2}$
- Fig. 18(\emph{c}) vs. Fig. 18(\emph{a}){]}. Similar outcomes can
be drawn {[}Fig. 17(\emph{b}){]} for the inhomogeneous profile in
Fig. 18(\emph{b}) ($\tau_{\Omega}^{\left(out\right)}=2$, $\tau_{\Omega}^{\left(int\right)}=4$,
$\upsilon=0.4$ - Tab. I), the dielectric profile inferred by the
\emph{SbD} being shown in Fig. 18(\emph{d}) {[}$\left.\Xi\right|_{SbD}=5.13\times10^{-2}$
- Fig. 17(\emph{b}){]}.

\noindent The extension to multiple objects (\emph{MO}) is dealt with
in the \emph{Test Case \#4} where two disconnected scatterers have
been considered. In this case, the $K=16$ unknowns are

\noindent \begin{equation}
\underline{\xi}_{MO}=\left\{ x_{\Omega}^{\left(1\right)},\, y_{\Omega}^{\left(1\right)},\, x_{\Omega}^{\left(2\right)},\, y_{\Omega}^{\left(2\right)},\,\Re\left(\tau_{\Omega}^{\left(1\right)}\right),\,\Im\left(\tau_{\Omega}^{\left(1\right)}\right),\,\Re\left(\tau_{\Omega}^{\left(2\right)}\right),\,\Im\left(\tau_{\Omega}^{\left(2\right)}\right),\,\underline{\rho}^{\left(1\right)},\,\underline{\rho}^{\left(2\right)}\right\} ,\label{eq:}\end{equation}
where the superscripts $\left(1\right)$/$\left(2\right)$ refer to
the two disconnected spline contours $\partial\Omega^{\left(1\right)}$/$\partial\Omega^{\left(2\right)}$
($Q=4$), and the \emph{SbD} has been run for $I_{SbD}=I_{GO}=60$
iterations starting from a training set with $S_{0}=5\times K=80$
I/O pairs. Despite the higher complexity of the \emph{ISP} problem
at hand, also related to a larger dimension of the solution space
as well as the non-negligible contrast of both scatterers ($\tau_{\Omega}^{\left(1\right)}=\tau_{\Omega}^{\left(2\right)}=4$),
the \emph{SbD} carefully images the investigation domain {[}i.e.,
$\frac{\left.\Xi\right|_{SbD}}{\left.\Xi\right|_{GO}}=1.05$ - Fig.
20(\emph{b}) vs. Fig. 20(\emph{c}) and $\frac{\left.\Xi\right|_{SbD}}{\left.\Xi\right|_{DO}}=8.9\times10^{-2}$
- Fig. 19(\emph{b}) vs. Fig. 19(\emph{d}){]} by reducing the inversion
time of about $\Delta t_{sav}=76.7\%$ (Fig. 19).

\noindent Finally (\emph{Test Case \#5}), the \emph{SbD}-based imaging
method has been assessed against laboratory-controlled experimental
data. With reference to the data provided by the Institut Fresnel
\cite{Geffrin 2005}, the {}``\emph{FoamDielInt}'' scattering scenario
has been selected as representative benchmark. It consists of a foam
cylinder with diameter $8.0\times10^{-2}$ {[}m{]} and contrast $\tau_{\Omega}^{\left(out\right)}=0.45$
that embeds a smaller, $3.1\times10^{-2}$ {[}m{]} in diameter, and
weaker, $\tau_{\Omega}^{\left(int\right)}=2.0$, dielectric cylinder
{[}Fig. 21(\emph{a}){]}. The acquisition system was composed by $V=8$
ridged-horn antennas working at $f=2$ {[}GHz{]} to probe a square
investigation domain $D$ of side $L_{D}=0.2$ {[}m{]}. The scattered
data have been collected in $M=241$ uniformly-spaced locations on
a circular observation domain $O$ with radius $\rho_{O}=1.67$ {[}m{]}
\cite{Geffrin 2005}. Because of the topology of the object at hand,
the exploration of the solution space defined by the \emph{DoF}s in
(\ref{eq:DD-DoFs}) has been carried out by letting $S_{0}=55$ and
$I_{SbD}=I_{GO}=85$ according to the previous examples. Figure 21(\emph{b})
shows the retrieved contrast distribution. Similarly to the \emph{GO}
image {[}Fig. 21(\emph{c}){]}, it is possible to detect the two-layers
scatterer with a reliable estimation of the outer support of the object,
$\partial\Omega^{\left(out\right)}$, as well as to infer the presence
of an inner scatterer/layer with higher permittivity. Once again,
it turns out that it is possible to address the problem of local minima
by exploiting the {}``hill-climbing'' features of an \emph{EA}-based
multiple-agent approach, but solving the arising global minimization
task with a remarkable time saving over a standard \emph{GO} implementation
(i.e., $\Delta t_{sav}=83.5\%$) by equalling the computational efficiency
of the \emph{DO} {[}Fig. 21(\emph{d}){]}.

\section{\noindent Conclusions}

An innovative strategy has been proposed to address the computationally-efficient
yet reliable solution of the fully non-linear \emph{ISP}. The inversion
method has been built by implementing the pillar concepts of the \emph{SbD}
framework \cite{Massa 2021} to allow an effective exploration of
the multi-modal landscape defined by the data mismatch cost function
with the same time cost of a standard deterministic local search. 

\noindent From a methodological point of view and to the best of the
authors' knowledge, the key advances of this research work with respect
to the state-of-the-art literature can be summarized as follows:

\begin{itemize}
\item \noindent a {}``smart'' and flexible minimum-dimensionality encoding
of complex-shaped scatterers yielded with a spline-based modeling
of the scattering profile (Sect. \ref{sub:Problem-Formulation}),
which not only {}``implements'' a more favorable {}``operating
environment'' for the underlying \emph{EA}-based \emph{GO} strategy,
but it also alleviates the {}``curse-of-dimensionality'' problem;
\item \noindent the use of a \emph{GP}-based \emph{LBE} approach for building
a fast and accurate \emph{DT} of the time-consuming \emph{FW} solver
that predicts the data mismatch cost function associated to each trial
solution, but also provides additional information on the associated
{}``confidence level'' of this latter;
\item \noindent the setup of a collaborative framework between the \emph{EA}
mechanisms and the \emph{DT} model that enables an effective exploration
of the solution space, which is adaptively sampled at selected and
promising points to increase the prediction accuracy of the \emph{DT}
model as well as to speed-up the converge towards the attraction basin
of the global-optimum/actual-solution.
\end{itemize}
\noindent Moreover, the main outcomes from the numerical and experimental
assessment (Sect. \ref{sec:Performance-Assessment}) are:

\begin{itemize}
\item \noindent the \emph{SbD}-based inversion method is a reliable tool
for reaching the attraction basin of the global optimum without being
trapped into local-minima/false-solutions also when highly nonlinear
cost functions/strong scatterers are at hand;
\item \noindent it exhibits the same computational efficiency of a \emph{DO},
breaking - for the first time to the authors' best knowledge - the
widely-diffused idea that solving an \emph{ISP} with an \emph{EA}-based
tool is generally computationally unaffordable;
\item \noindent the range of a reliable and effective application of the
\emph{SbD} inversion method extends from weak to strong simple as
well as complex and multiple objects in harsh environmental conditions,
as well, subject to a suitable choice of the \emph{SbD} building blocks
according to the {}``no-free lunch'' theorems \cite{Wolpert 1997};
\item \noindent the \emph{SbD} inversion is able to effectively and efficiently
process synthetic as well as real laboratory-controlled scattering
data.
\end{itemize}
\noindent Future works, beyond the scope of this paper, will be aimed
at extending the proposed \emph{SbD}-based method to other applicative
contexts (e.g., \emph{NDT/NDE}, \emph{GPR} investigations, biomedical
imaging, or food quality assessment) involving - for instance - \emph{differential}
formulations of the \emph{ISP} to embed the \emph{a-priori} knowledge
on a reference/healthy background scenario.

\part*{Appendix I\label{par:Appendix-I}}

The \emph{LHS} strategy is implemented through the following procedure:

\begin{itemize}
\item Uniformly divide the admissible range $\mathbb{A}_{k}=\left[\xi_{k}^{\min},\,\xi_{k}^{\max}\right]$
of each $k$-th ($k=1,...,K$) \emph{DoF} into $S_{i}$ intervals
$\left\{ \mathbb{I}_{k}^{\left(s\right)};\, s=1,...,S_{i}\right\} $
such that $\mathbb{A}_{k}=\cup_{s=1,...,S_{i}}\mathbb{I}_{k}^{\left(s\right)}$;
\item For each $k$-th ($k=1,...,K$) variable, randomly choose one value
$\varkappa_{k}^{\left(s\right)}$ within each $s$-th ($s=1,...,S_{i}$)
interval, $\mathbb{I}_{k}^{\left(s\right)}$, and form the corresponding
set $\mathbb{S}_{k}=\left\{ \varkappa_{k}^{\left(s\right)}\in\mathbb{I}_{k}^{\left(s\right)};\, s=1,...,S_{i}\right\} $;
\item Until $s=S_{i}$, form the $s$-th $K$-dimensional sample $\underline{\xi}^{\left(s\right)}$
($\underline{\xi}^{\left(s\right)}=\left\{ \xi_{k}^{\left(s\right)};\, k=1,...,K\right\} $)
by letting $\xi_{k}^{\left(s\right)}=\mathcal{R}\left(\mathbb{S}_{k}\right)$
($k=1,...,K$) where the operator $\mathcal{R}\left(\,.\,\right)$
outputs the value of one randomly-chosen entry of $\mathbb{S}_{k}$,
which is then removed from it. Update the index $s$ {[}$s\leftarrow\left(s+1\right)${]}
and repeat.
\end{itemize}

\section*{Acknowledgements}

\noindent This work has been partially supported by the Italian Ministry
of Education, University, and Research within the Program PRIN 2017
(CUP: E64I19002530001) for the Project CYBER-PHYSICAL ELECTROMAGNETIC
VISION: Context-Aware Electromagnetic Sensing and Smart Reaction (EMvisioning)
(Grant no. 2017HZJXSZ) and benefited from the networking activities
carried out within the Project {}``SPEED'' (Grant No. 61721001)
funded by National Science Foundation of China under the Chang-Jiang
Visiting Professorship Program, the Project 'Inversion Design Method
of Structural Factors of Conformal Load-bearing Antenna Structure
based on Desired EM Performance Interval' (Grant no. 2017HZJXSZ) funded
by the National Natural Science Foundation of China, and the Project
'Research on Uncertainty Factors and Propagation Mechanism of Conformal
Loab-bearing Antenna Structure' (Grant No. 2021JZD-003) funded by
the Department of Science and Technology of Shaanxi Province within
the Program Natural Science Basic Research Plan in Shaanxi Province.
A. Massa wishes to thank E. Vico for her never-ending inspiration,
support, guidance, and help.

~\\
\emph{This work has been submitted to the IEEE for possible publication.
Copyright may be transferred without notice, after which this version
may no longer be accessible.}

\newpage
\section*{FIGURE CAPTIONS}

\begin{itemize}
\item \textbf{Figure 1.} Block scheme of the \emph{SbD}-based inversion
method.
\item \textbf{Figure 2.} Pictorial sketch of the spline-based scatterer
modeling.
\item \textbf{Figure 3.} \emph{SbD-SSE} update rules for (\emph{a}) the
personal best position of each $p$-th ($p=1,...,P$) particle, $\underline{\zeta}_{i}^{\left(p\right)}$,
and (\emph{b}) the global best, $\underline{\zeta}_{i}$, at the $i$-th
iteration ($i=1,...,I_{SbD}$).
\item \textbf{Figure 4.} \emph{Numerical Assessment} (\emph{Test Case \#1}:
$\tau_{\Omega}=4.0$, $V=M=18$, Noiseless Data; $K=8$) - Maps of
(\emph{a}) the actual and (\emph{b})-(\emph{d}) retrieved contrast
distributions with (\emph{b}) the \emph{SbD}, (\emph{c}) the \emph{GO},
and (\emph{d}) the \emph{DO} methods.
\item \textbf{Figure 5.} \emph{Numerical Assessment} (\emph{Test Case \#1}:
$\tau_{\Omega}=4.0$, $V=M=18$, Noiseless Data; $K=8$) - Prediction
error of the \emph{DT}, $\eta$, and time saving, $\Delta t_{sav}$,
versus the ratio $S_{0}/K$ between the number of initial training
samples $S_{0}$, and the number of unknowns/\emph{SbD-DoF}s, $K$.
\item \textbf{Figure 6.} \emph{Numerical Assessment} (\emph{Test Case \#1}:
$\tau_{\Omega}=4.0$, $V=M=18$, Noiseless Data; $K=8$) - Evolution
of the optimal value of the cost function, $\Phi_{i}$, versus the
iteration index, $i$.
\item \textbf{Figure 7.} \emph{Numerical Assessment} (\emph{Test Case \#1}:
$\tau_{\Omega}=4.0$, $V=M=18$, Noiseless Data; $K=8$) - Plot of
the functional (\ref{eq:2D-cost}) (\emph{a})(\emph{b}) in the ranges
$-1.5\le a\le0.5$ and $-0.5\le b\le1.5$ when setting $\underline{\xi}^{\left(1\right)}=\underline{\xi}^{\left(SbD\right)}$
and (\emph{a}) $\underline{\xi}^{\left(2\right)}=\underline{\xi}^{\left(DO\right)}$
or (\emph{b}) $\underline{\xi}^{\left(2\right)}=\underline{\xi}^{\left(GO\right)}$
or (\emph{c}) along the lines passing through ($\underline{\xi}^{\left(SbD\right)}$,
$\underline{\xi}^{\left(act\right)}$), ($\underline{\xi}^{\left(DO\right)}$,
$\underline{\xi}^{\left(act\right)}$), and ($\underline{\xi}^{\left(GO\right)}$,
$\underline{\xi}^{\left(act\right)}$).
\item \textbf{Figure 8.} \emph{Numerical Assessment} (\emph{Test Case \#1}:
$\tau_{\Omega}=4.0$, $V=M=18$, Noiseless Data; $K=8$) - Values
of the reconstruction error, $\Xi$, and total inversion time, $\Delta t$.
\item \textbf{Figure 9.} \emph{Numerical Assessment} (\emph{Test Case \#1}:
$\tau_{\Omega}=4.0$, $V=M=18$; $K=8$) - Plot of (\emph{a}) the
evolution of the optimal value of the cost function, $\Phi_{i}$,
versus the iteration index, $i$, and of (\emph{b}) the reconstruction
error, $\Xi$, and the execution time, $\Delta t$, versus the \emph{SNR}
value of the scattered data.
\item \textbf{Figure 10.} \emph{Numerical Assessment} (\emph{Test Case \#1}:
$\tau_{\Omega}=4.0$, $V=M=18$; $K=8$) - Plot of the functional
(\ref{eq:2D-cost}) (\emph{a})(\emph{b}) in the ranges $-1.5\le a\le0.5$
and $-0.5\le b\le1.5$ when setting $\underline{\xi}^{\left(1\right)}=\underline{\xi}^{\left(SbD\right)}$
and (\emph{a})(\emph{c})(\emph{e}) $\underline{\xi}^{\left(2\right)}=\underline{\xi}^{\left(DO\right)}$
or (\emph{b})(\emph{d})(\emph{f}) $\underline{\xi}^{\left(2\right)}=\underline{\xi}^{\left(GO\right)}$
for noisy scattered data with (\emph{a})(\emph{b}) $SNR=20$ {[}dB{]},
(\emph{c})(\emph{d}) $SNR=10$ {[}dB{]}, and (\emph{e})(\emph{f})
$SNR=5$ {[}dB{]}.
\item \textbf{Figure 11.} \emph{Numerical Assessment} (\emph{Test Case \#1}:
$\tau_{\Omega}=4.0$, $V=M=18$; $K=8$) - \textbf{}Reconstructions
of the contrast profile in $D$ obtained by (\emph{a})-(\emph{c})
the \emph{SbD}, (\emph{d})-(\emph{f}) the \emph{GO}, and (\emph{g})-(\emph{i})
the \emph{DO} when processing noisy data with (\emph{a})(\emph{d})(\emph{g})
$SNR=20$ {[}dB{]}, (\emph{b})(\emph{e})(\emph{h}) $SNR=10$ {[}dB{]},
and (\emph{c})(\emph{f})(\emph{i}) $SNR=5$ {[}dB{]}.
\item \textbf{Figure 12.} \emph{Numerical Assessment} (\emph{Test Case \#1}:
$V=M=18$, $SNR=5$ {[}dB{]}; $K=8$) - Plot of (\emph{a}) the evolution
of the optimal value of the cost function, $\Phi_{i}$, versus the
iteration index, $i$, and of (\emph{b}) the reconstruction error,
$\Xi$, and the execution time, $\Delta t$, versus the value of the
contrast of the scatterer, $\tau_{\Omega}$.
\item \textbf{Figure 13.} \emph{Numerical Assessment} (\emph{Test Case \#1}:
$V=M=18$, $SNR=5$ {[}dB{]}; $K=8$) - Reconstructions of the contrast
profile in $D$ obtained by (\emph{a})-(\emph{c}) the \emph{SbD},
(\emph{d})-(\emph{f}) the \emph{GO}, and (\emph{g})-(\emph{i}) the
\emph{DO} when the actual value of the contrast of the scatterer is
(\emph{a})(\emph{d})(\emph{g}) $\tau_{\Omega}=1$, (\emph{b})(\emph{e})(\emph{h})
$\tau_{\Omega}=2$, and (\emph{c})(\emph{f})(\emph{i}) $\tau_{\Omega}=10$.
\item \textbf{Figure 14.} \emph{Numerical Assessment} (\emph{Test Case \#}2:
$\tau_{\Omega}=4.0$, $V=M=18$, $SNR=10$ {[}dB{]}; $K=12$) - Plot
of (\emph{a}) the evolution of the optimal value of the cost function,
$\Phi_{i}$, versus the iteration index, $i$, and color maps of the
functional (\ref{eq:2D-cost}) (\emph{c})(\emph{d}) in the ranges
$-1.5\le a\le0.5$ and $-1.5\le b\le0.5$ when setting $\underline{\xi}^{\left(1\right)}=\underline{\xi}^{\left(SbD\right)}$
and (\emph{c}) $\underline{\xi}^{\left(2\right)}=\underline{\xi}^{\left(DO\right)}$
or (\emph{d}) $\underline{\xi}^{\left(2\right)}=\underline{\xi}^{\left(GO\right)}$.
\item \textbf{Figure 15.} \emph{Numerical Assessment} (\emph{Test Case \#2}:
$\tau_{\Omega}=4.0$, $V=M=18$, $SNR=10$ {[}dB{]}; $K=12$) - Values
of the reconstruction error, $\Xi$, and total inversion time, $\Delta t$.
\item \textbf{Figure 16.} \emph{Numerical Assessment} (\emph{Test Case \#2}:
$\tau_{\Omega}=4.0$, $V=M=18$, $SNR=10$ {[}dB{]}; $K=12$) - Maps
of (\emph{a}) the actual and (\emph{b})-(\emph{d}) the retrieved contrast
distributions with (\emph{b}) the \emph{SbD}, (\emph{c}) the \emph{GO},
and (\emph{d}) the \emph{DO} methods.
\item \textbf{Figure 17.} \emph{Numerical Assessment} (\emph{Test Case \#3}:
$V=M=18$, $SNR=10$ {[}dB{]}; $K=11$) - Values of the reconstruction
error, $\Xi$, and total inversion time, $\Delta t$, for the scattering
scenario in Fig. 18(a) when (\emph{a}) ($\tau_{\Omega}^{\left(out\right)}=3$,
$\tau_{\Omega}^{\left(int\right)}=0$) and (\emph{b}) ($\tau_{\Omega}^{\left(out\right)}=2$,
$\tau_{\Omega}^{\left(int\right)}=4$).
\item \textbf{Figure 18.} \emph{Numerical Assessment} (\emph{Test Case \#3}:
$V=M=18$, $SNR=10$ {[}dB{]}; $K=11$) - Maps of (\emph{a})(\emph{b})
the actual and (\emph{b})(\emph{d}) the \emph{SbD}-retrieved contrast
distributions when (\emph{a})(\emph{c}) ($\tau_{\Omega}^{\left(out\right)}=3$,
$\tau_{\Omega}^{\left(int\right)}=0$) and (\emph{b})(\emph{d}) ($\tau_{\Omega}^{\left(out\right)}=2$,
$\tau_{\Omega}^{\left(int\right)}=4$).
\item \textbf{Figure 19.} \emph{Numerical Assessment} (\emph{Test Case \#4}:
$\tau_{\Omega}^{\left(1\right)}=\tau_{\Omega}^{\left(2\right)}=4$,
$V=M=18$, $SNR=10$ {[}dB{]}; $K=16$) - Values of the reconstruction
error, $\Xi$, and total inversion time, $\Delta t$.
\item \textbf{Figure 20.} \emph{Numerical Assessment} (\emph{Test Case \#4}:
$\tau_{\Omega}^{\left(1\right)}=\tau_{\Omega}^{\left(2\right)}=4$,
$V=M=18$, $SNR=10$ {[}dB{]}; $K=16$) - Maps of (\emph{a}) the actual
and (\emph{b})-(\emph{d}) the retrieved contrast distributions with
(\emph{b}) the \emph{SbD}, (\emph{c}) the \emph{GO}, and (\emph{d})
the \emph{DO} methods.
\item \textbf{Figure 21.} \emph{Experimental Assessment} (\emph{Test Case
\#5} : $f=2$ {[}GHz{]}, $\tau_{\Omega}^{\left(out\right)}=0.45$,
$\tau_{\Omega}^{\left(int\right)}=2$, $V=8$, $M=241$; $K=11$)
- Maps of (\emph{a}) the actual {}``\emph{FoamDielInt}'' \cite{Geffrin 2005}
and (\emph{b})-(\emph{d}) the retrieved contrast distributions with
(\emph{b}) the \emph{SbD}, (\emph{c}) the \emph{GO}, and (\emph{d})
the \emph{DO} methods.
\end{itemize}

\section*{TABLE CAPTIONS}

\begin{itemize}
\item \textbf{Table I.} \emph{Performance} \textbf{}\emph{Assessment} ($V=M=18$)
- Test cases description.
\end{itemize}
\newpage
\begin{center}~\vfill\end{center}

\begin{center}\begin{tabular}{cc}
\multicolumn{2}{c}{\includegraphics[%
  width=0.90\columnwidth]{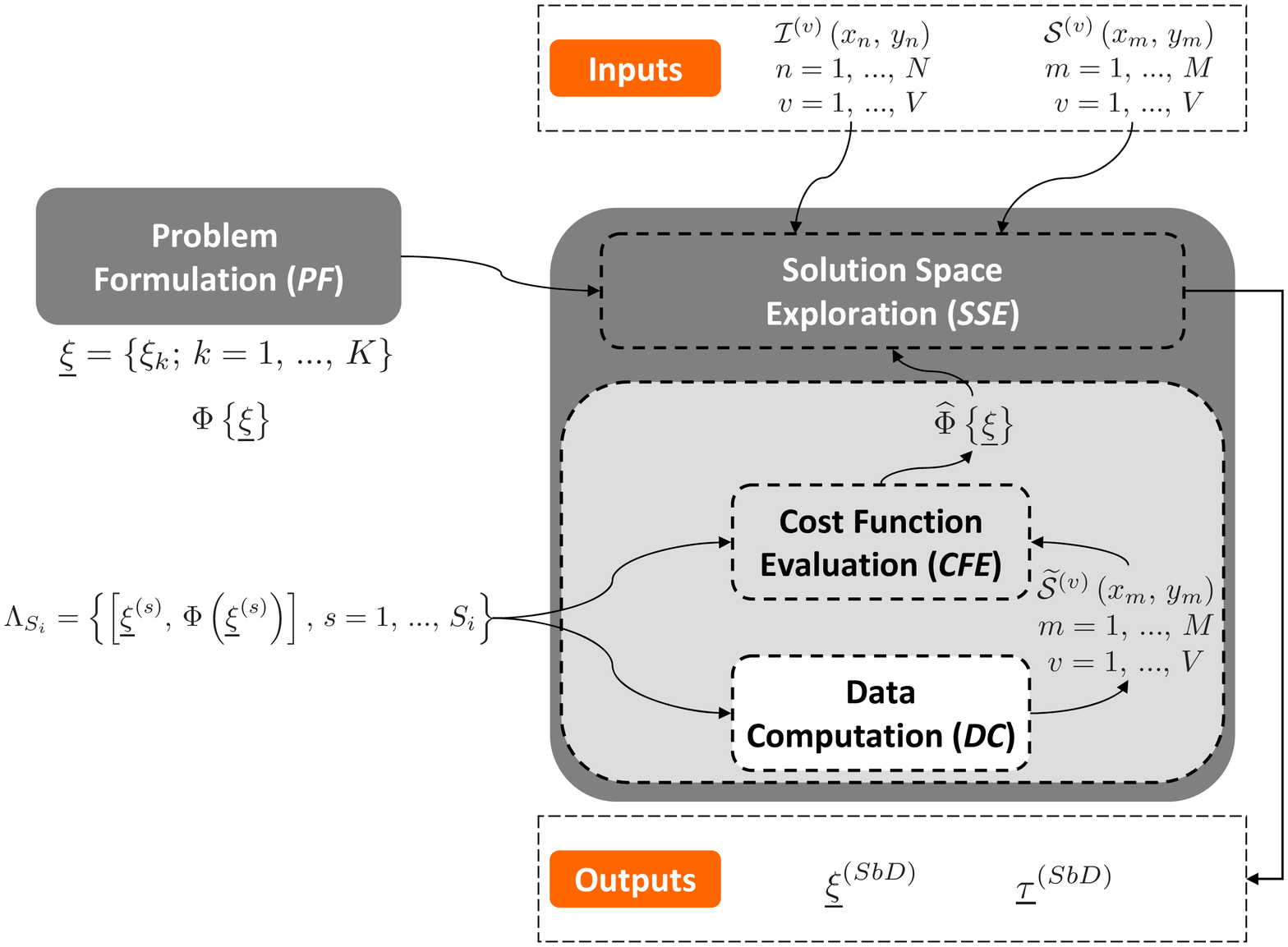}}\tabularnewline
\end{tabular}\end{center}

\begin{center}~\vfill\end{center}

\begin{center}\textbf{Fig. 1 - M. Salucci} \textbf{\emph{et al.}}\textbf{,}
\textbf{\emph{{}``}}Learned Global Optimization ...''\end{center}

\newpage
\begin{center}~\vfill\end{center}

\begin{center}\begin{tabular}{cc}
\multicolumn{2}{c}{\includegraphics[%
  width=0.90\columnwidth]{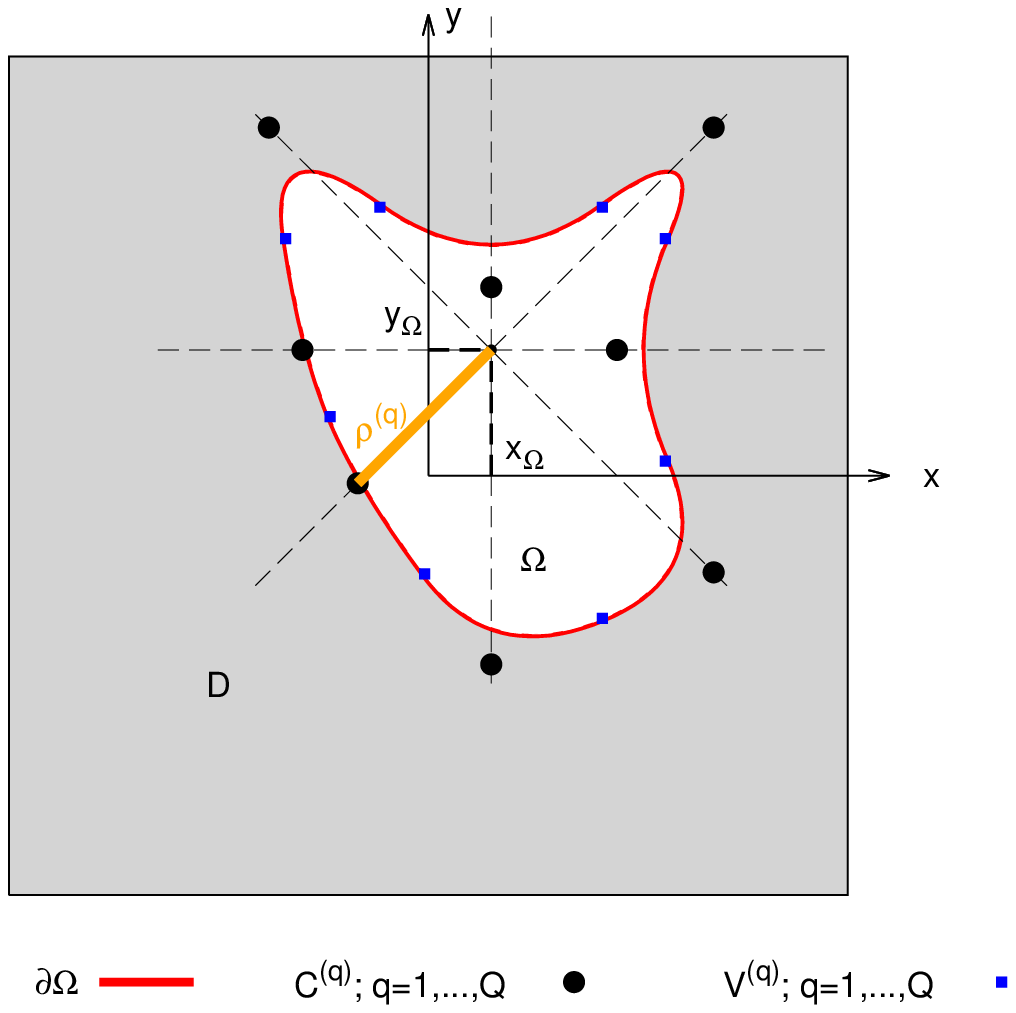}}\tabularnewline
\end{tabular}\end{center}

\begin{center}~\vfill\end{center}

\begin{center}\textbf{Fig. 2 - M. Salucci} \textbf{\emph{et al.}}\textbf{,}
\textbf{\emph{{}``}}Learned Global Optimization ...''\end{center}

\newpage
\begin{center}~\vfill\end{center}

\begin{center}\begin{tabular}{cc}
\multicolumn{2}{c}{\includegraphics[%
  width=0.90\columnwidth]{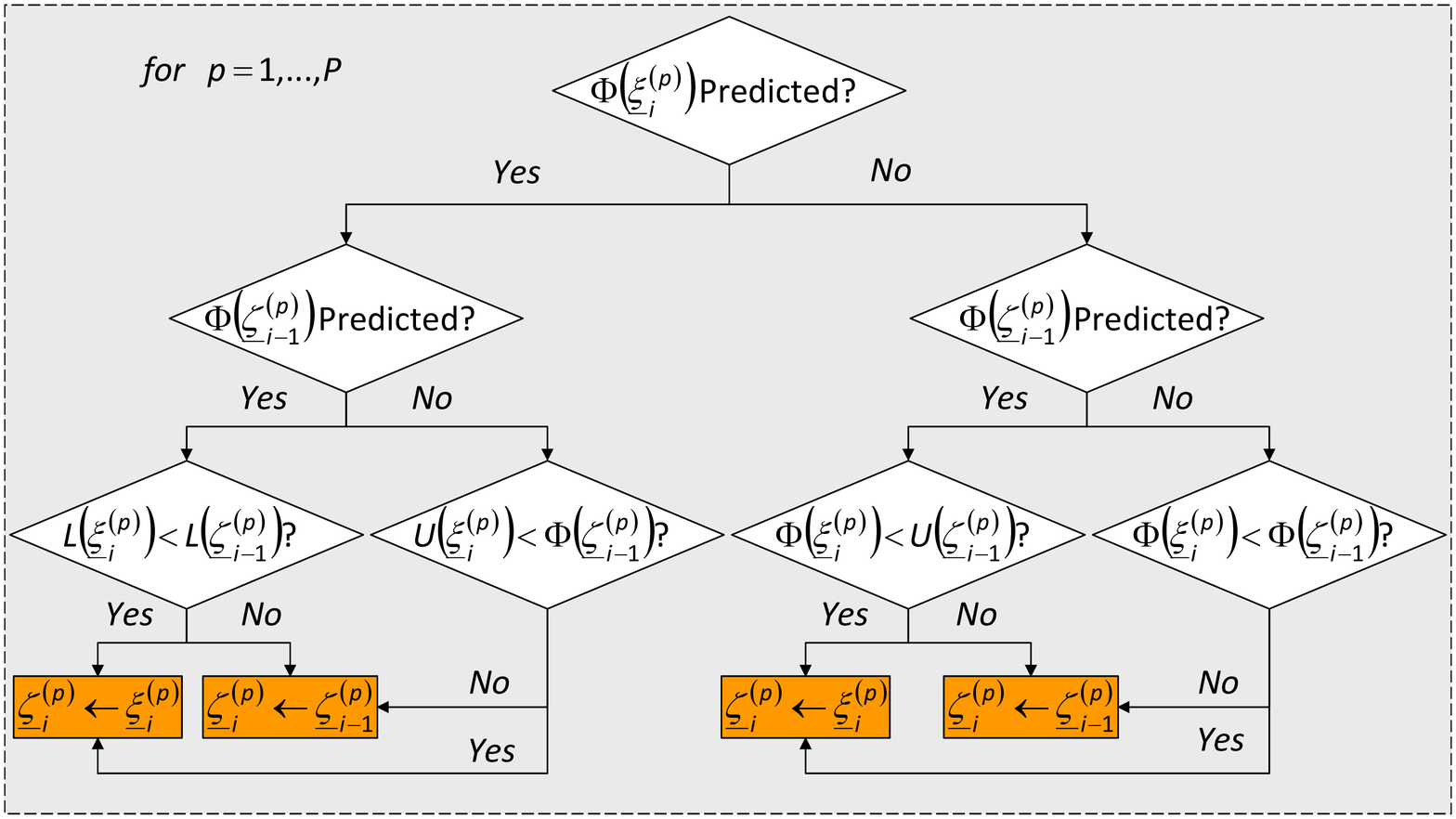}}\tabularnewline
\multicolumn{2}{c}{(\emph{a})}\tabularnewline
\multicolumn{2}{c}{}\tabularnewline
\multicolumn{2}{c}{\includegraphics[%
  width=0.65\columnwidth]{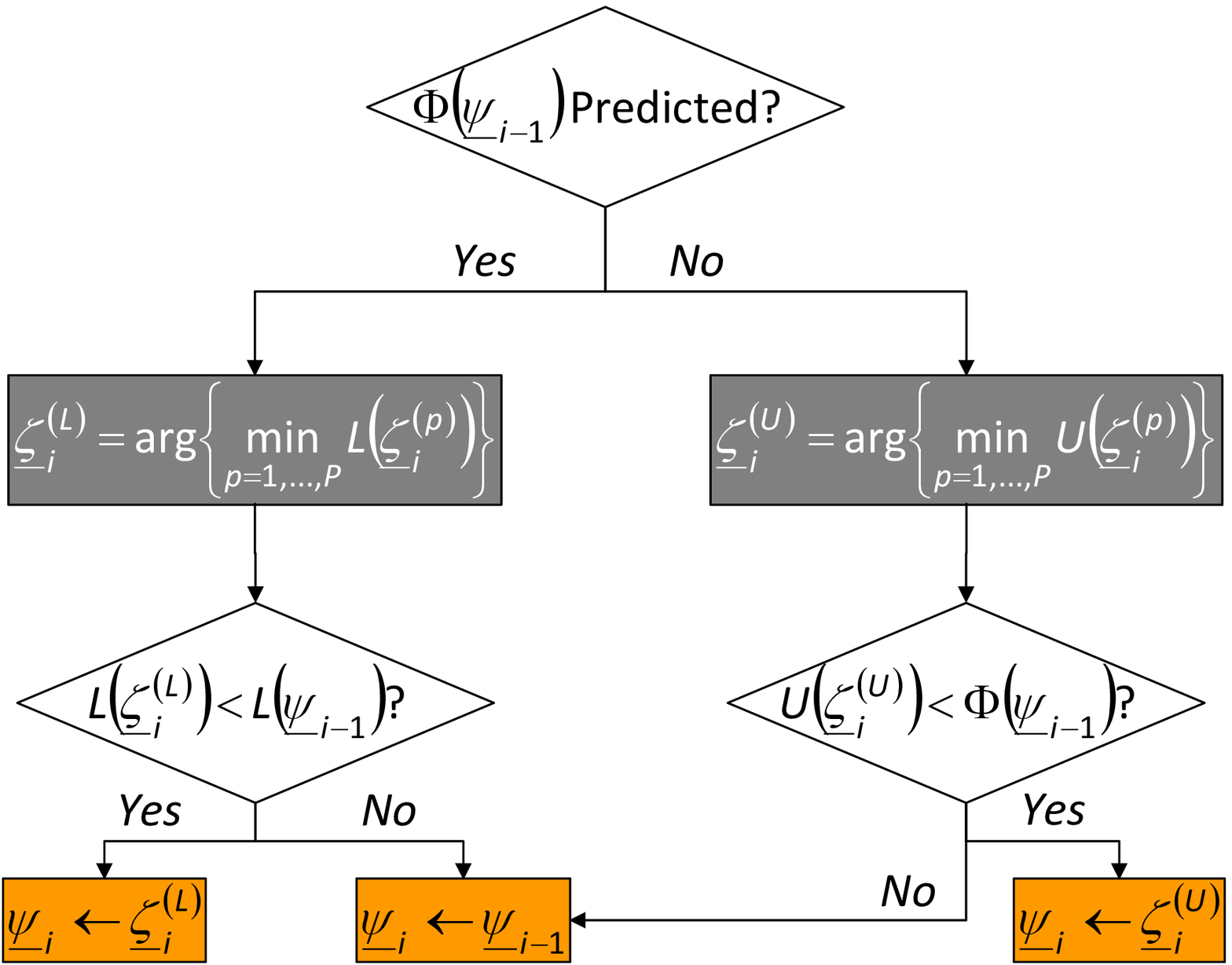}}\tabularnewline
\multicolumn{2}{c}{(\emph{b})}\tabularnewline
\end{tabular}\end{center}

\begin{center}~\vfill\end{center}

\begin{center}\textbf{Fig. 3 - M. Salucci} \textbf{\emph{et al.}}\textbf{,}
\textbf{\emph{{}``}}Learned Global Optimization ...''\end{center}

\newpage
\begin{center}~\vfill\end{center}

\begin{center}\begin{tabular}{cc}
\includegraphics[%
  width=0.50\columnwidth]{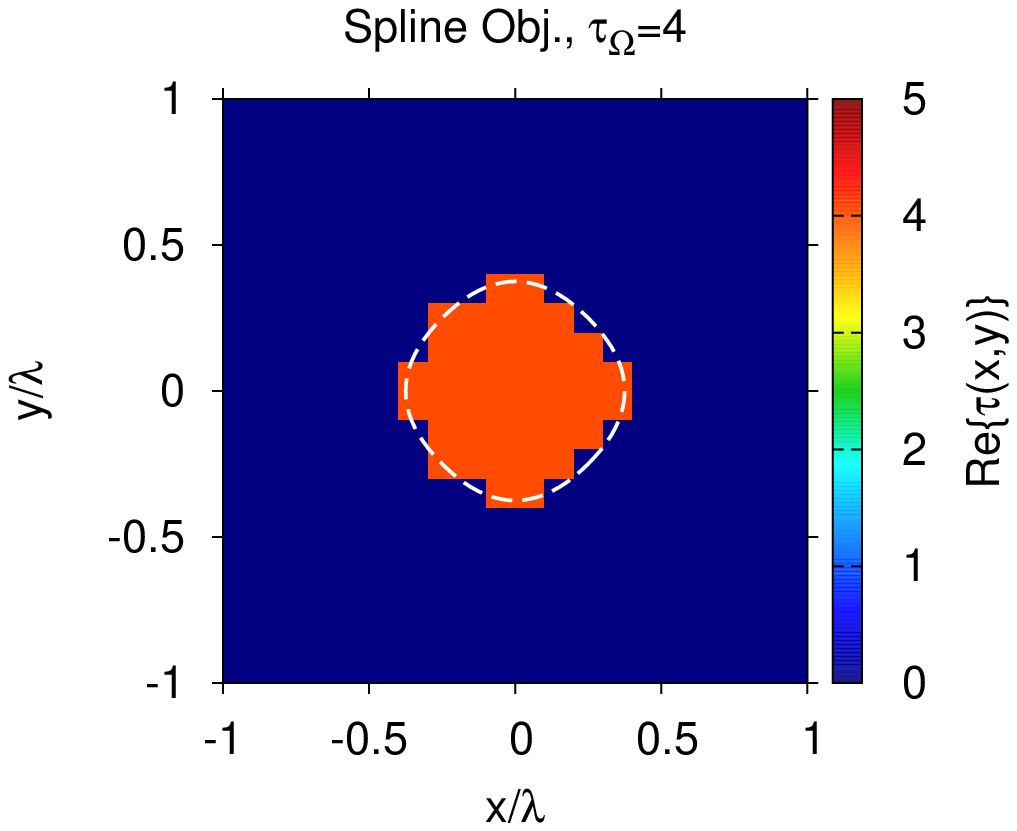}&
\includegraphics[%
  width=0.50\columnwidth]{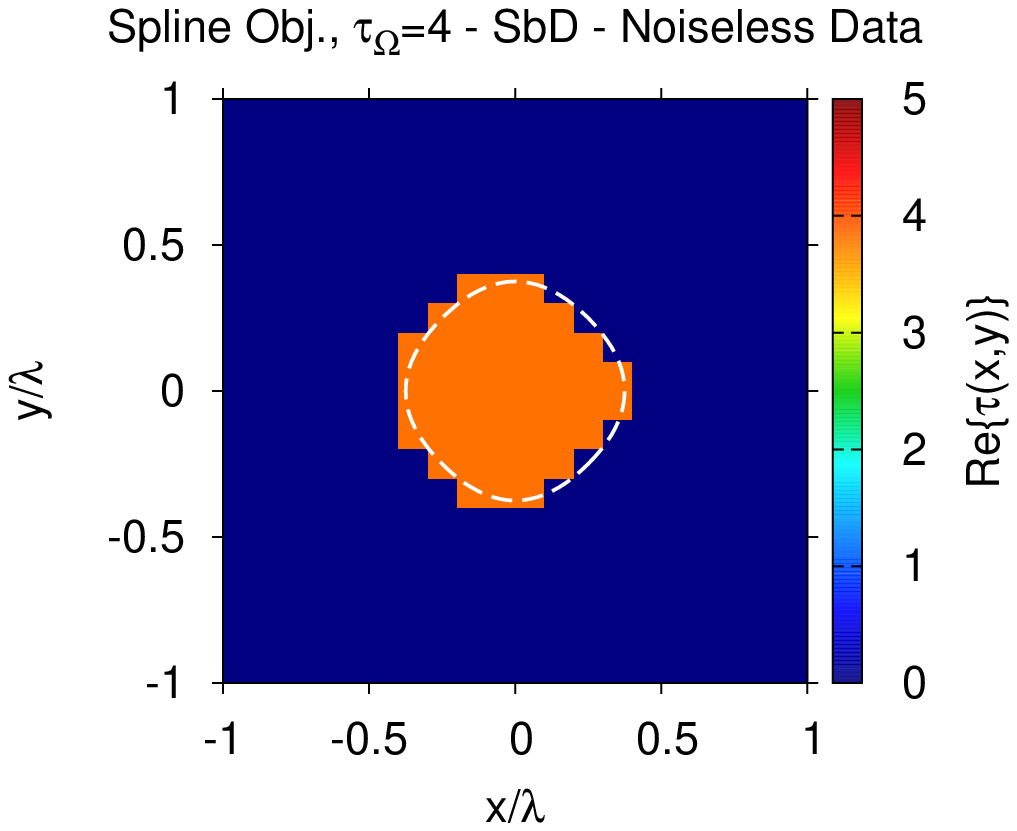}\tabularnewline
(\emph{a})&
(\emph{b})\tabularnewline
\multicolumn{1}{c}{}&
\tabularnewline
\multicolumn{1}{c}{\includegraphics[%
  width=0.50\columnwidth]{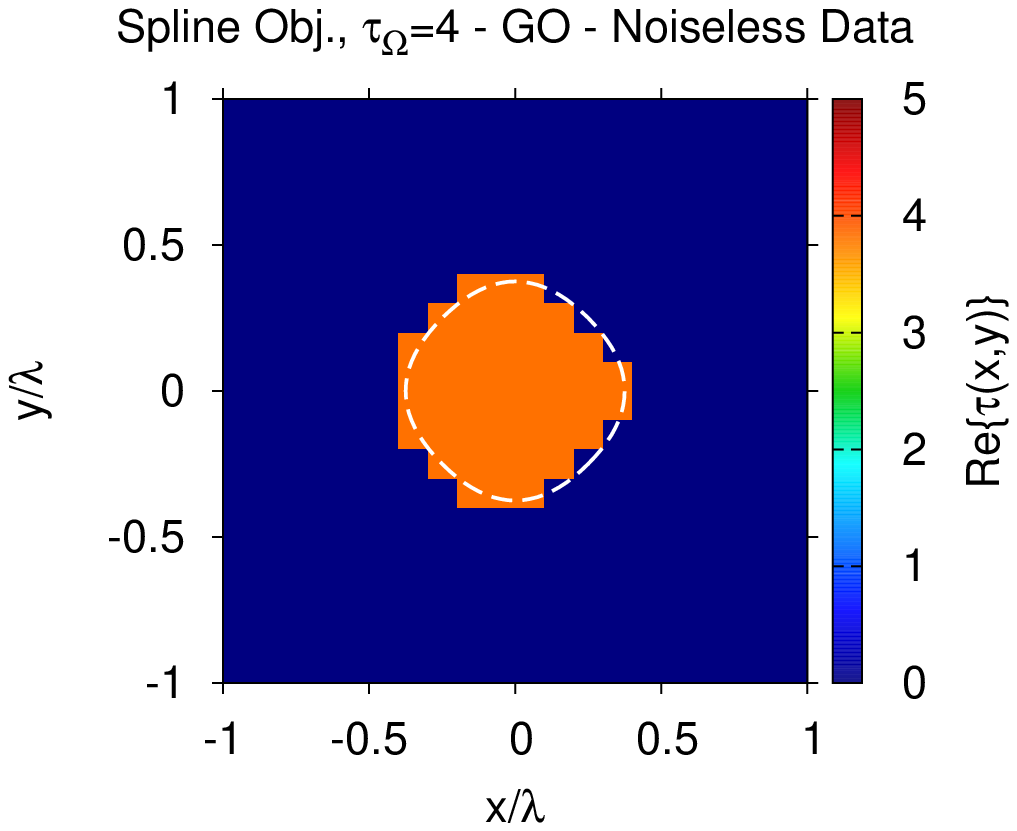}}&
\includegraphics[%
  width=0.50\columnwidth]{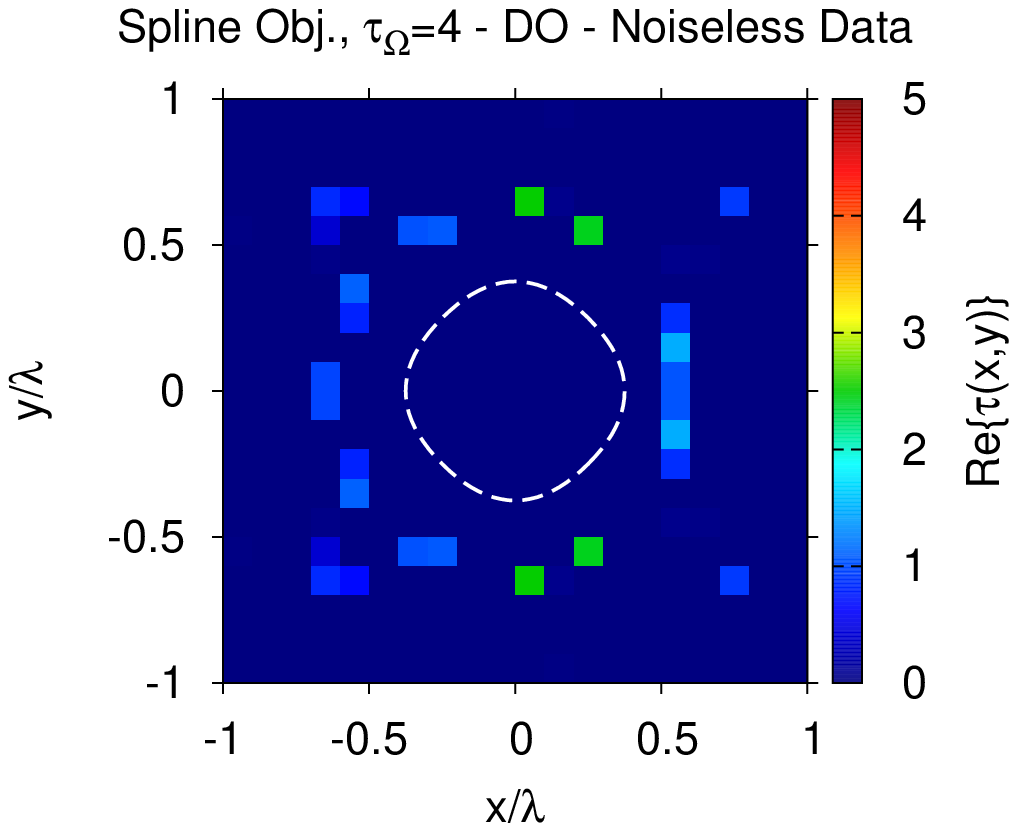}\tabularnewline
\multicolumn{1}{c}{(\emph{c})}&
(\emph{d})\tabularnewline
\end{tabular}\end{center}

\begin{center}~\vfill\end{center}

\begin{center}\textbf{Fig. 4 - M. Salucci} \textbf{\emph{et al.}}\textbf{,}
\textbf{\emph{{}``}}Learned Global Optimization ...''\end{center}

\newpage
\begin{center}~\vfill\end{center}

\begin{center}\begin{tabular}{cc}
\multicolumn{2}{c}{\includegraphics[%
  width=0.90\columnwidth]{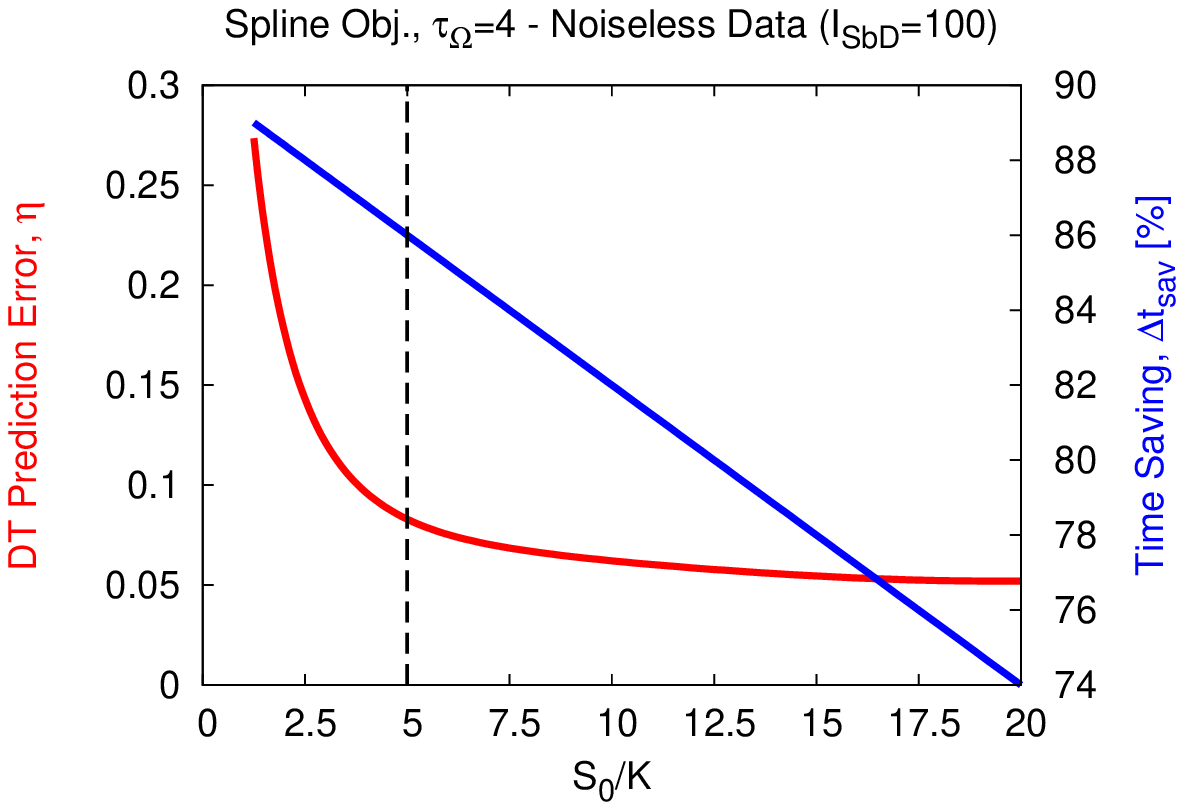}}\tabularnewline
\end{tabular}\end{center}

\begin{center}~\vfill\end{center}

\begin{center}\textbf{Fig. 5 - M. Salucci} \textbf{\emph{et al.}}\textbf{,}
\textbf{\emph{{}``}}Learned Global Optimization ...''\end{center}

\newpage
\begin{center}~\vfill\end{center}

\begin{center}\begin{tabular}{c}
\multicolumn{1}{c}{\includegraphics[%
  width=0.90\columnwidth]{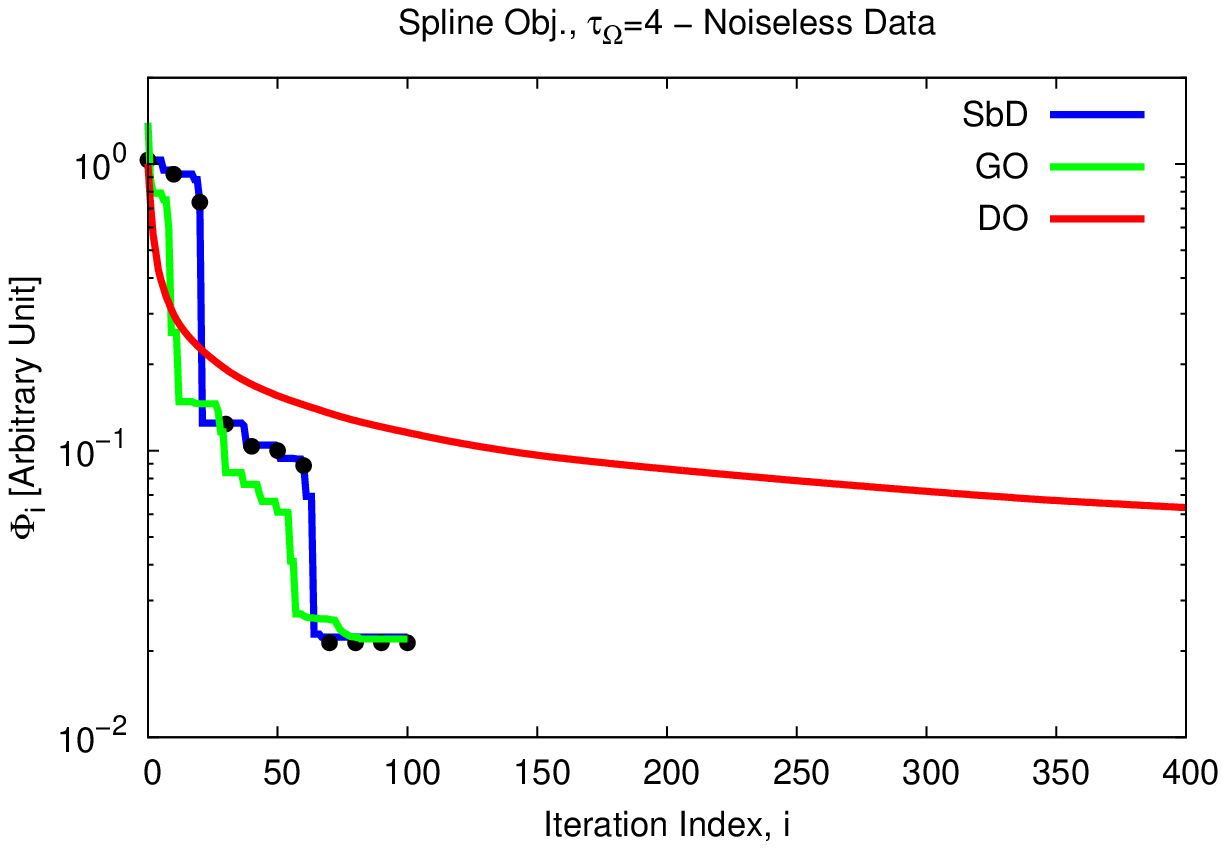}}\tabularnewline
\end{tabular}\end{center}

\begin{center}~\vfill\end{center}

\begin{center}\textbf{Fig. 6 - M. Salucci} \textbf{\emph{et al.}}\textbf{,}
\textbf{\emph{{}``}}Learned Global Optimization ...''\end{center}

\newpage
\begin{center}~\vfill\end{center}

\begin{center}\begin{tabular}{cc}
\multicolumn{1}{c}{\includegraphics[%
  width=0.45\columnwidth]{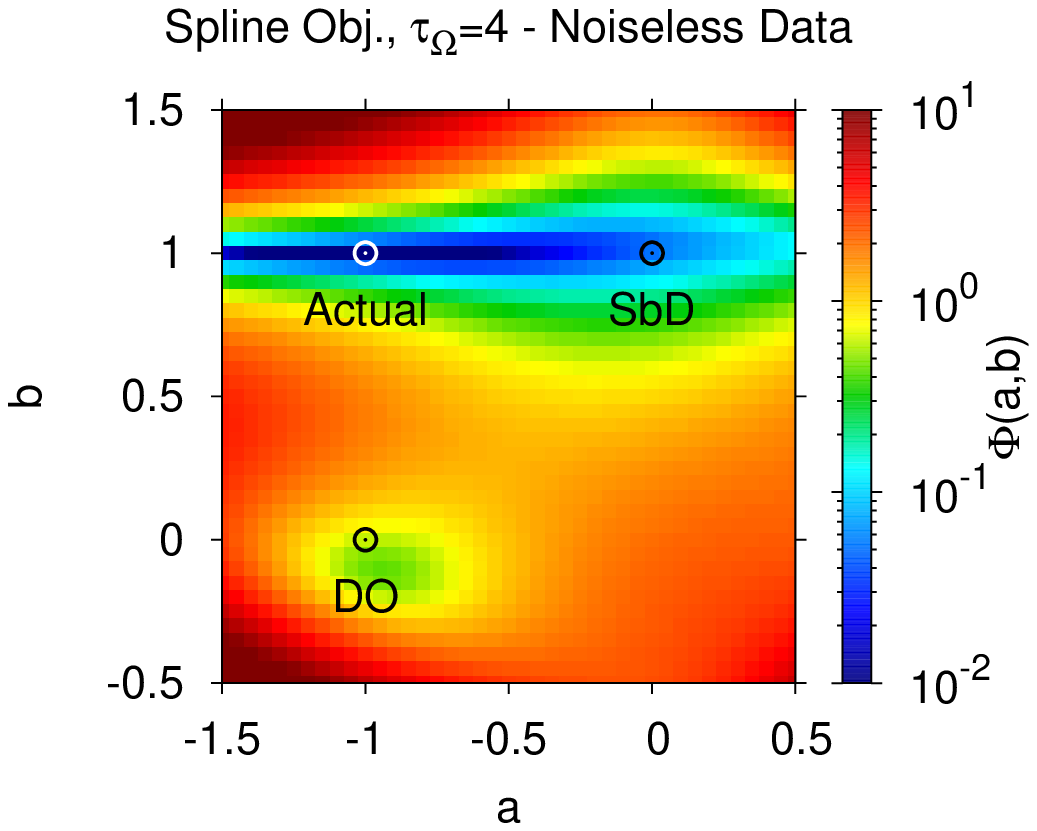}}&
\includegraphics[%
  width=0.45\columnwidth]{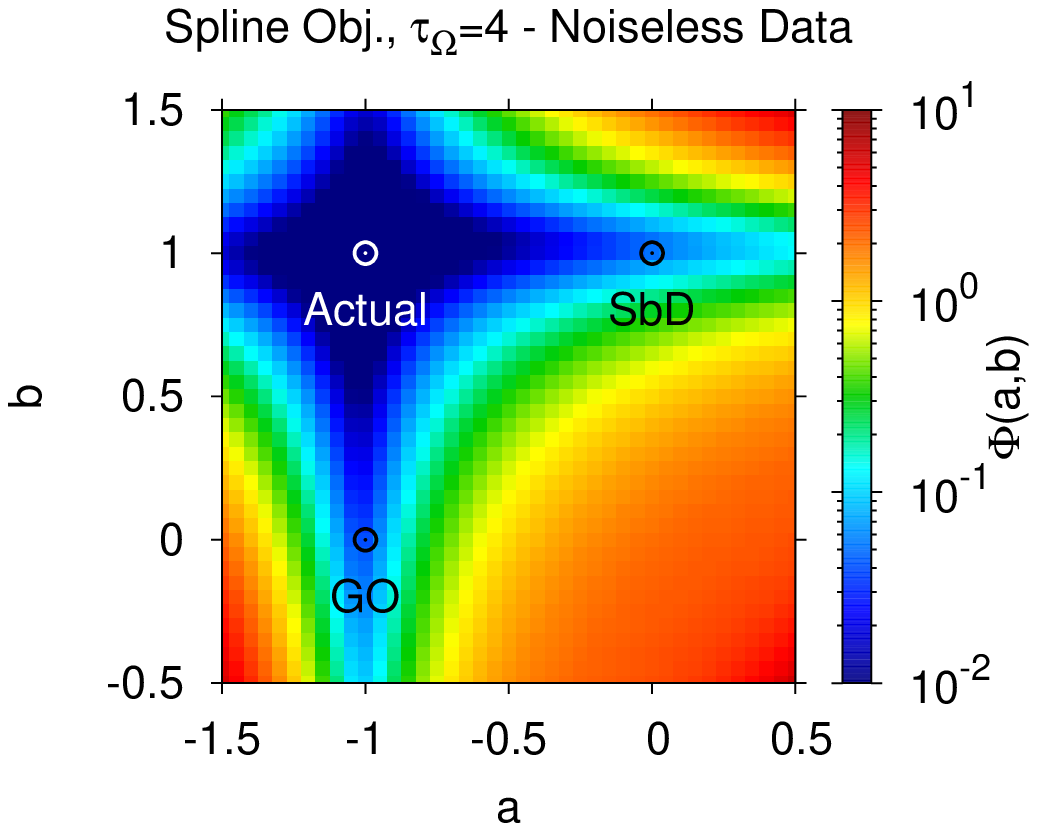}\tabularnewline
\multicolumn{1}{c}{(\emph{a})}&
(\emph{b})\tabularnewline
\multicolumn{1}{c}{}&
\tabularnewline
\multicolumn{2}{c}{\includegraphics[%
  width=0.90\columnwidth]{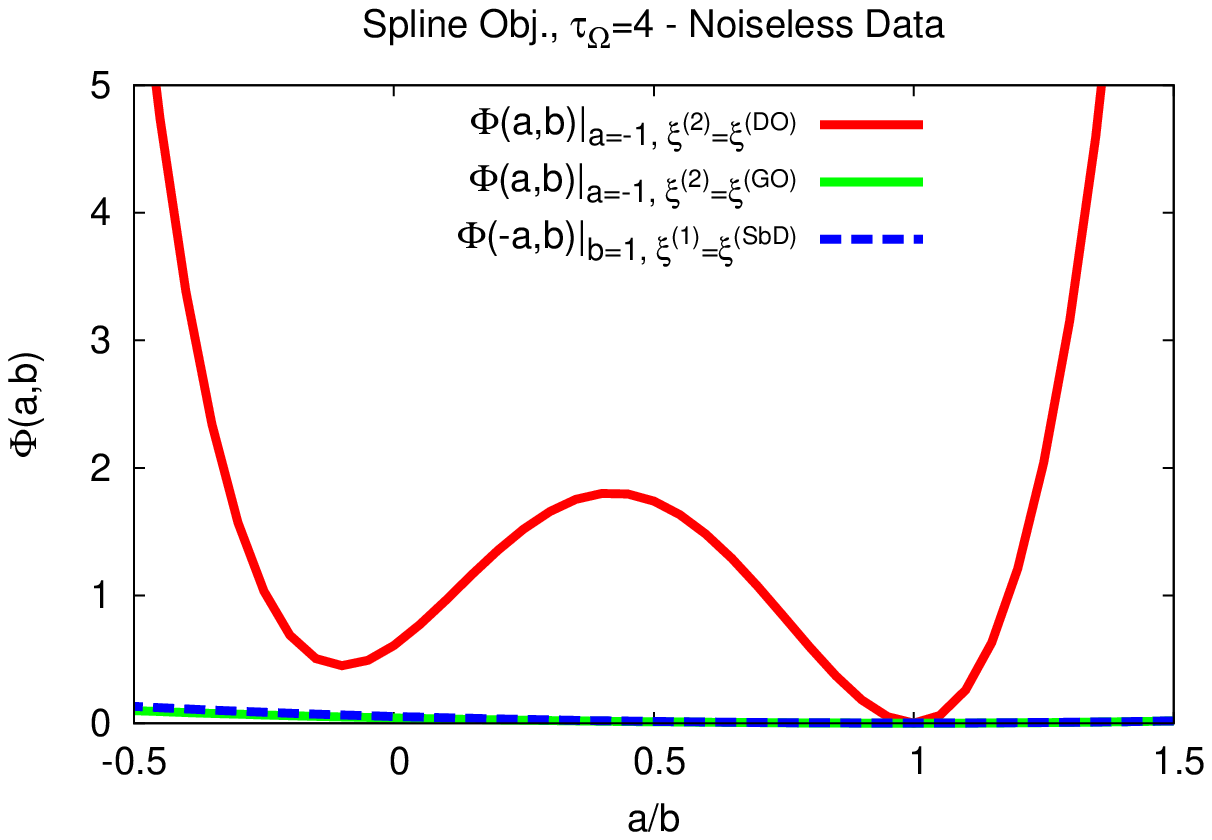}}\tabularnewline
\multicolumn{2}{c}{(\emph{c})}\tabularnewline
\end{tabular}\end{center}

\begin{center}~\vfill\end{center}

\begin{center}\textbf{Fig. 7 - M. Salucci} \textbf{\emph{et al.}}\textbf{,}
\textbf{\emph{{}``}}Learned Global Optimization ...''\end{center}

\newpage
\begin{center}~\vfill\end{center}

\begin{center}\begin{tabular}{c}
\multicolumn{1}{c}{\includegraphics[%
  width=0.90\columnwidth]{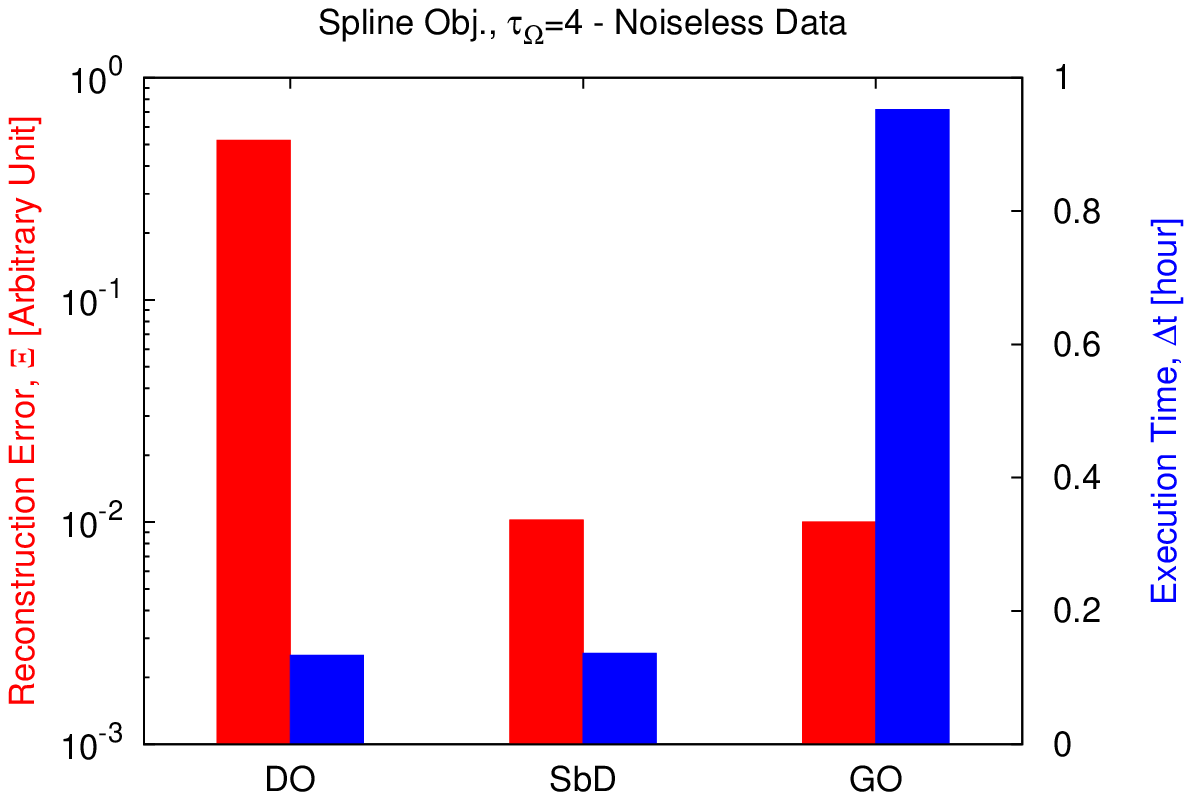}}\tabularnewline
\end{tabular}\end{center}

\begin{center}~\vfill\end{center}

\begin{center}\textbf{Fig. 8 - M. Salucci} \textbf{\emph{et al.}}\textbf{,}
\textbf{\emph{{}``}}Learned Global Optimization ...''\end{center}

\newpage
\begin{center}~\vfill\end{center}

\begin{center}\begin{tabular}{c}
\multicolumn{1}{c}{\includegraphics[%
  width=0.75\columnwidth]{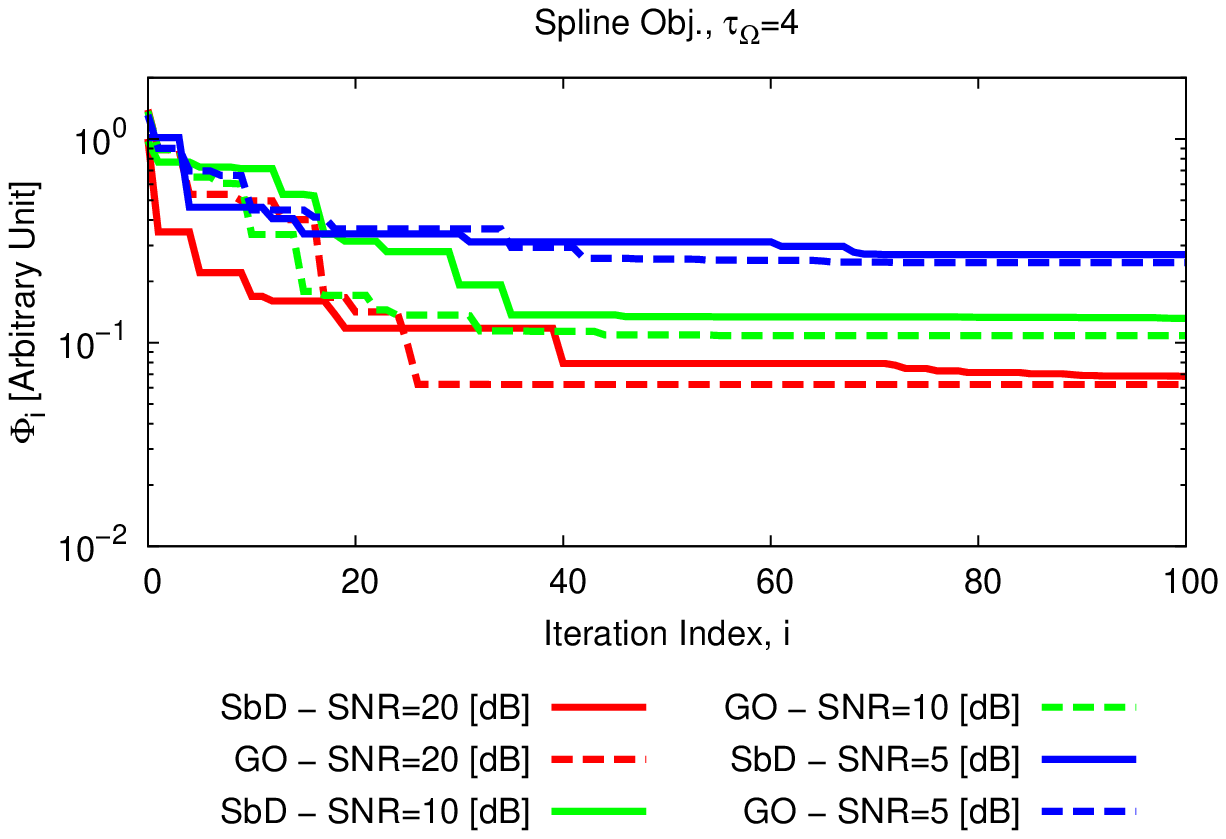}}\tabularnewline
\multicolumn{1}{c}{(\emph{a})}\tabularnewline
\multicolumn{1}{c}{}\tabularnewline
\multicolumn{1}{c}{\includegraphics[%
  width=0.75\columnwidth]{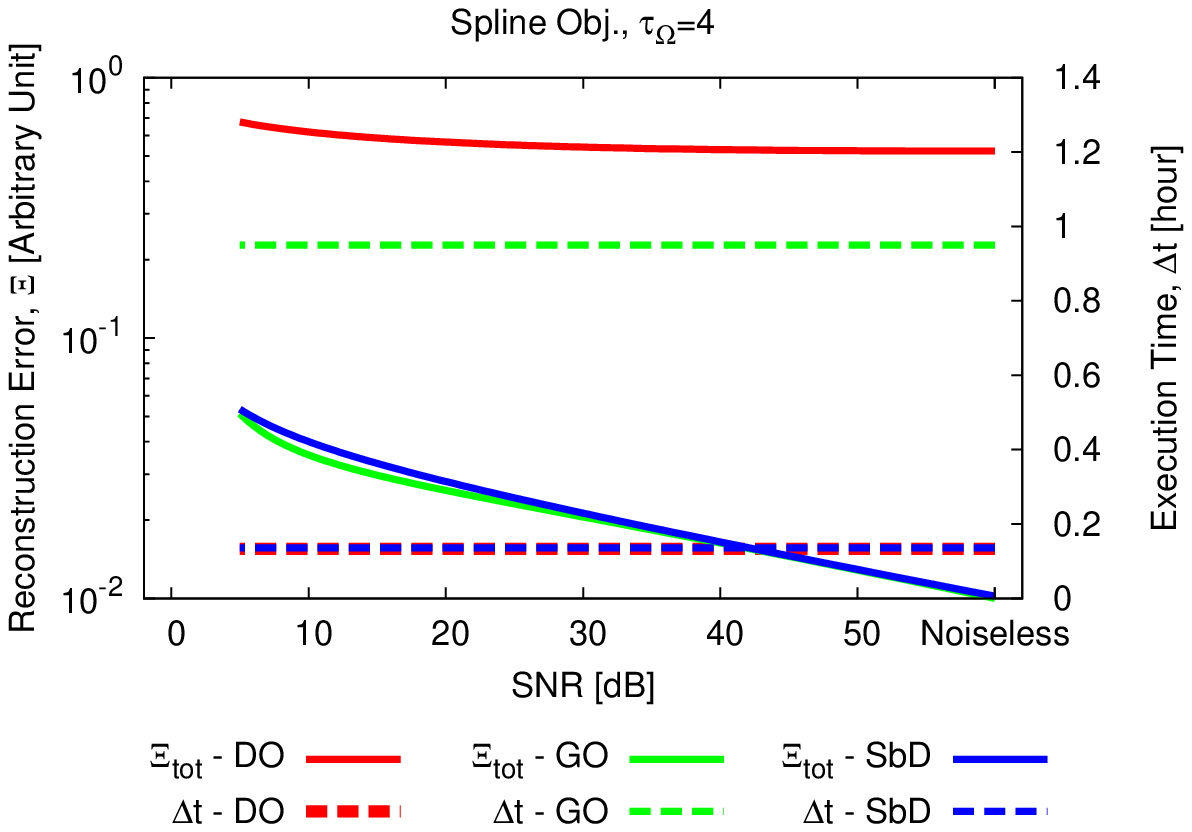}}\tabularnewline
\multicolumn{1}{c}{(\emph{b})}\tabularnewline
\end{tabular}\end{center}

\begin{center}~\vfill\end{center}

\begin{center}\textbf{Fig. 9 - M. Salucci} \textbf{\emph{et al.}}\textbf{,}
\textbf{\emph{{}``}}Learned Global Optimization ...''\end{center}

\newpage
\begin{center}~\vfill\end{center}

\begin{center}\begin{tabular}{ccc}
\begin{sideways}
\emph{~~~~~~~~~~~~}$SNR=20$ {[}dB{]}%
\end{sideways}&
\includegraphics[%
  width=0.45\columnwidth]{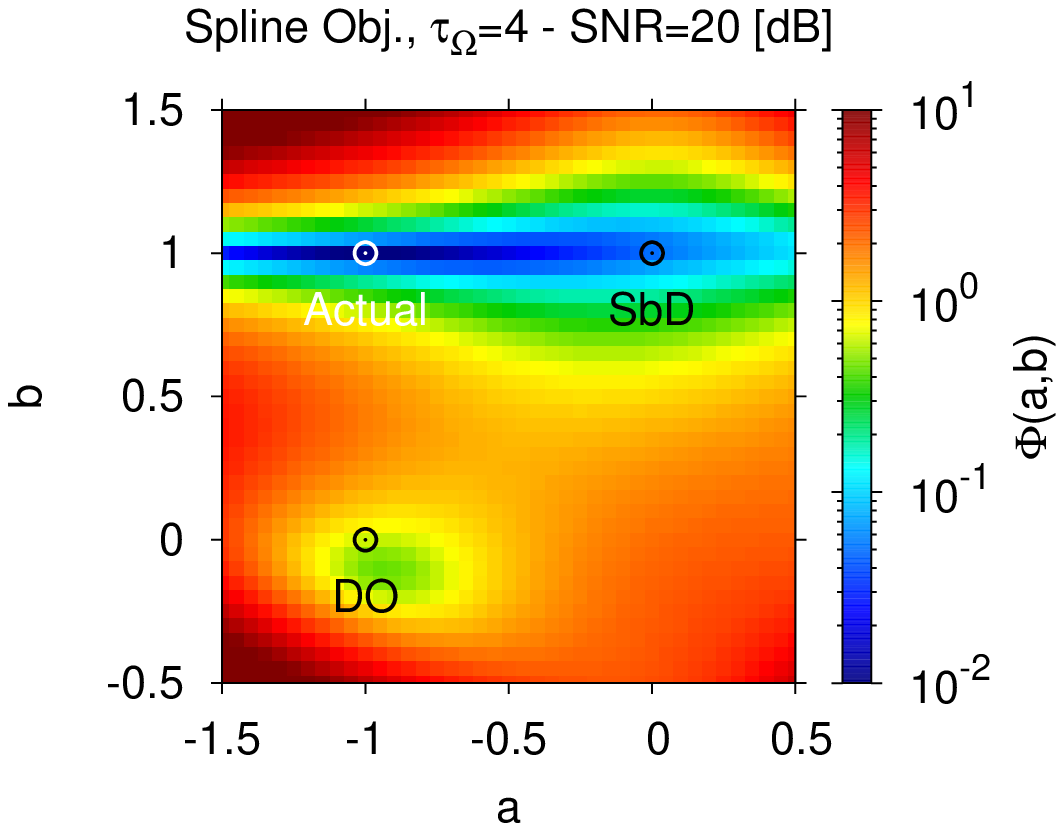}&
\includegraphics[%
  width=0.45\columnwidth]{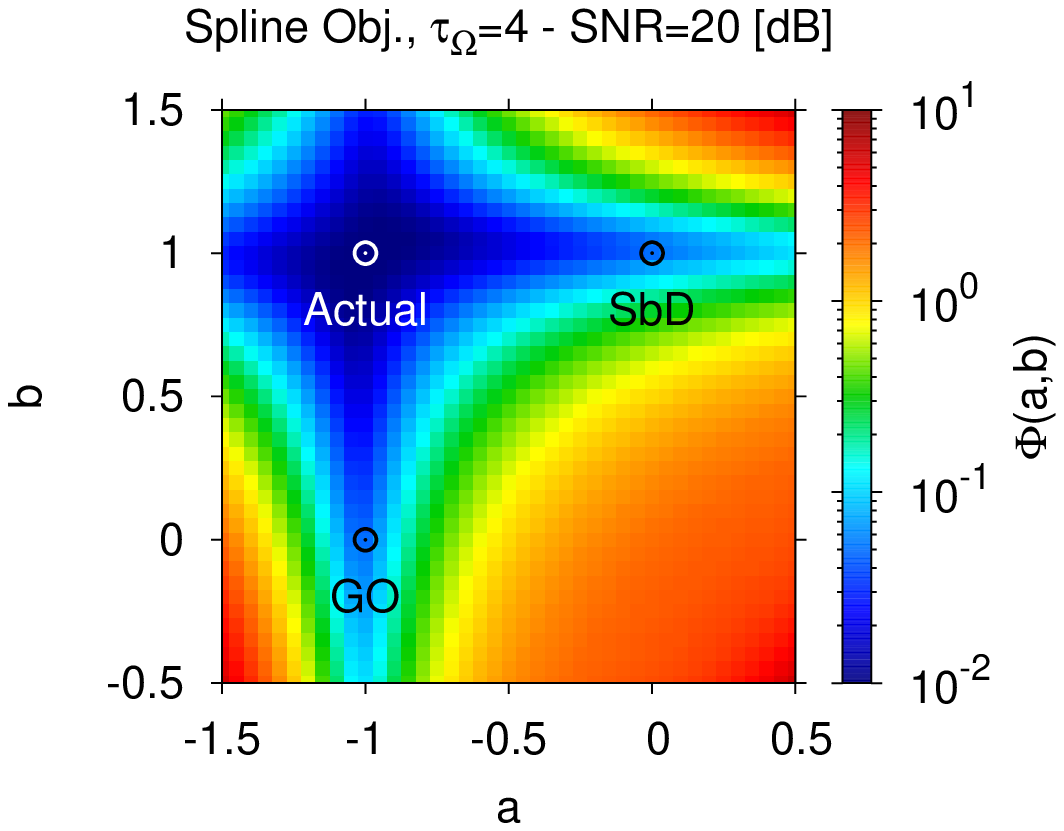}\tabularnewline
\begin{sideways}
\end{sideways}&
(\emph{a})&
(\emph{b})\tabularnewline
\begin{sideways}
\emph{~~~~~~~~~~~~}$SNR=10$ {[}dB{]}%
\end{sideways}&
\includegraphics[%
  width=0.45\columnwidth]{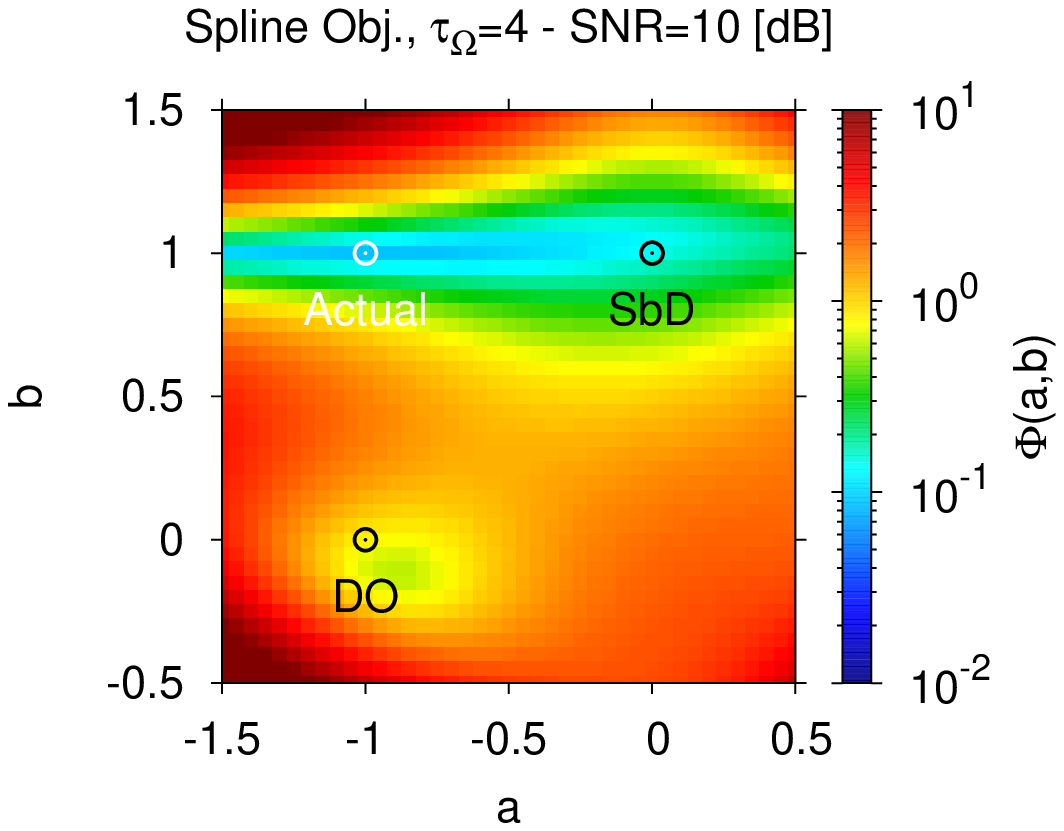}&
\includegraphics[%
  width=0.45\columnwidth]{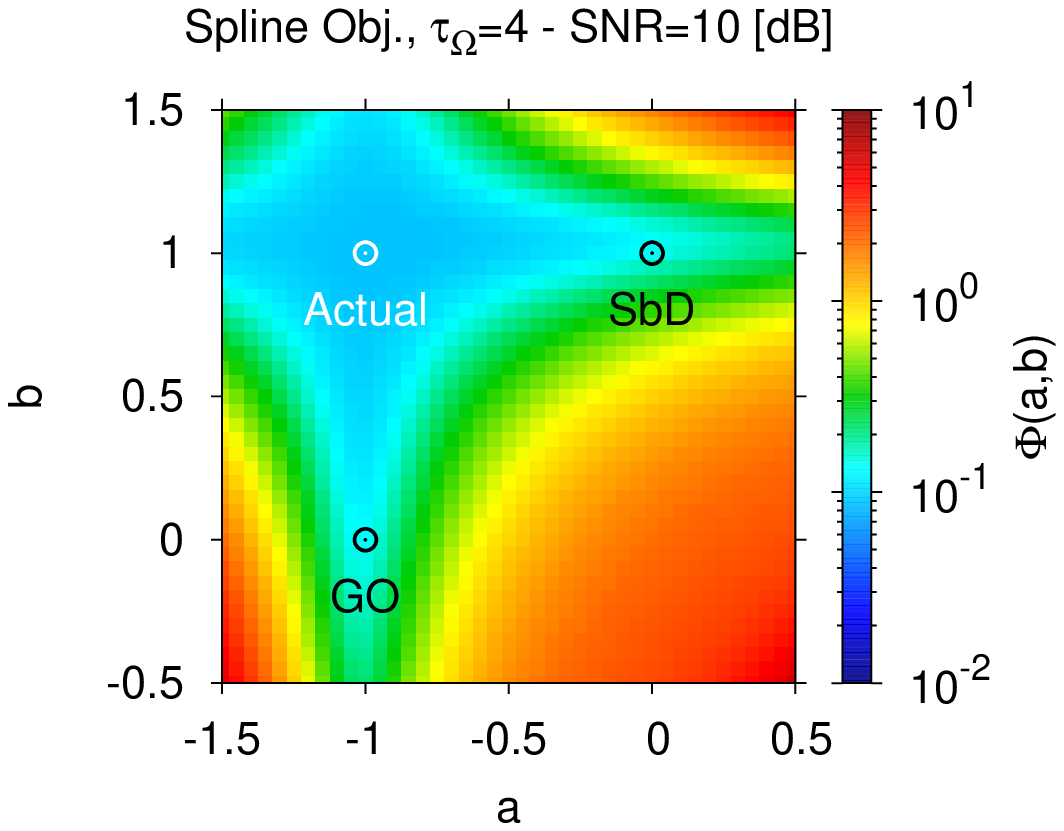}\tabularnewline
\begin{sideways}
\end{sideways}&
(\emph{c})&
(\emph{d})\tabularnewline
\begin{sideways}
\emph{~~~~~~~~~~~~}$SNR=5$ {[}dB{]}%
\end{sideways}&
\includegraphics[%
  width=0.45\columnwidth]{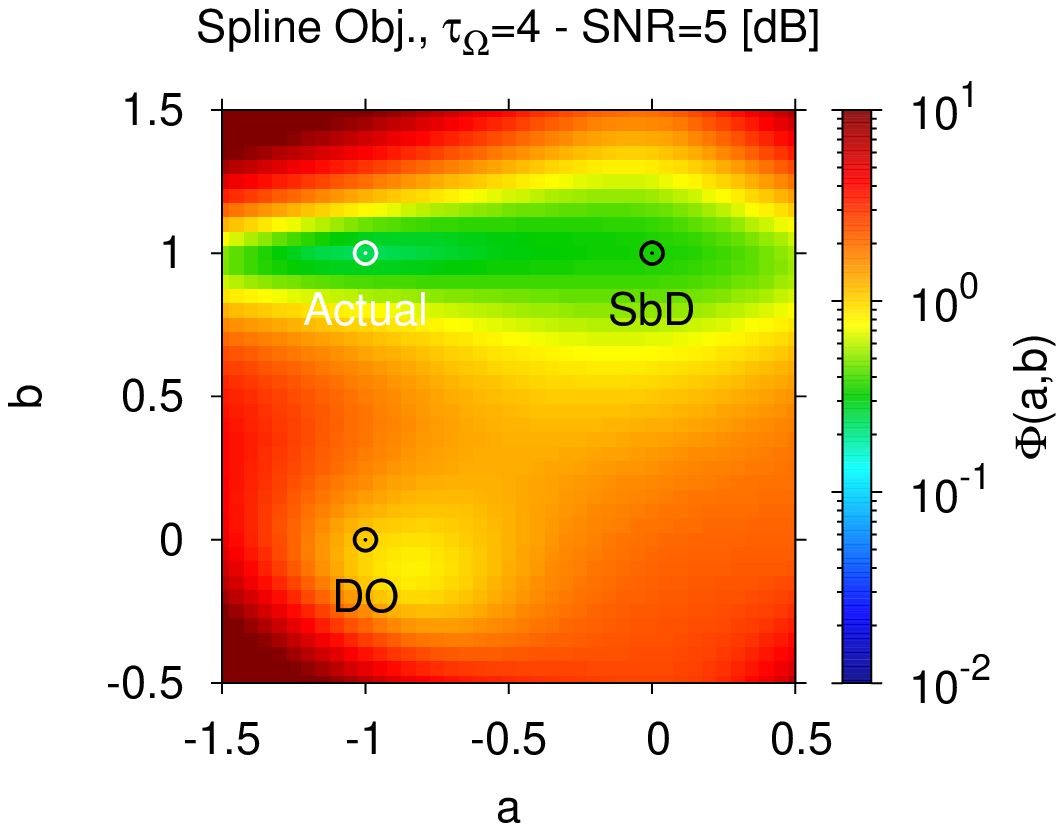}&
\includegraphics[%
  width=0.45\columnwidth]{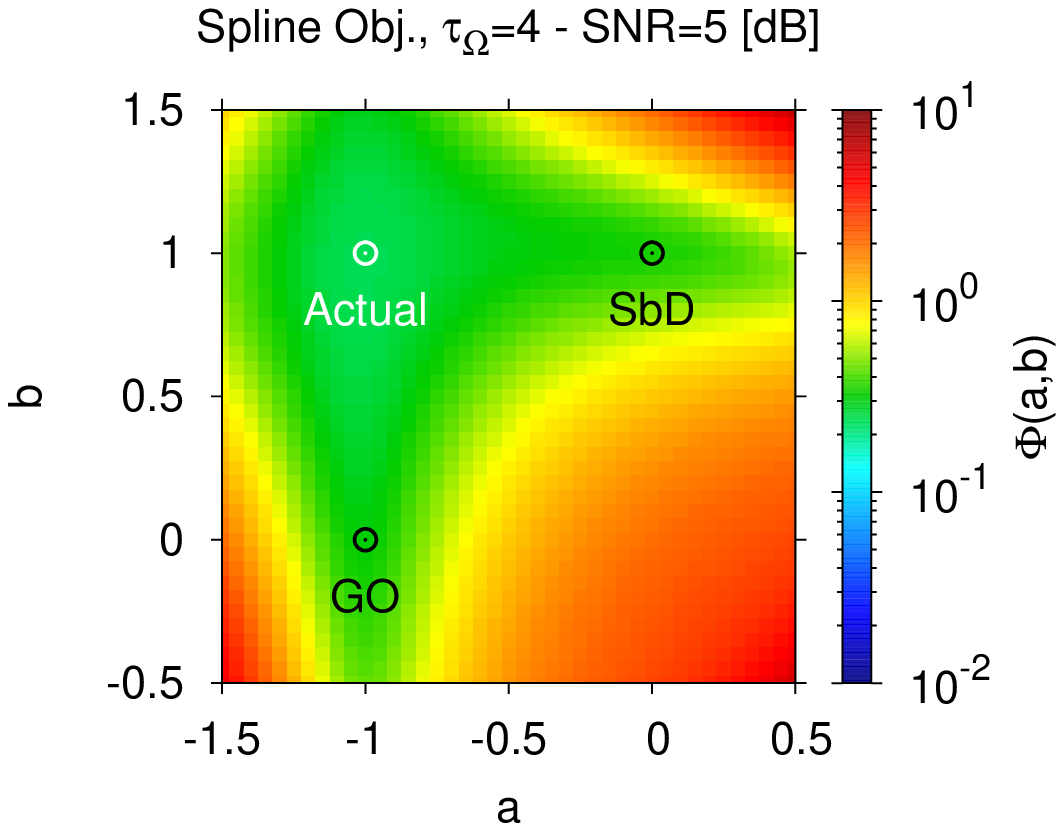}\tabularnewline
\begin{sideways}
\end{sideways}&
(\emph{e})&
(\emph{f})\tabularnewline
\end{tabular}\end{center}

\begin{center}~\vfill\end{center}

\begin{center}\textbf{Fig. 10 - M. Salucci} \textbf{\emph{et al.}}\textbf{,}
\textbf{\emph{{}``}}Learned Global Optimization ...''\end{center}

\newpage
\begin{center}~\vfill\end{center}

\begin{center}\begin{tabular}{cccc}
\begin{sideways}
\end{sideways}&
$SNR=20$ {[}dB{]}&
$SNR=10$ {[}dB{]}&
$SNR=5$ {[}dB{]}\tabularnewline
\begin{sideways}
\emph{~~~~~~~~~~~~~~~~~~SbD}%
\end{sideways}&
\includegraphics[%
  width=0.33\columnwidth]{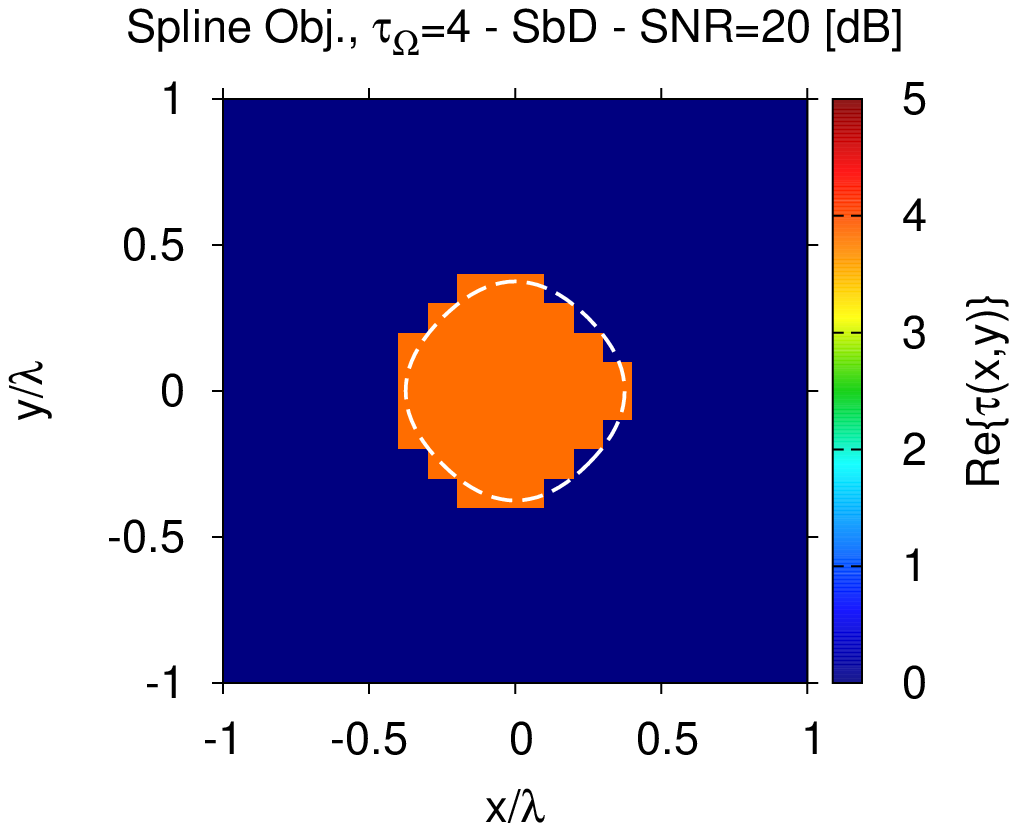}&
\includegraphics[%
  width=0.33\columnwidth]{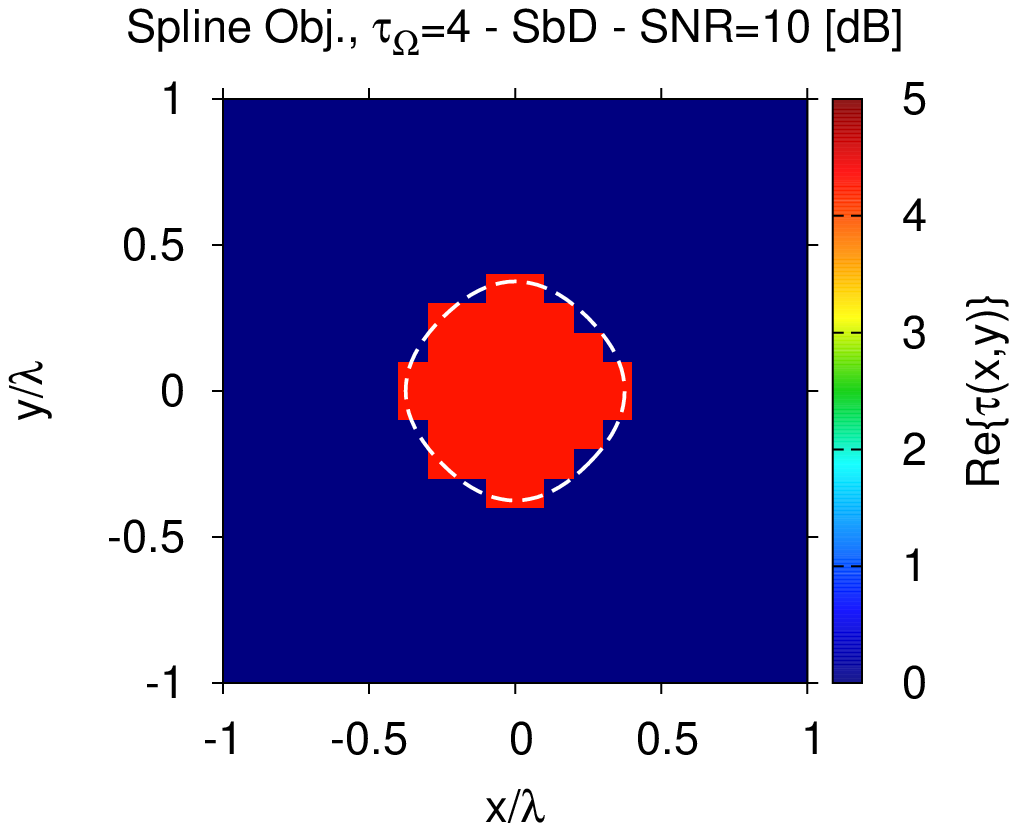}&
\includegraphics[%
  width=0.33\columnwidth]{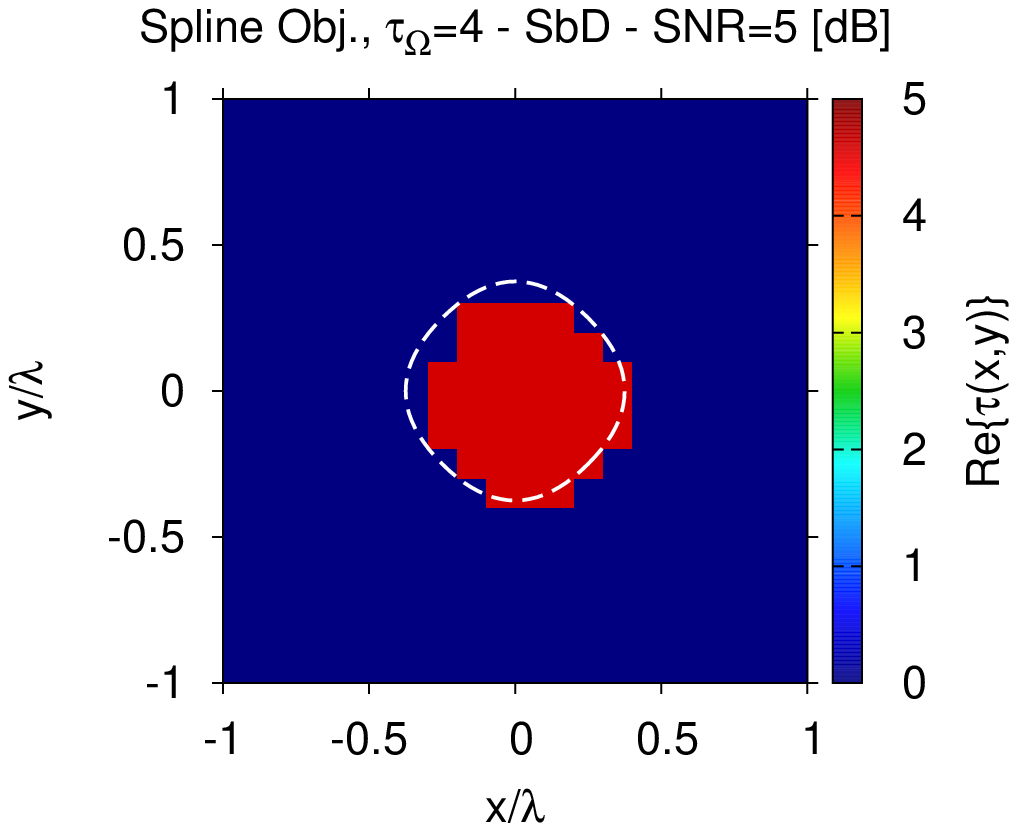}\tabularnewline
\begin{sideways}
\end{sideways}&
(\emph{a})&
(\emph{b})&
(\emph{c})\tabularnewline
\begin{sideways}
\end{sideways}&
&
&
\tabularnewline
\begin{sideways}
\emph{~~~~~~~~~~~~~~~~~~GO}%
\end{sideways}&
\includegraphics[%
  width=0.33\columnwidth]{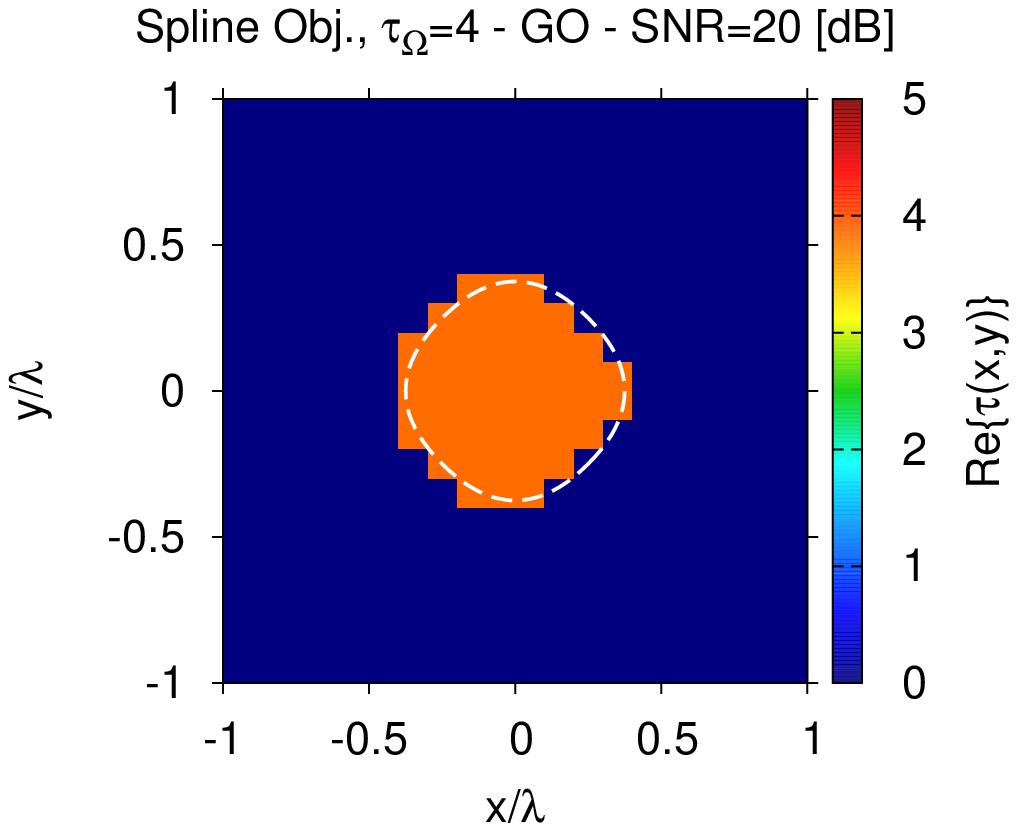}&
\includegraphics[%
  width=0.33\columnwidth]{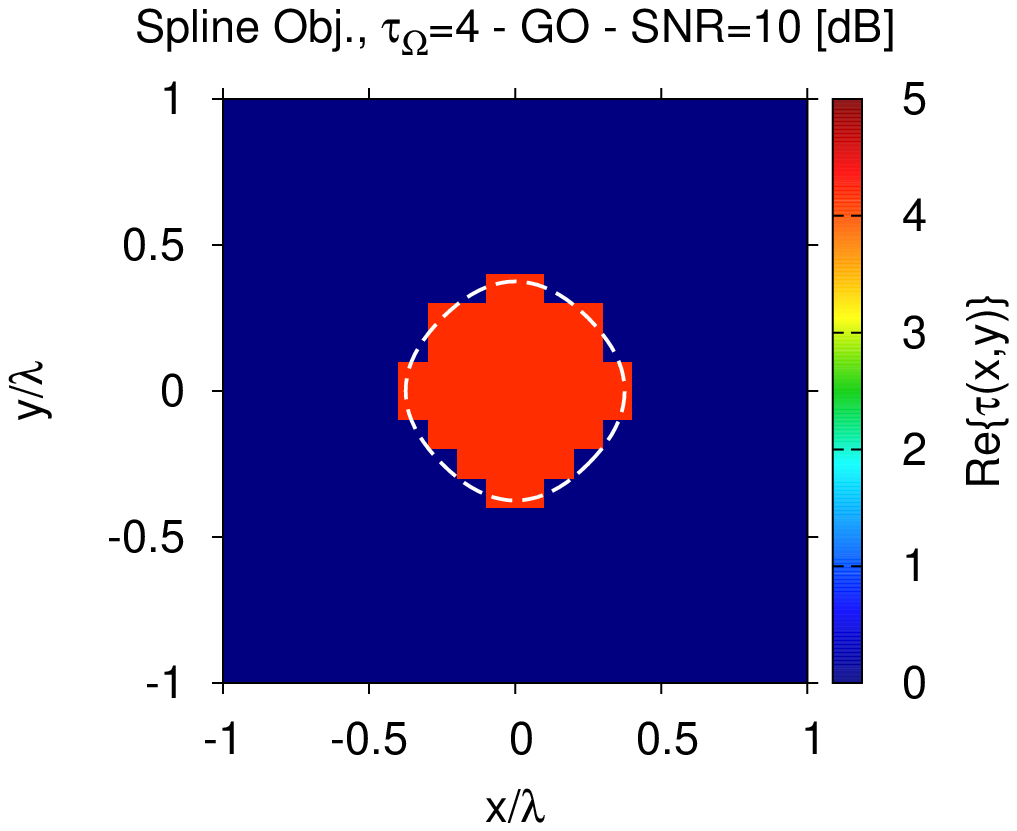}&
\includegraphics[%
  width=0.33\columnwidth]{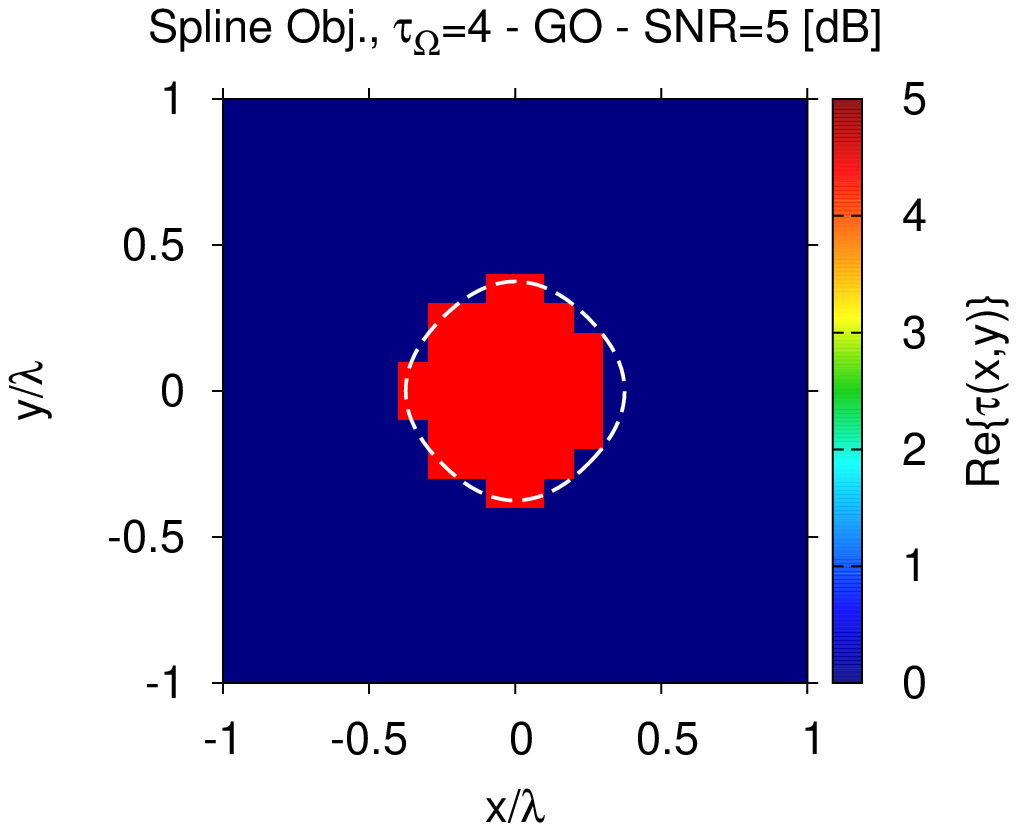}\tabularnewline
\begin{sideways}
\end{sideways}&
(\emph{d})&
(\emph{e})&
(\emph{f})\tabularnewline
\begin{sideways}
\end{sideways}&
&
&
\tabularnewline
\begin{sideways}
\emph{~~~~~~~~~~~~~~~~~~DO}%
\end{sideways}&
\includegraphics[%
  width=0.33\columnwidth]{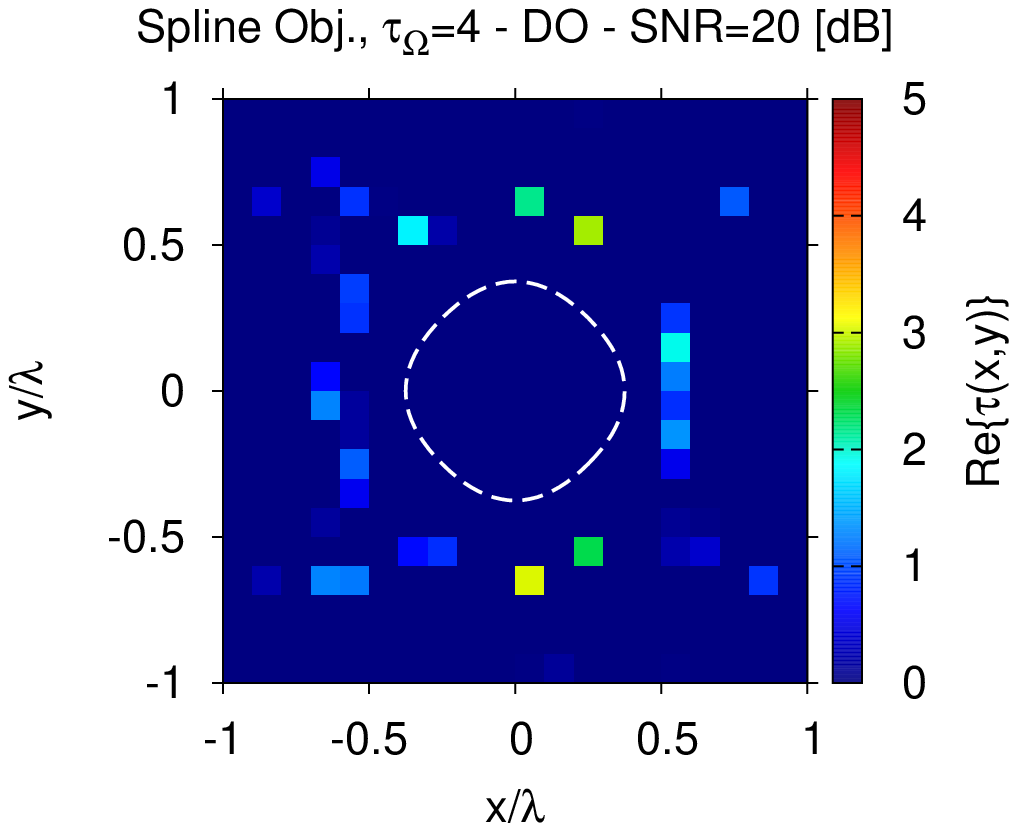}&
\includegraphics[%
  width=0.33\columnwidth]{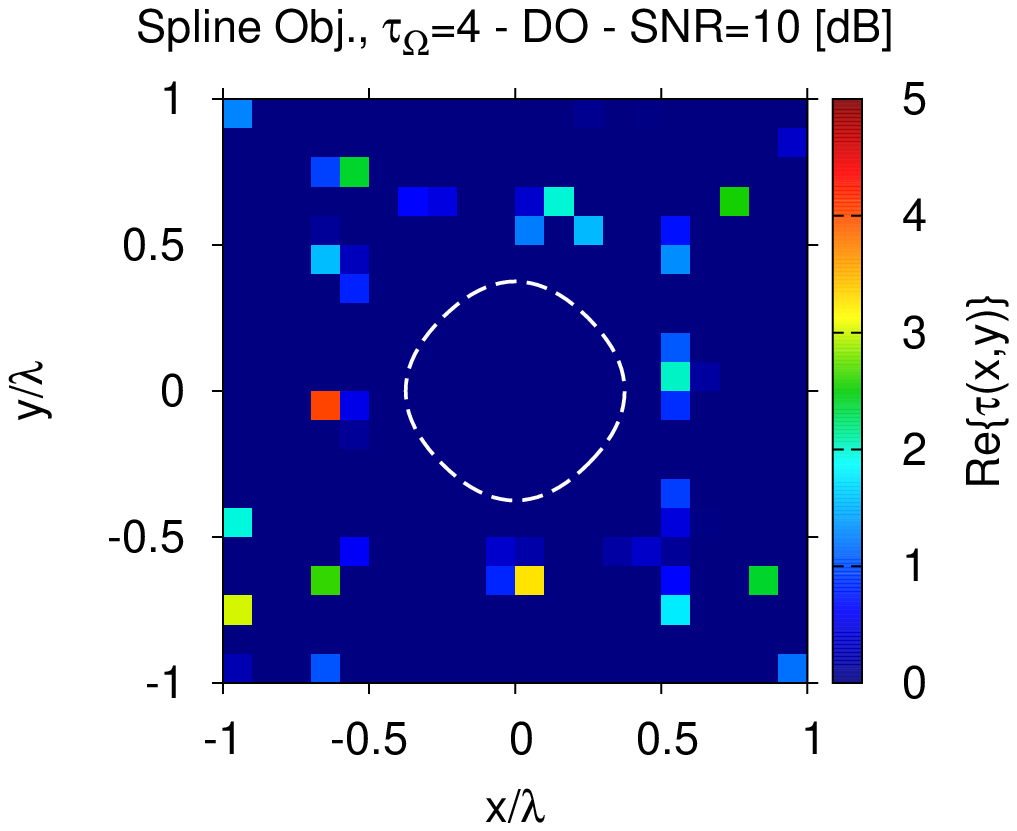}&
\includegraphics[%
  width=0.33\columnwidth]{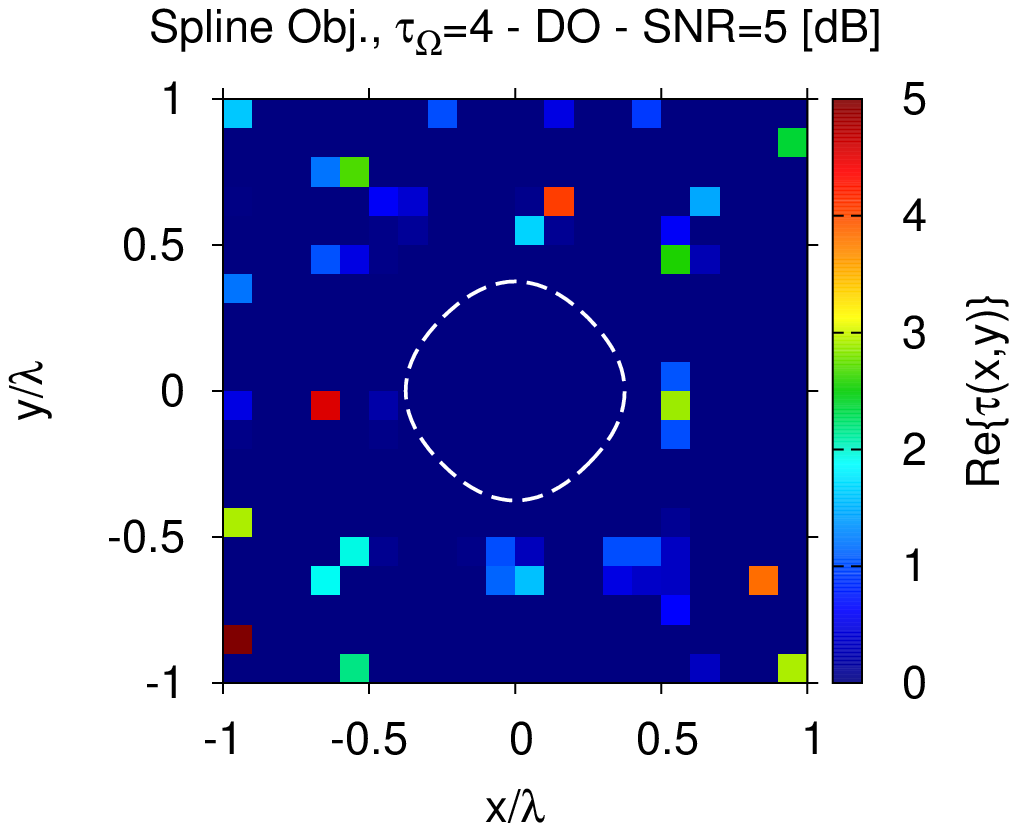}\tabularnewline
\begin{sideways}
\end{sideways}&
(\emph{g})&
(\emph{h})&
(\emph{i})\tabularnewline
\end{tabular}\end{center}

\begin{center}~\vfill\end{center}

\begin{center}\textbf{Fig. 11 - M. Salucci} \textbf{\emph{et al.}}\textbf{,}
\textbf{\emph{{}``}}Learned Global Optimization ...''\end{center}

\newpage
\begin{center}~\vfill\end{center}

\begin{center}\begin{tabular}{c}
\multicolumn{1}{c}{\includegraphics[%
  width=0.75\columnwidth]{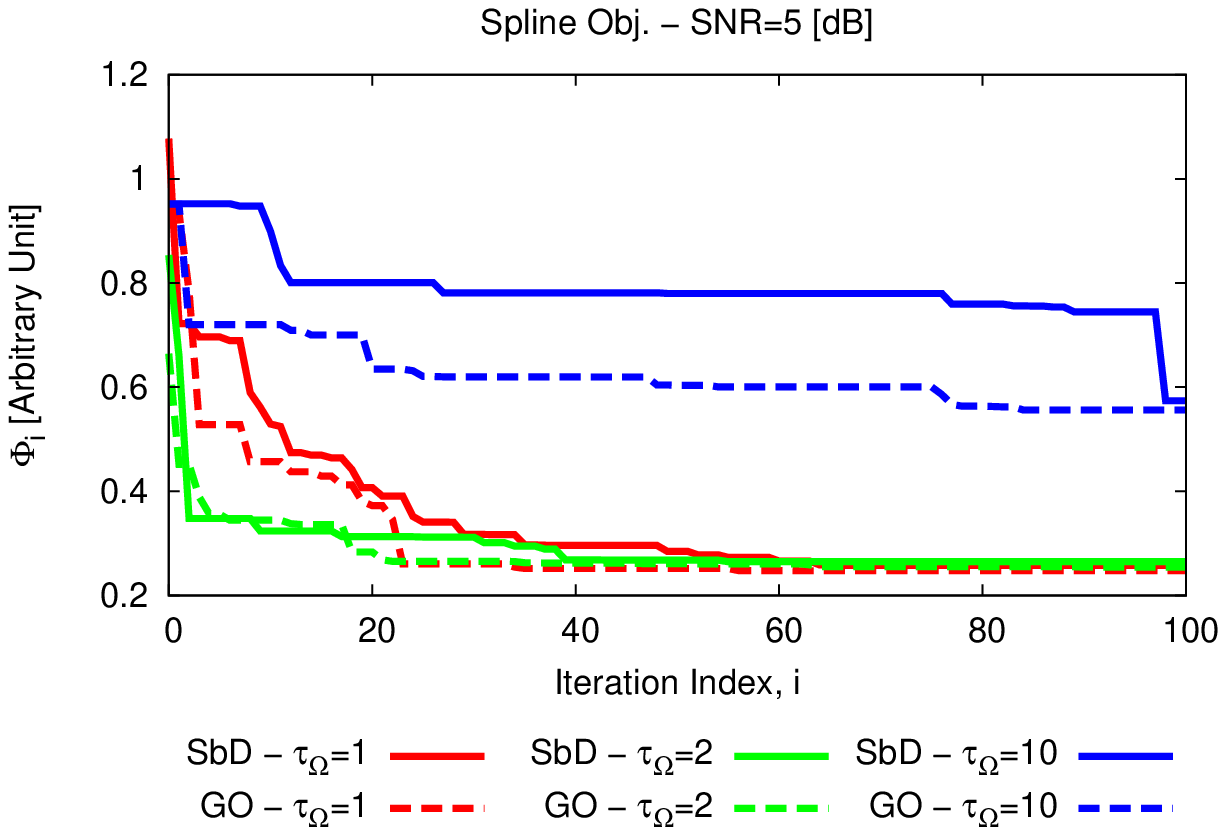}}\tabularnewline
\multicolumn{1}{c}{(\emph{a})}\tabularnewline
\multicolumn{1}{c}{}\tabularnewline
\multicolumn{1}{c}{\includegraphics[%
  width=0.75\columnwidth]{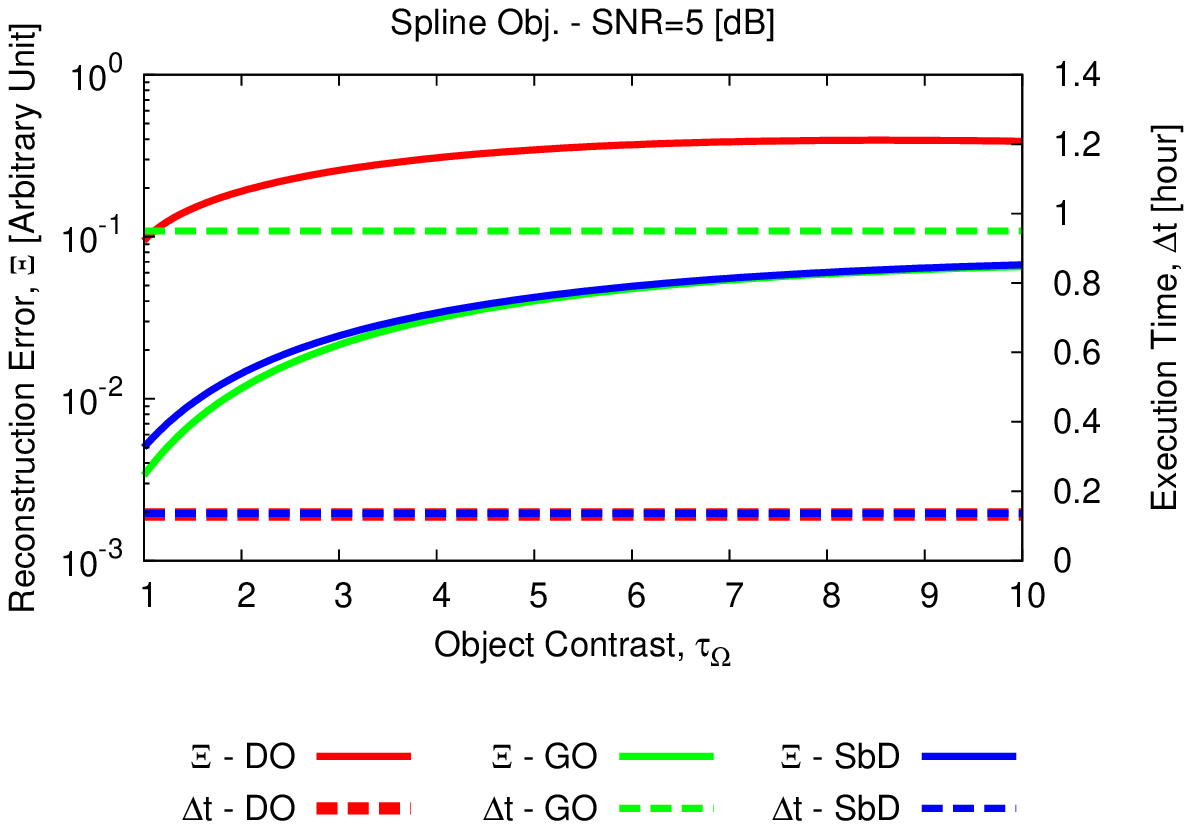}}\tabularnewline
\multicolumn{1}{c}{(\emph{b})}\tabularnewline
\end{tabular}\end{center}

\begin{center}~\vfill\end{center}

\begin{center}\textbf{Fig. 12- M. Salucci} \textbf{\emph{et al.}}\textbf{,}
\textbf{\emph{{}``}}Learned Global Optimization ...''\end{center}

\newpage
\begin{center}~\vfill\end{center}

\begin{center}\begin{tabular}{cccc}
\begin{sideways}
\end{sideways}&
$\tau_{\Omega}=1$&
$\tau_{\Omega}=2$&
$\tau_{\Omega}=10$\tabularnewline
\begin{sideways}
\emph{~~~~~~~~~~~~~~~~~~SbD}%
\end{sideways}&
\includegraphics[%
  width=0.33\columnwidth]{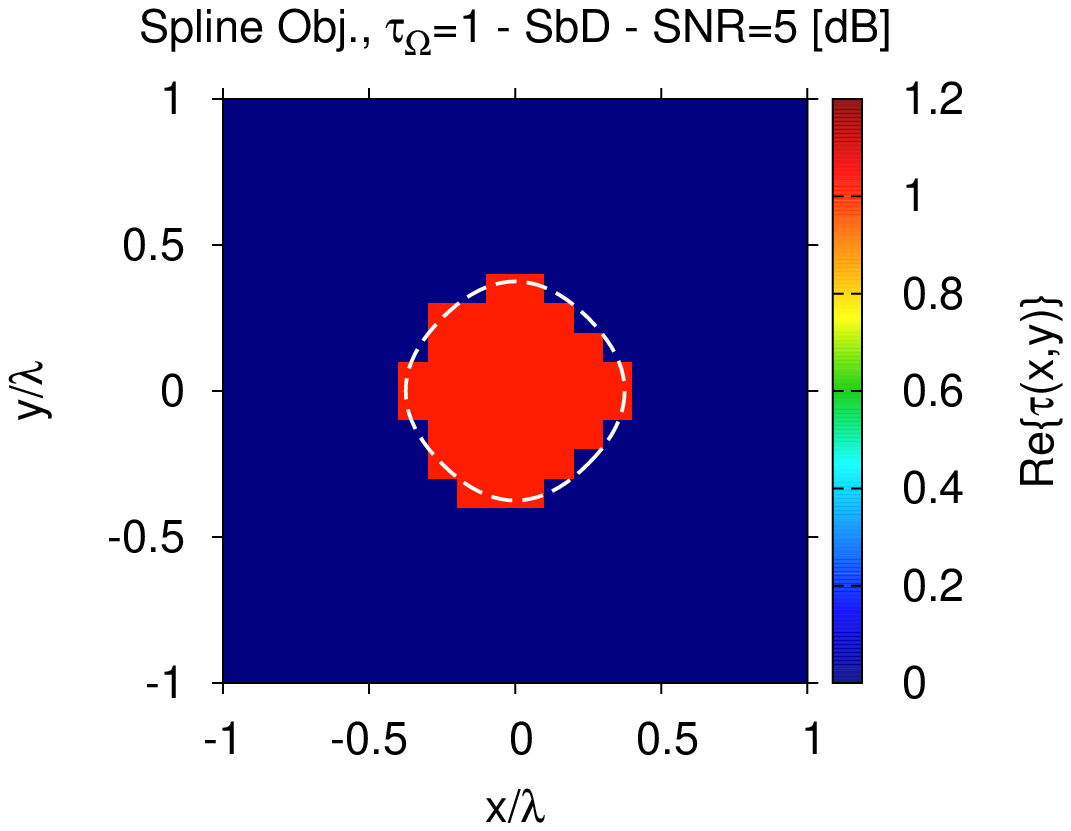}&
\includegraphics[%
  width=0.33\columnwidth]{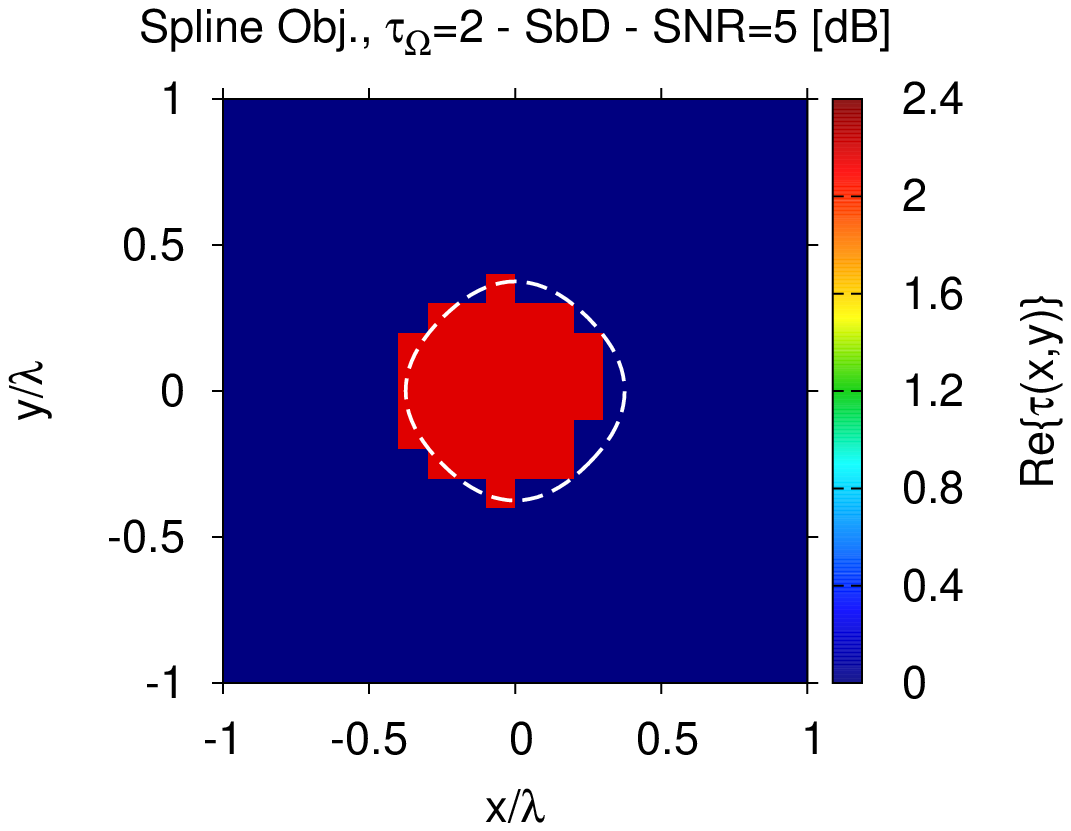}&
\includegraphics[%
  width=0.33\columnwidth]{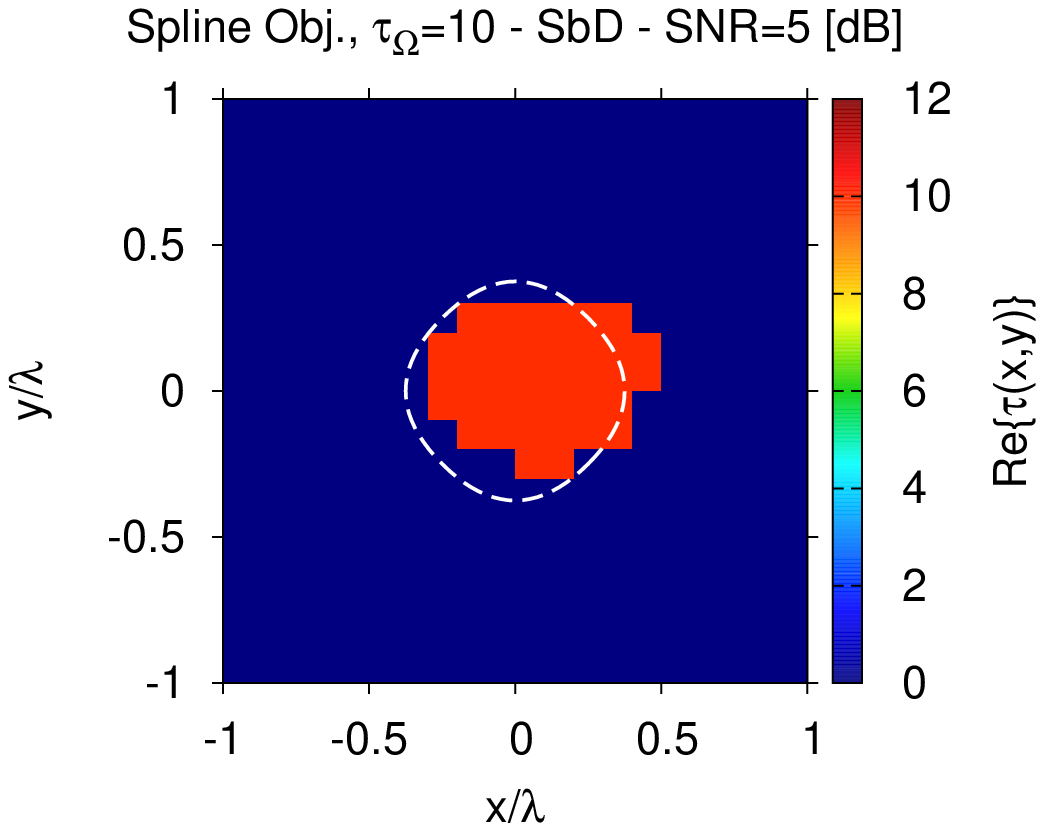}\tabularnewline
\begin{sideways}
\end{sideways}&
(\emph{a})&
(\emph{b})&
(\emph{c})\tabularnewline
\begin{sideways}
\end{sideways}&
&
&
\tabularnewline
\begin{sideways}
\emph{~~~~~~~~~~~~~~~~~~GO}%
\end{sideways}&
\includegraphics[%
  width=0.33\columnwidth]{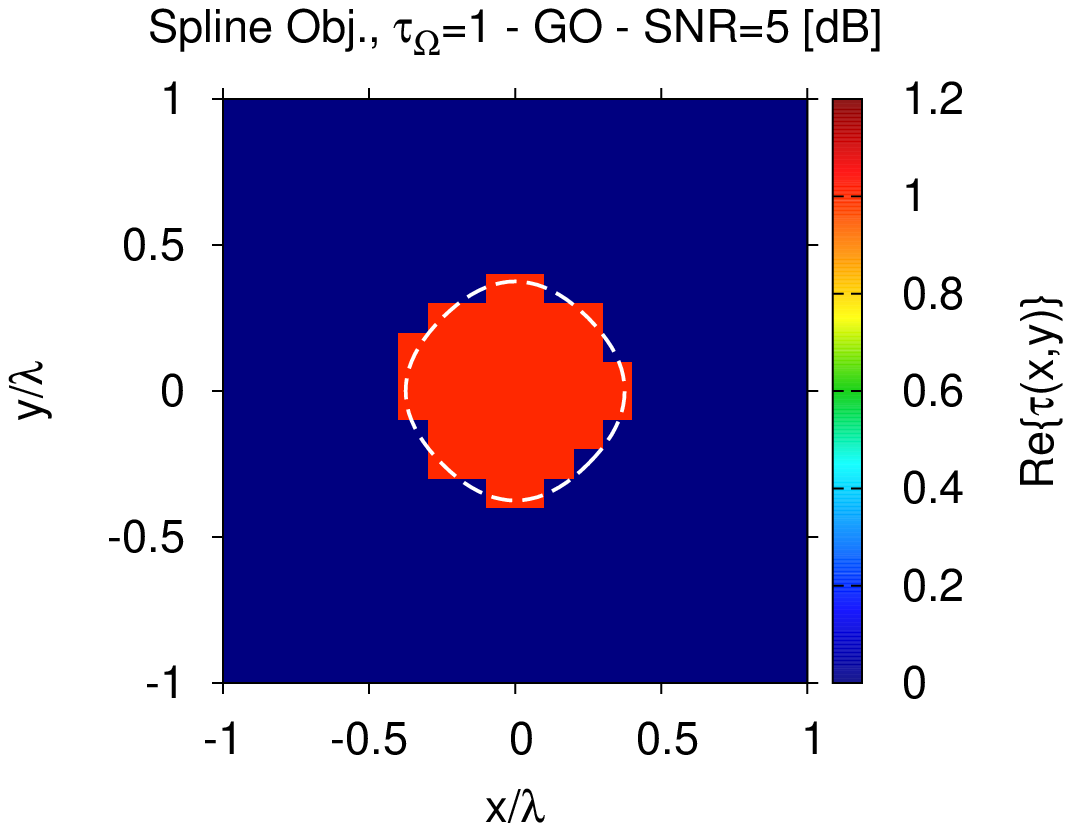}&
\includegraphics[%
  width=0.33\columnwidth]{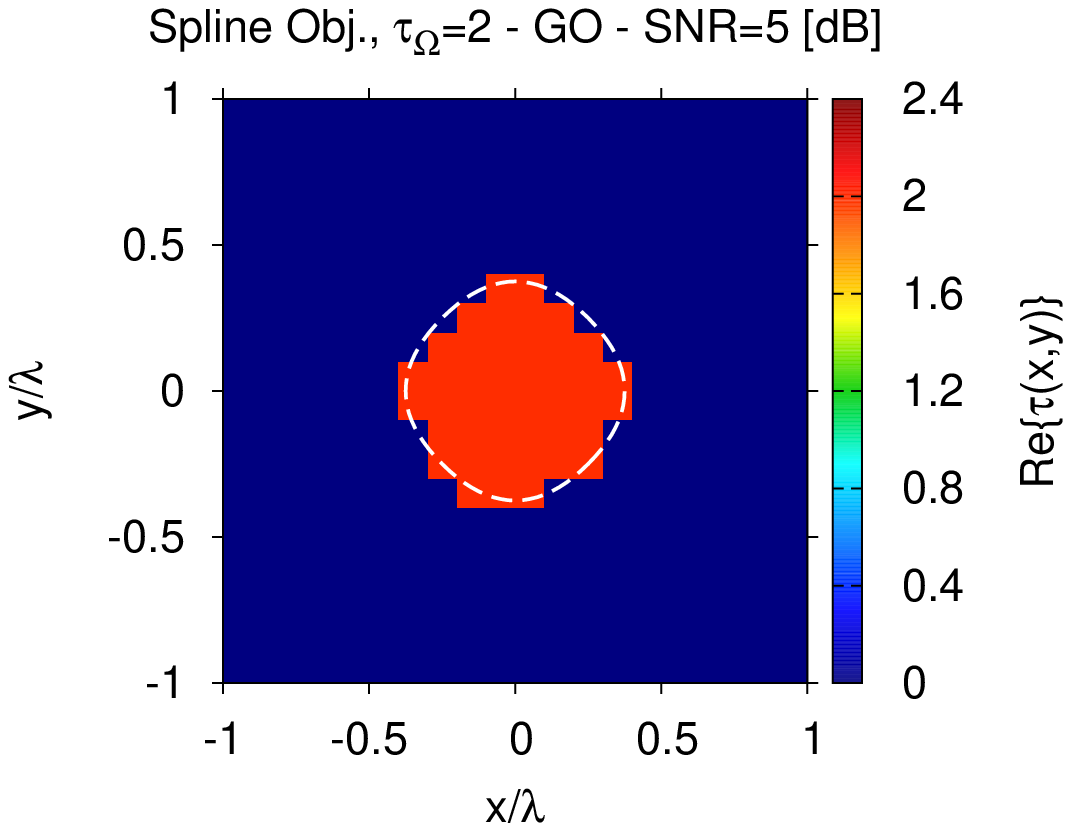}&
\includegraphics[%
  width=0.33\columnwidth]{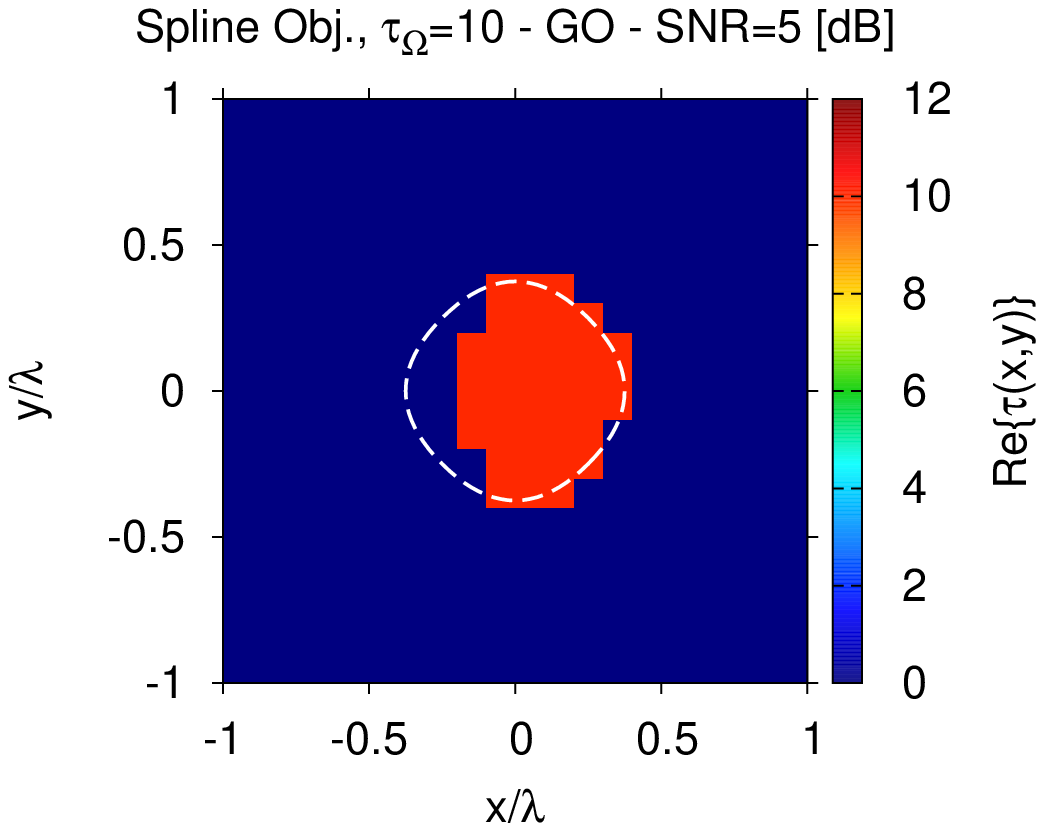}\tabularnewline
\begin{sideways}
\end{sideways}&
(\emph{d})&
(\emph{e})&
(\emph{f})\tabularnewline
\begin{sideways}
\end{sideways}&
&
&
\tabularnewline
\begin{sideways}
\emph{~~~~~~~~~~~~~~~~~~DO}%
\end{sideways}&
\includegraphics[%
  width=0.33\columnwidth]{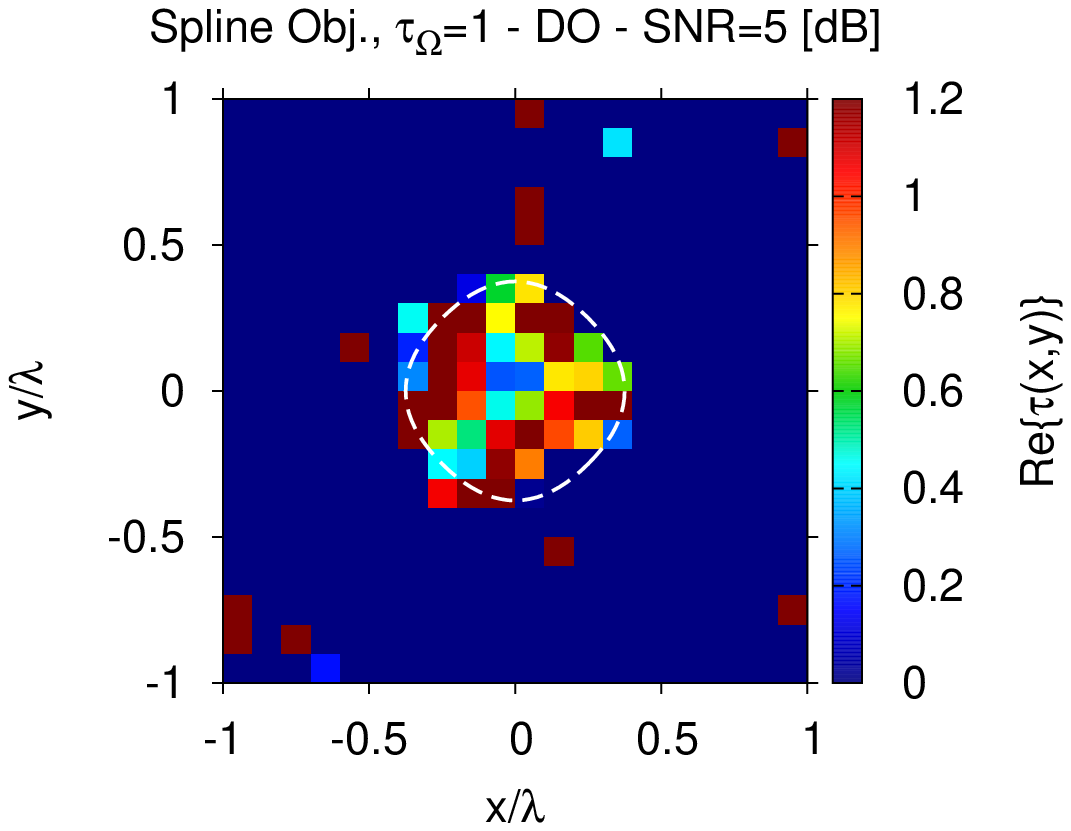}&
\includegraphics[%
  width=0.33\columnwidth]{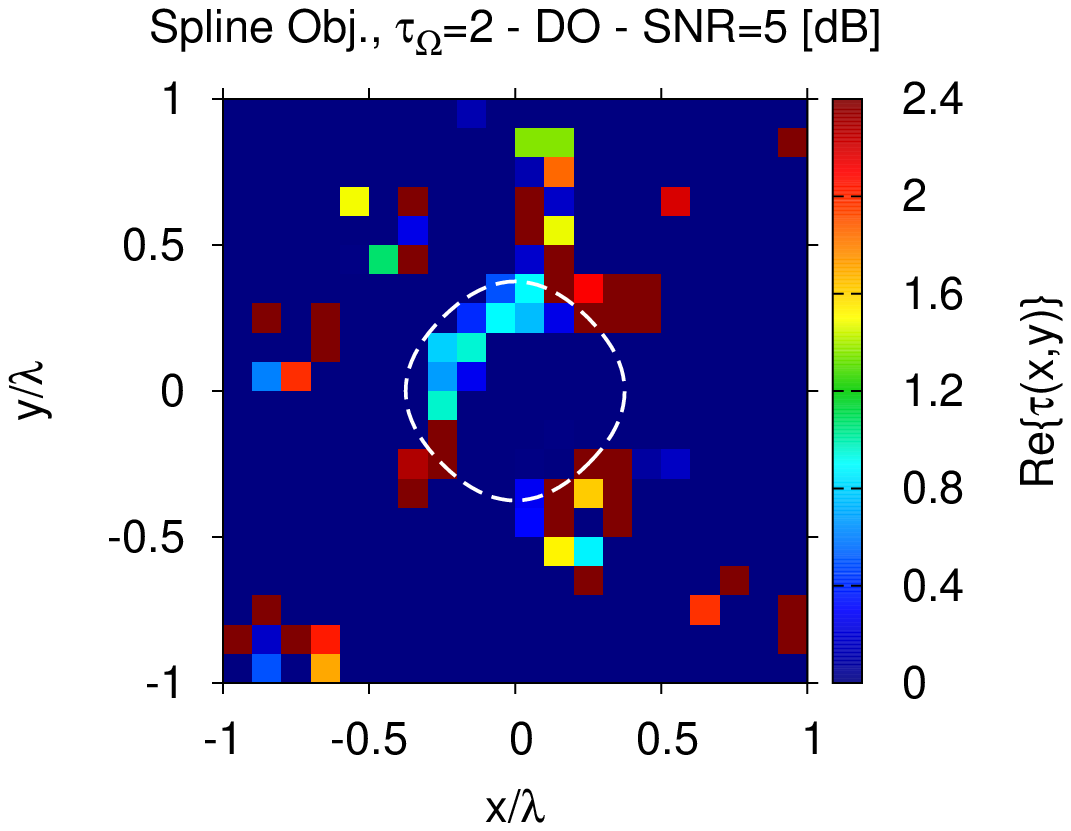}&
\includegraphics[%
  width=0.33\columnwidth]{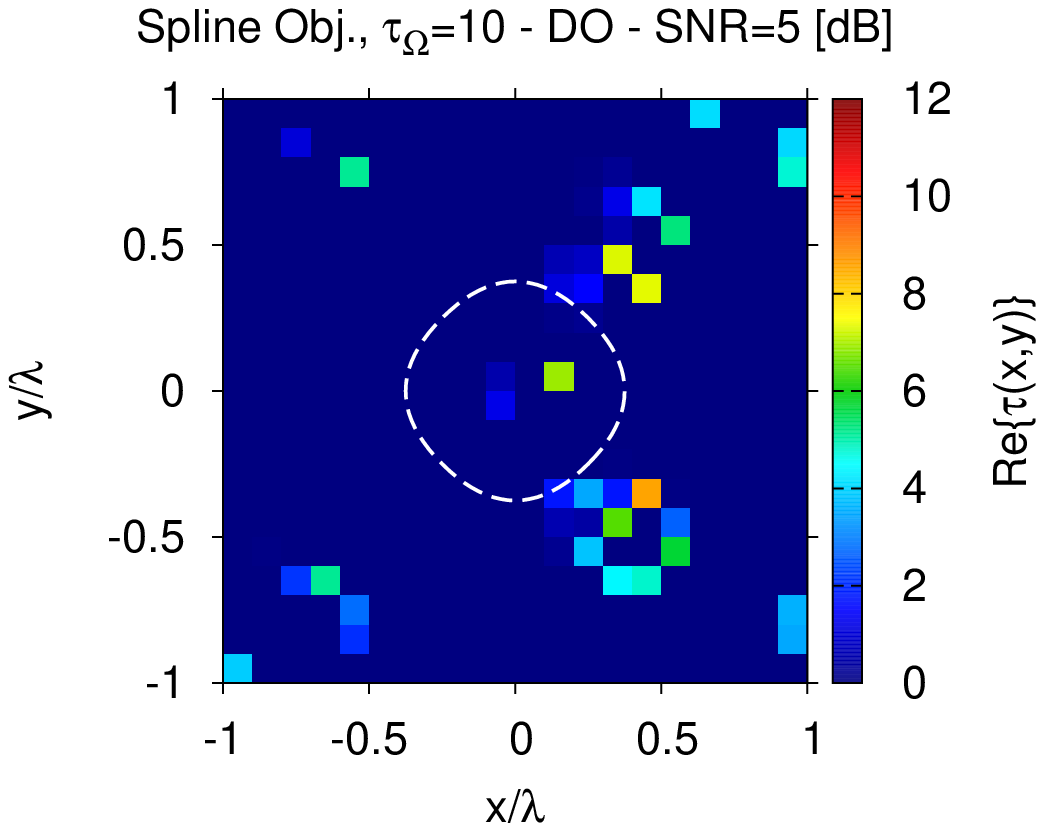}\tabularnewline
\begin{sideways}
\end{sideways}&
(\emph{g})&
(\emph{h})&
(\emph{i})\tabularnewline
\end{tabular}\end{center}

\begin{center}~\vfill\end{center}

\begin{center}\textbf{Fig. 13 - M. Salucci} \textbf{\emph{et al.}}\textbf{,}
\textbf{\emph{{}``}}Learned Global Optimization ...''\end{center}

\newpage
\begin{center}~\vfill\end{center}

\begin{center}\begin{tabular}{cc}
\multicolumn{2}{c}{\includegraphics[%
  width=0.60\columnwidth]{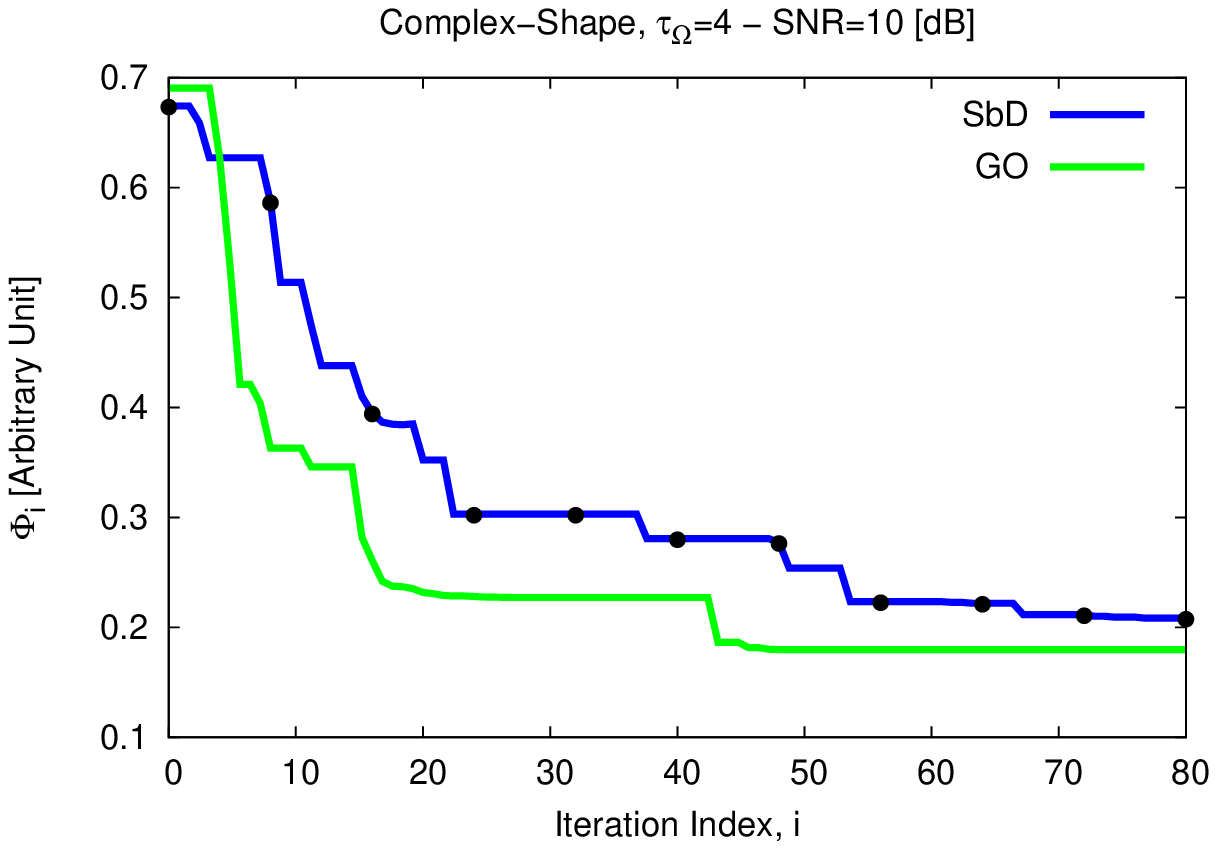}}\tabularnewline
\multicolumn{2}{c}{(\emph{a})}\tabularnewline
\multicolumn{1}{c}{\includegraphics[%
  width=0.45\columnwidth]{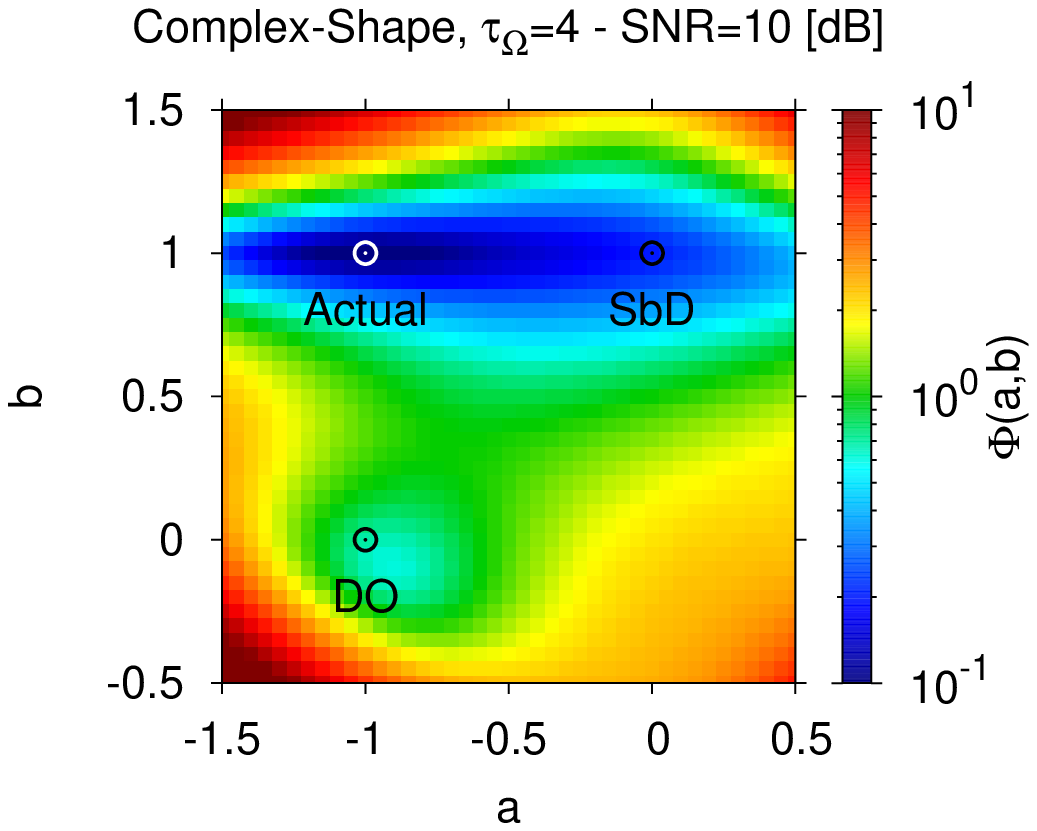}}&
\includegraphics[%
  width=0.45\columnwidth]{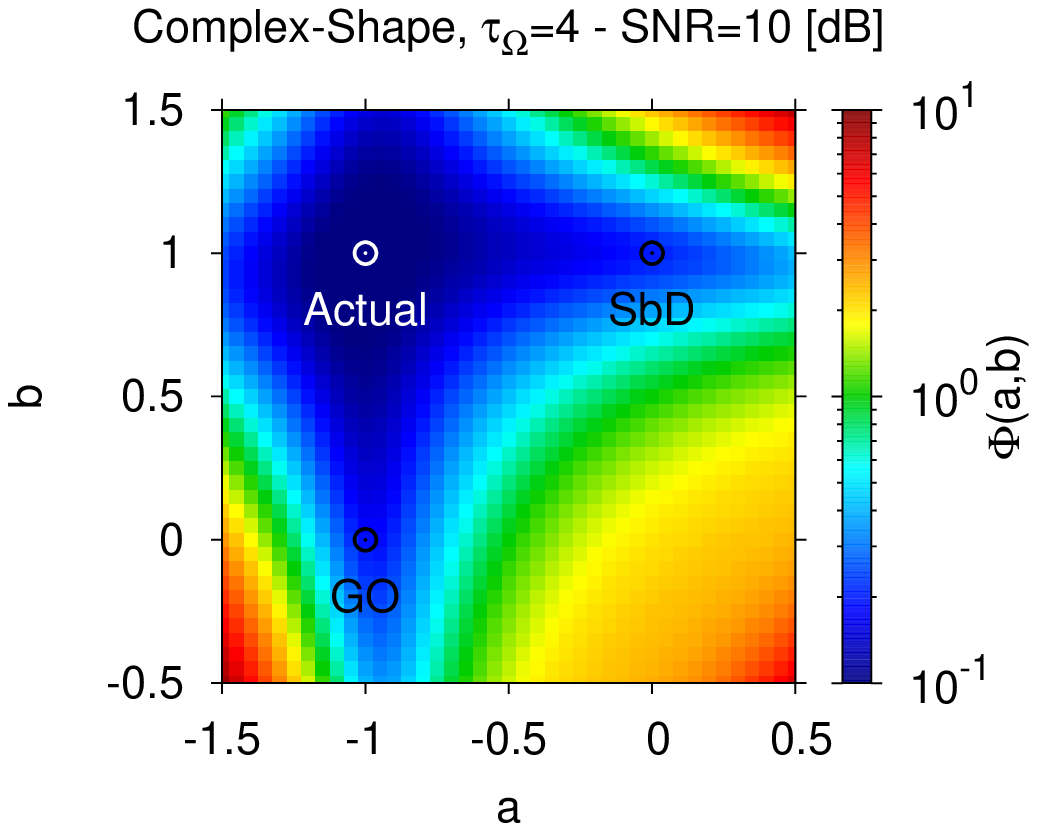}\tabularnewline
\multicolumn{1}{c}{(\emph{b})}&
(\emph{c})\tabularnewline
\end{tabular}\end{center}

\begin{center}~\vfill\end{center}

\begin{center}\textbf{Fig. 14 - M. Salucci} \textbf{\emph{et al.}}\textbf{,}
\textbf{\emph{{}``}}Learned Global Optimization ...''\end{center}

\newpage
\begin{center}~\vfill\end{center}

\begin{center}\begin{tabular}{ccc}
\multicolumn{3}{c}{\includegraphics[%
  width=0.75\columnwidth]{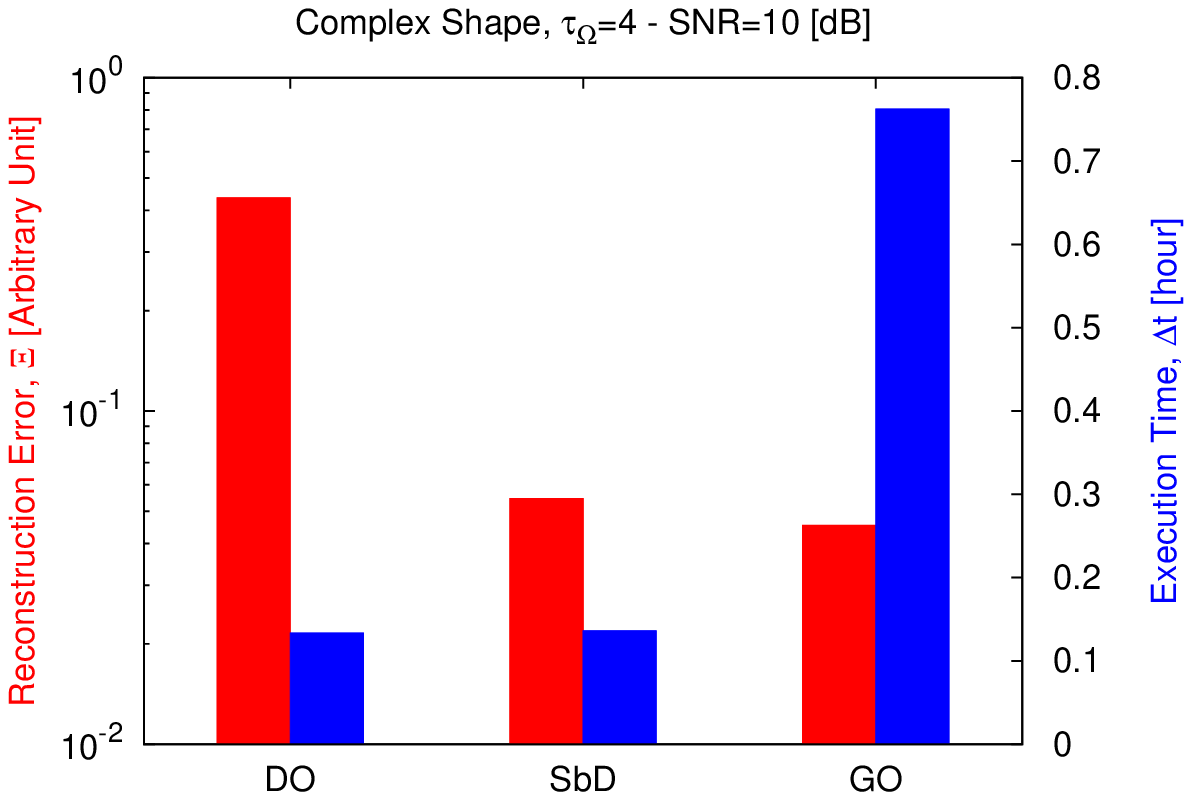}}\tabularnewline
\end{tabular}\end{center}

\begin{center}~\vfill\end{center}

\begin{center}\textbf{Fig. 15 - M. Salucci} \textbf{\emph{et al.}}\textbf{,}
\textbf{\emph{{}``}}Learned Global Optimization ...''\end{center}

\newpage
\begin{center}~\vfill\end{center}

\begin{center}\begin{tabular}{cc}
\multicolumn{1}{c}{\includegraphics[%
  width=0.50\columnwidth]{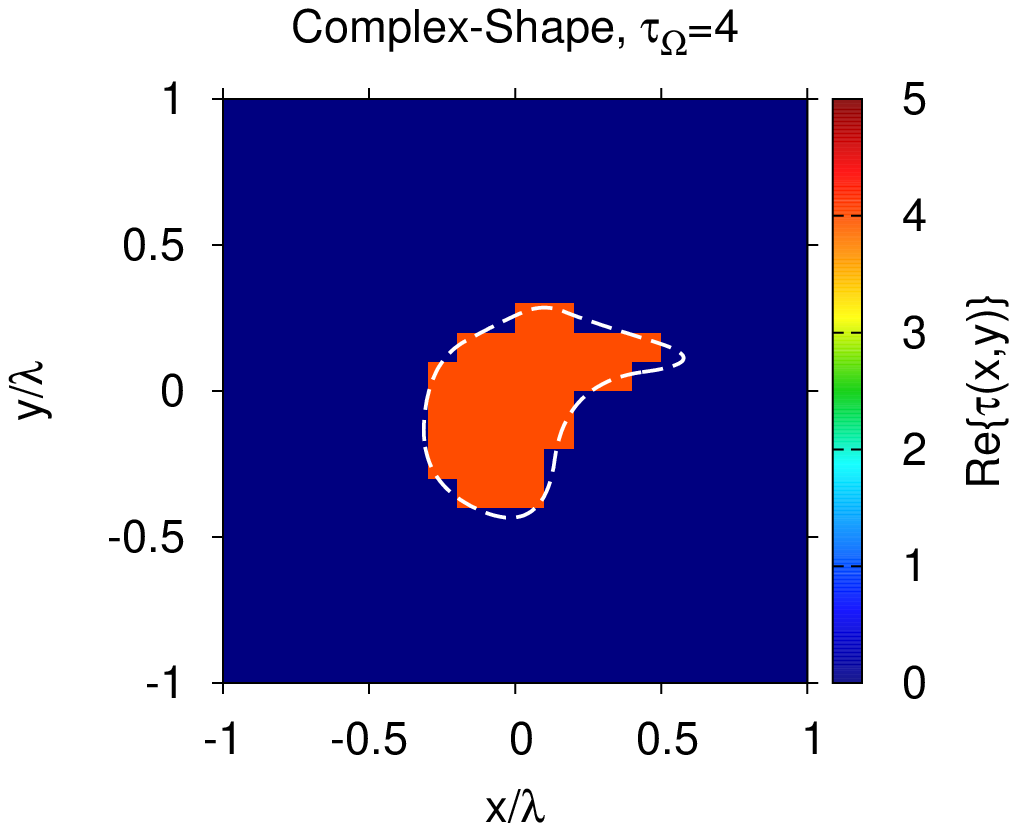}}&
\includegraphics[%
  width=0.50\columnwidth]{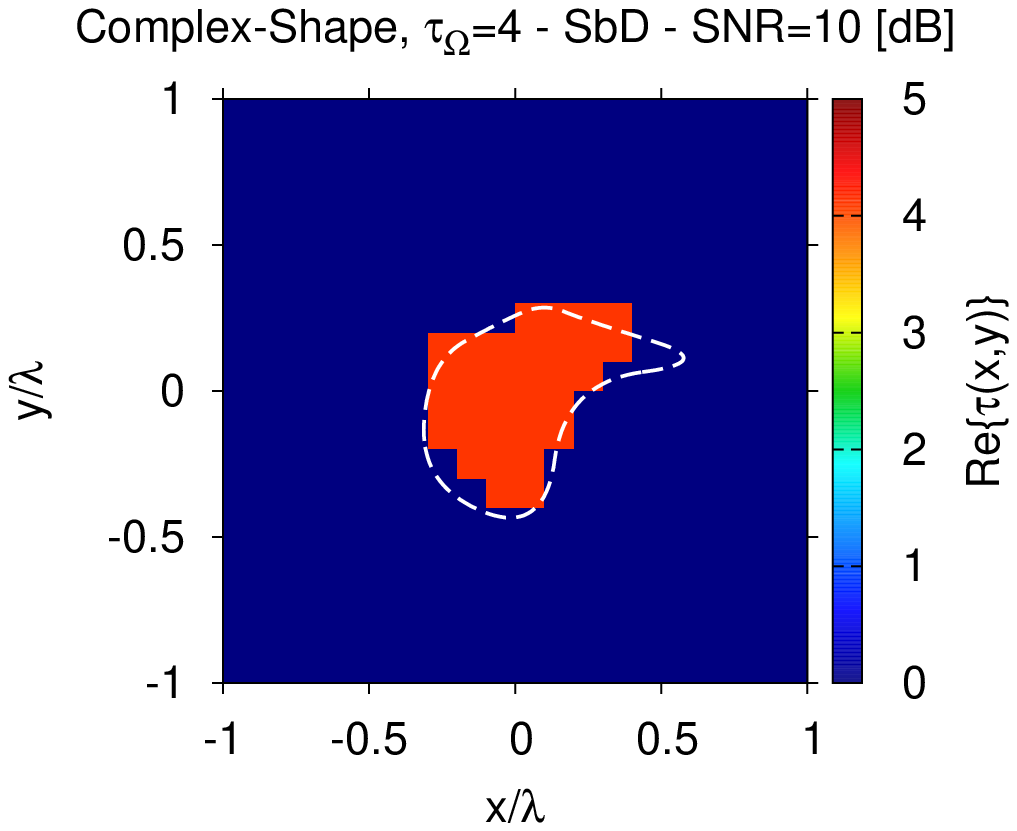}\tabularnewline
\multicolumn{1}{c}{(\emph{a})}&
(\emph{b})\tabularnewline
\multicolumn{1}{c}{}&
\tabularnewline
\multicolumn{1}{c}{\includegraphics[%
  width=0.50\columnwidth]{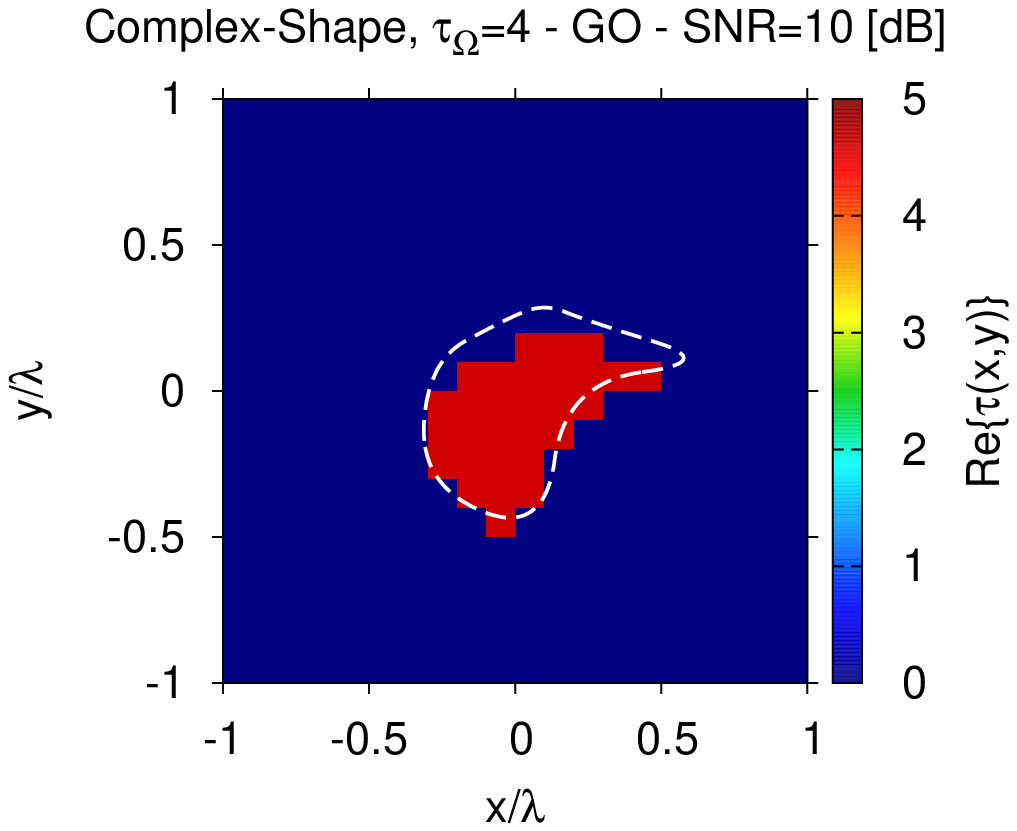}}&
\includegraphics[%
  width=0.50\columnwidth]{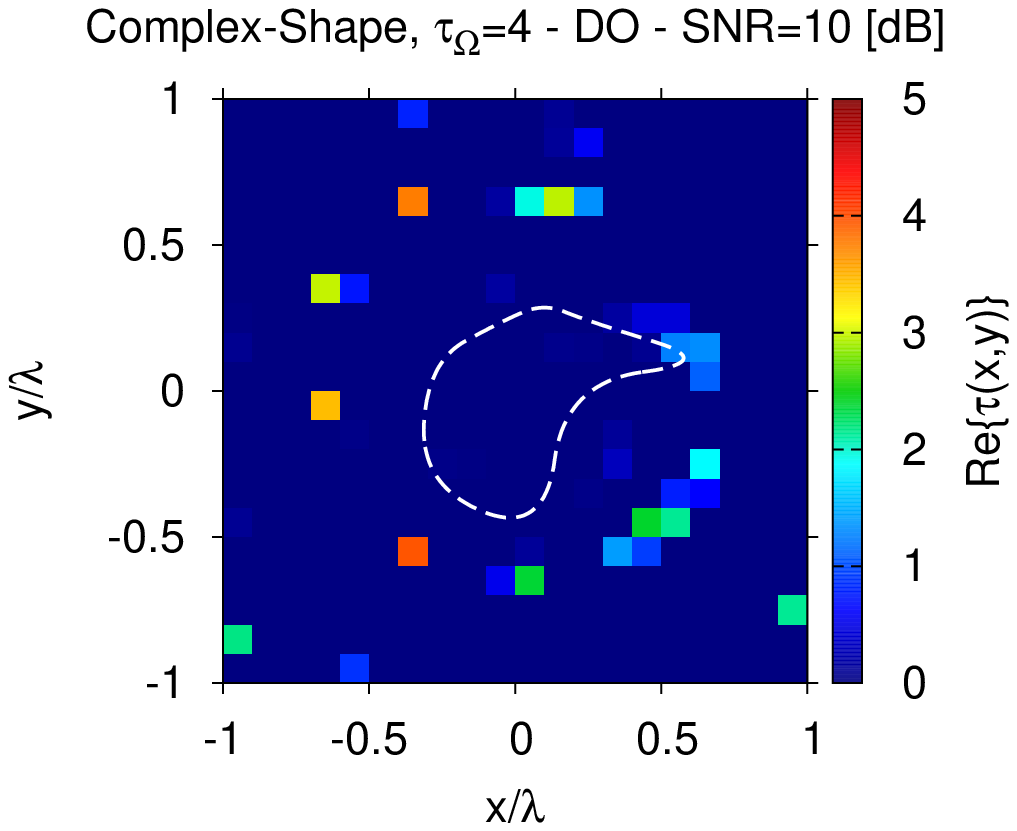}\tabularnewline
\multicolumn{1}{c}{(\emph{c})}&
(\emph{d})\tabularnewline
\end{tabular}\end{center}

\begin{center}~\vfill\end{center}

\begin{center}\textbf{Fig. 16 - M. Salucci} \textbf{\emph{et al.}}\textbf{,}
\textbf{\emph{{}``}}Learned Global Optimization ...''\end{center}

\newpage
\begin{center}~\vfill\end{center}

\begin{center}\begin{tabular}{c}
\includegraphics[%
  width=0.65\columnwidth]{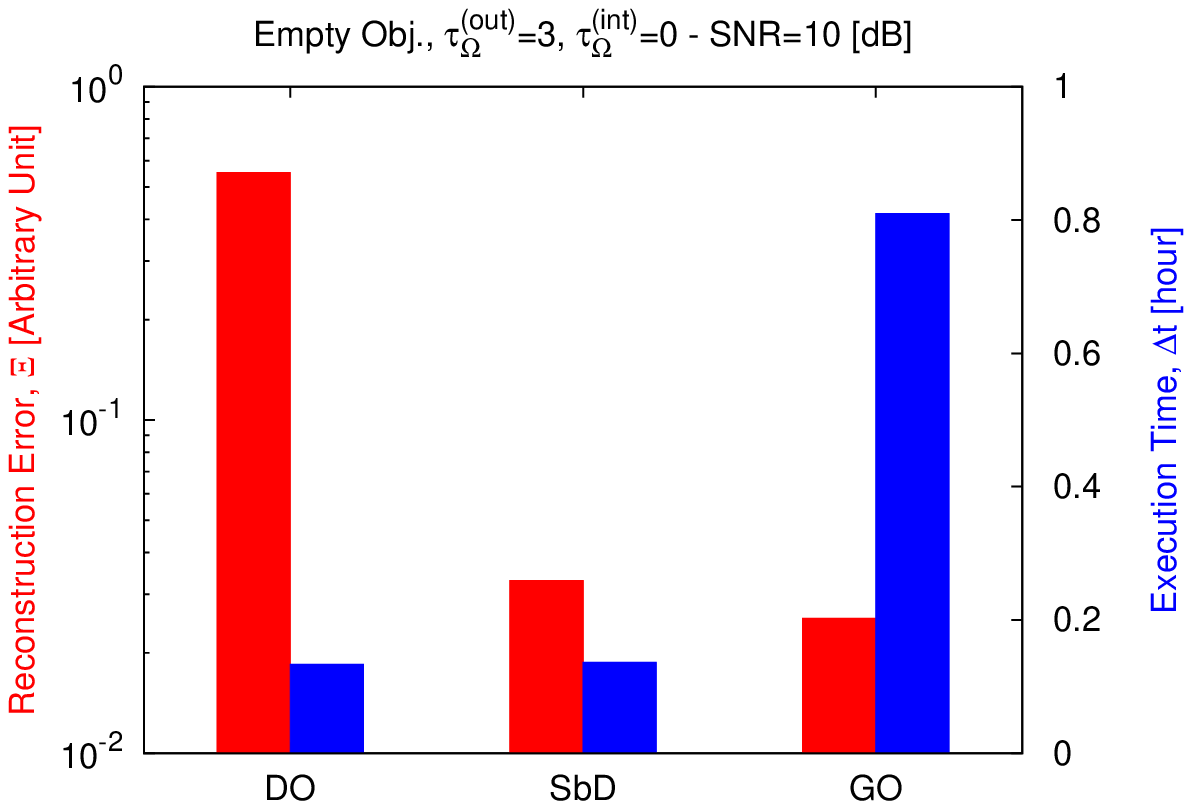}\tabularnewline
(\emph{a})\tabularnewline
\tabularnewline
\includegraphics[%
  width=0.65\columnwidth]{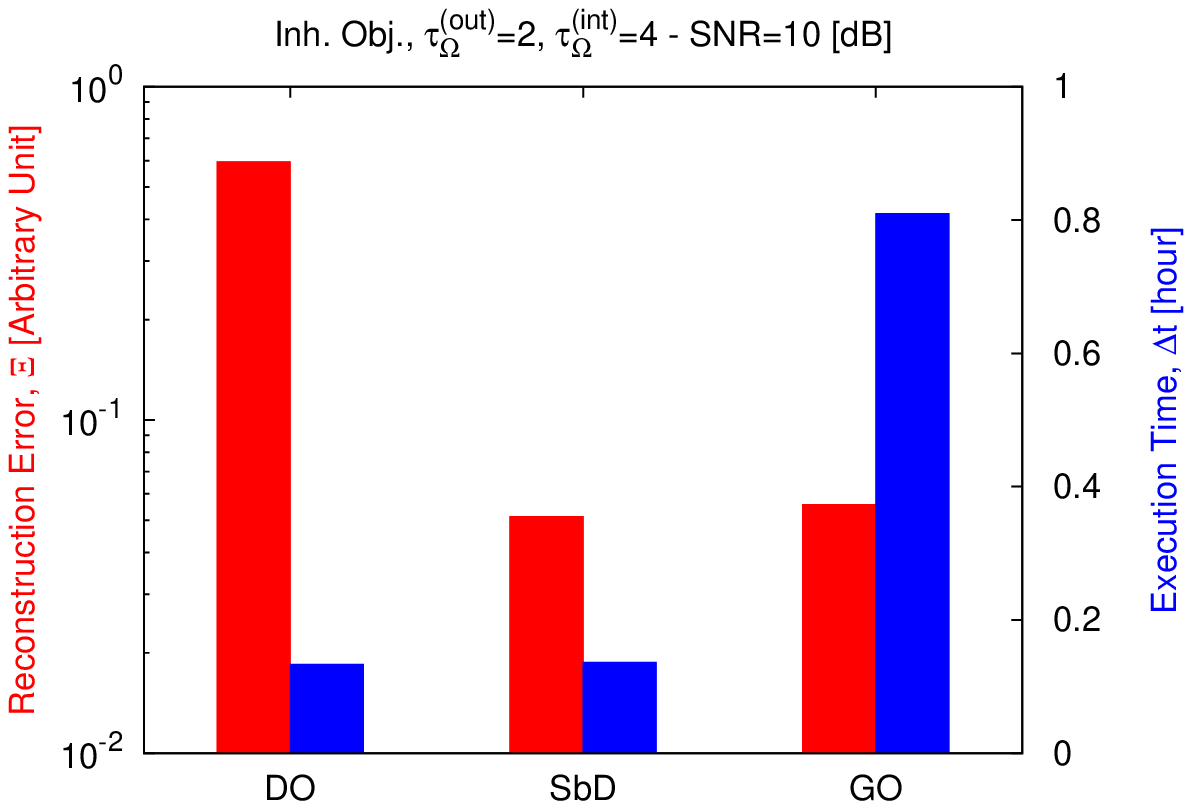}\tabularnewline
(\emph{b})\tabularnewline
\end{tabular}\end{center}

\begin{center}~\vfill\end{center}

\begin{center}\textbf{Fig. 17 - M. Salucci} \textbf{\emph{et al.}}\textbf{,}
\textbf{\emph{{}``}}Learned Global Optimization ...''\end{center}

\newpage
\begin{center}~\vfill\end{center}

\begin{center}\begin{tabular}{cc}
\includegraphics[%
  width=0.45\columnwidth]{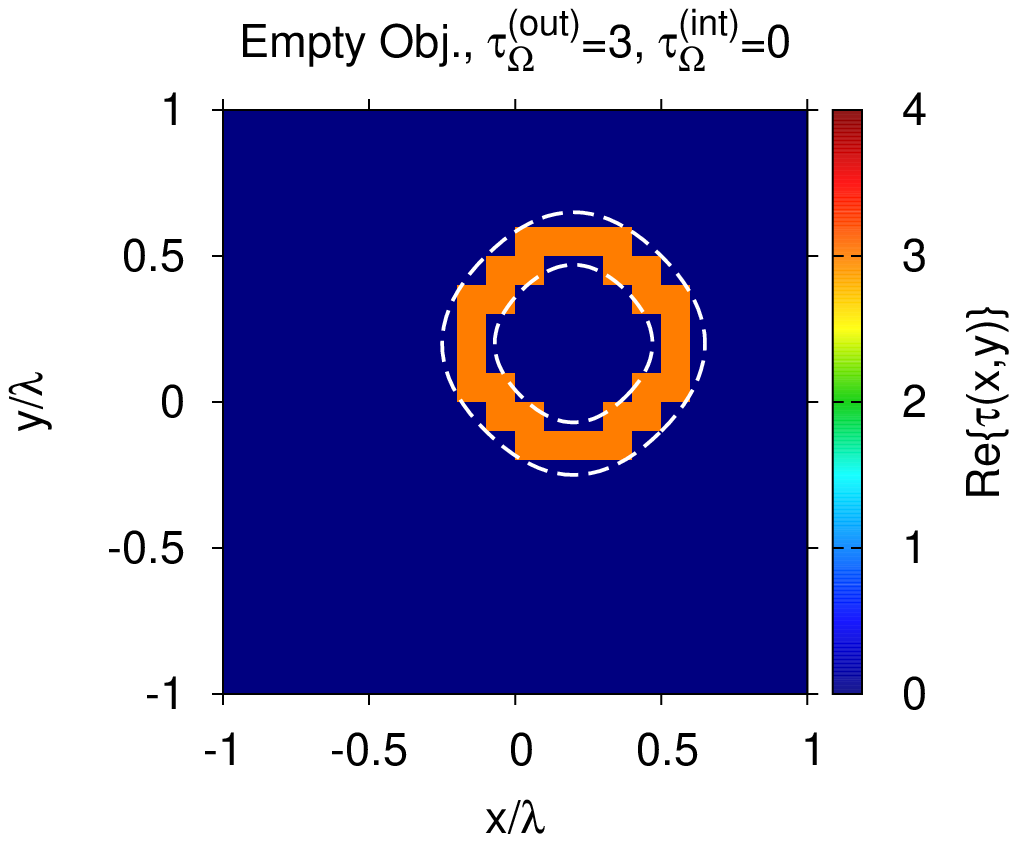}&
\includegraphics[%
  width=0.45\columnwidth]{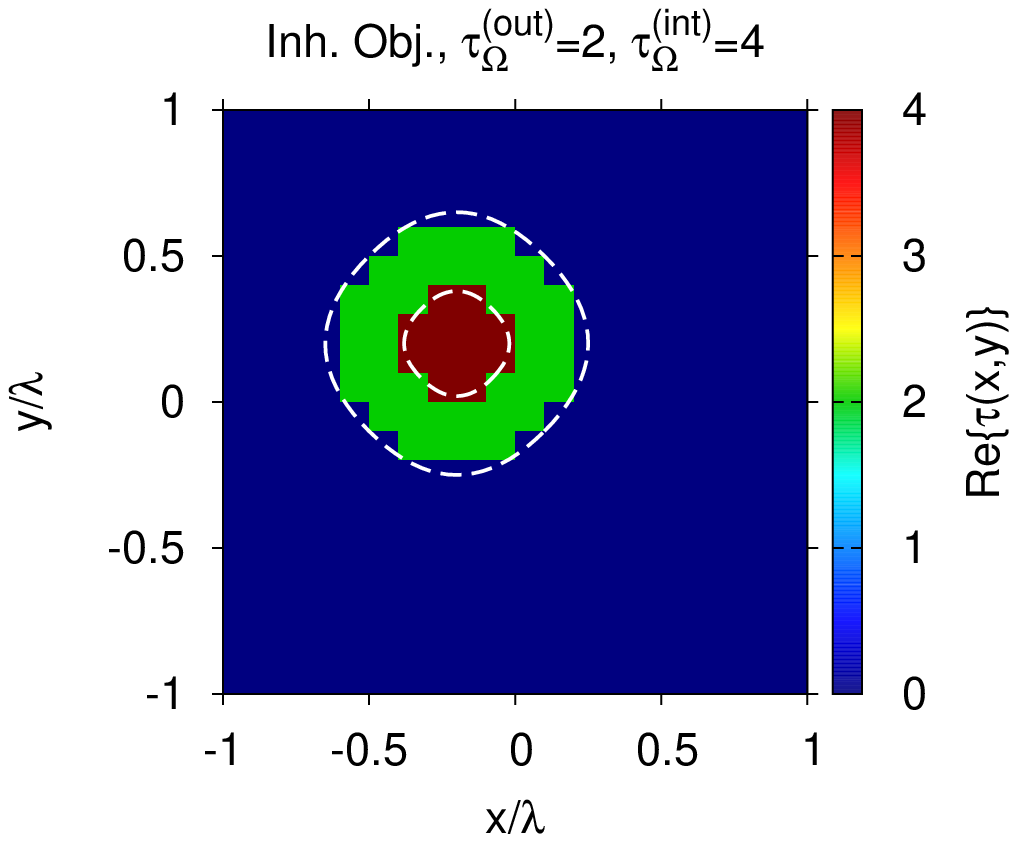}\tabularnewline
(\emph{a})&
(\emph{b})\tabularnewline
\includegraphics[%
  width=0.45\columnwidth]{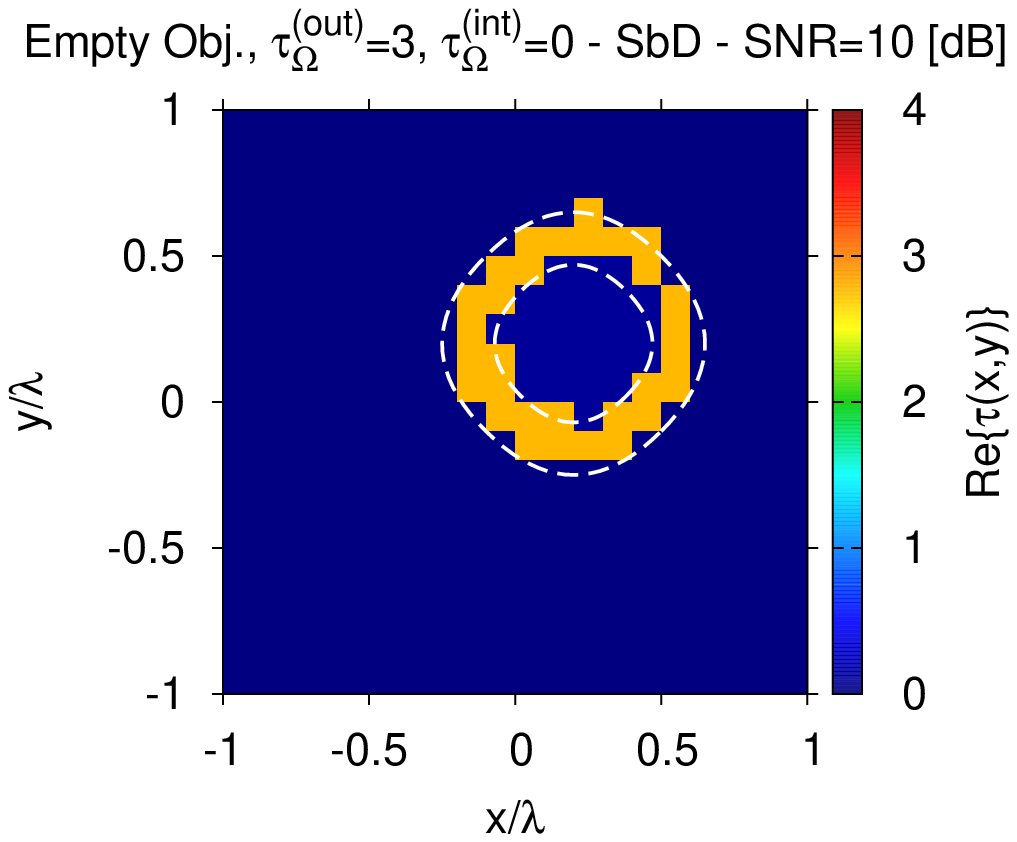}&
\includegraphics[%
  width=0.45\columnwidth]{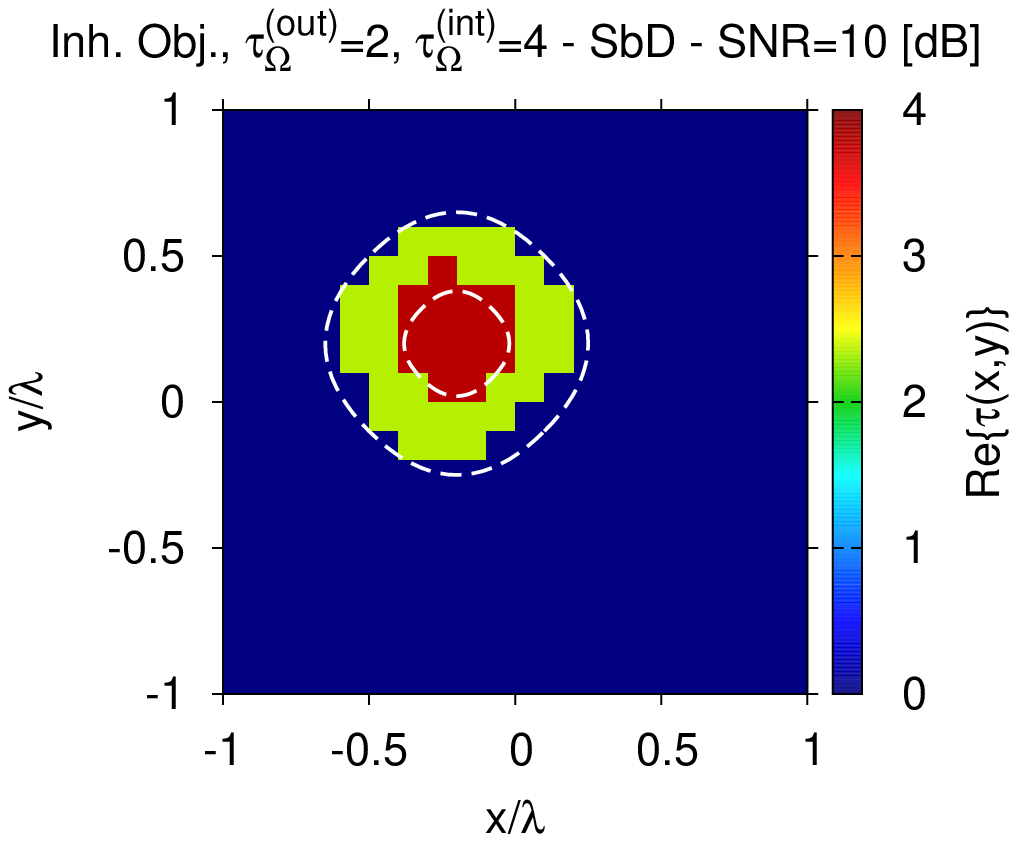}\tabularnewline
(\emph{c})&
(\emph{d})\tabularnewline
\end{tabular}\end{center}

\begin{center}~\vfill\end{center}

\begin{center}\textbf{Fig. 18 - M. Salucci} \textbf{\emph{et al.}}\textbf{,}
\textbf{\emph{{}``}}Learned Global Optimization ...''\end{center}

\newpage
\begin{center}~\vfill\end{center}

\begin{center}\begin{tabular}{cc}
\multicolumn{2}{c}{\includegraphics[%
  width=0.75\columnwidth]{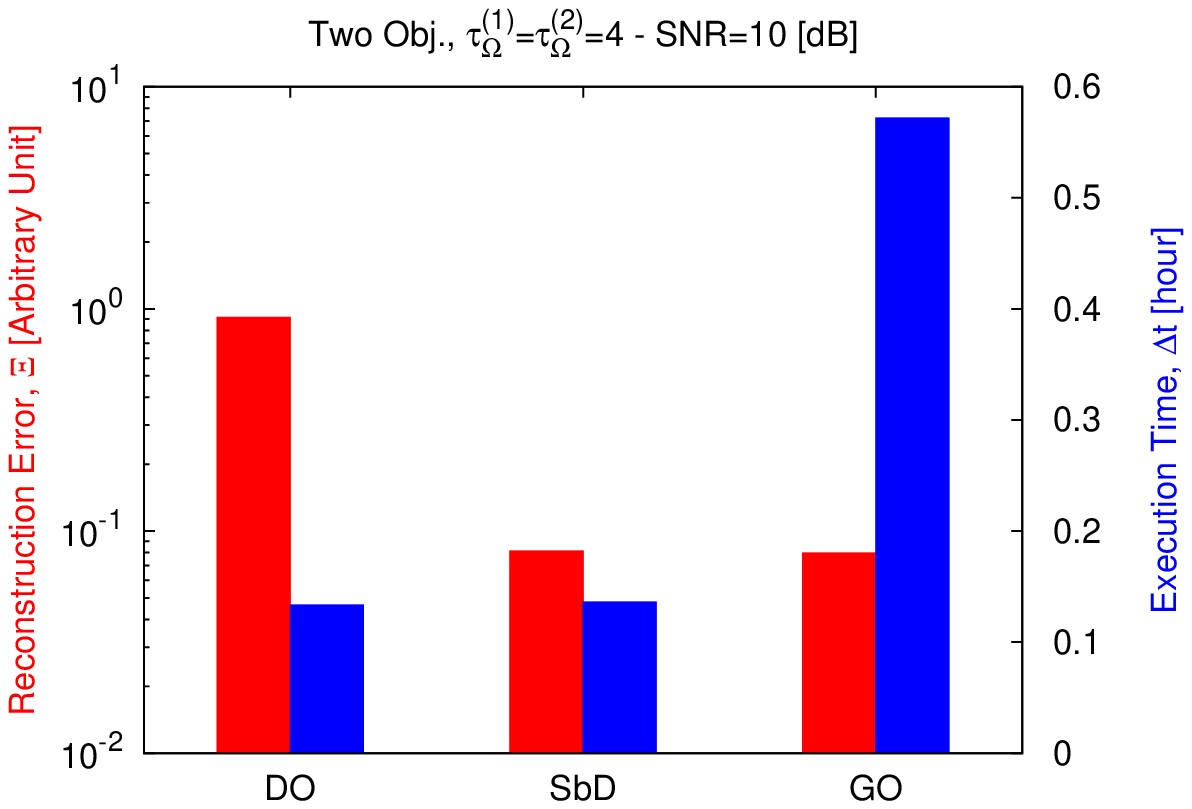}}\tabularnewline
\end{tabular}\end{center}

\begin{center}~\vfill\end{center}

\begin{center}\textbf{Fig. 19 - M. Salucci} \textbf{\emph{et al.}}\textbf{,}
\textbf{\emph{{}``}}Learned Global Optimization ...''\end{center}

\newpage
\begin{center}~\vfill\end{center}

\begin{center}\begin{tabular}{cc}
\multicolumn{1}{c}{\includegraphics[%
  width=0.50\columnwidth]{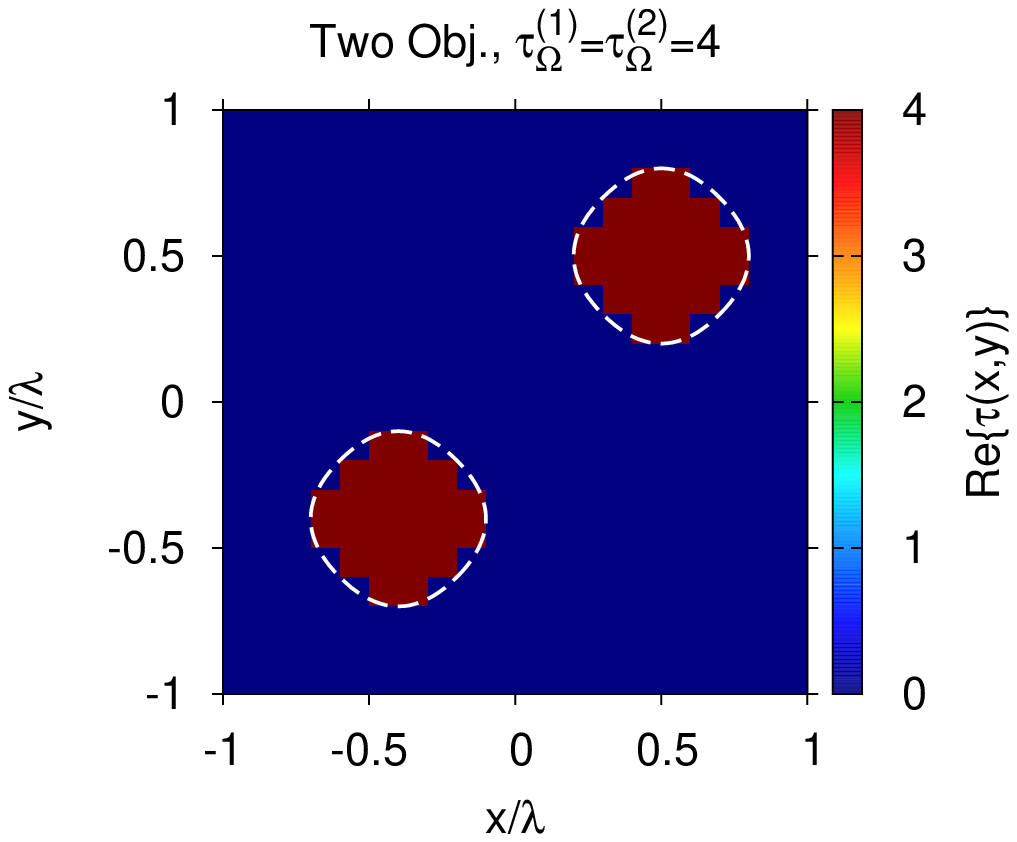}}&
\includegraphics[%
  width=0.50\columnwidth]{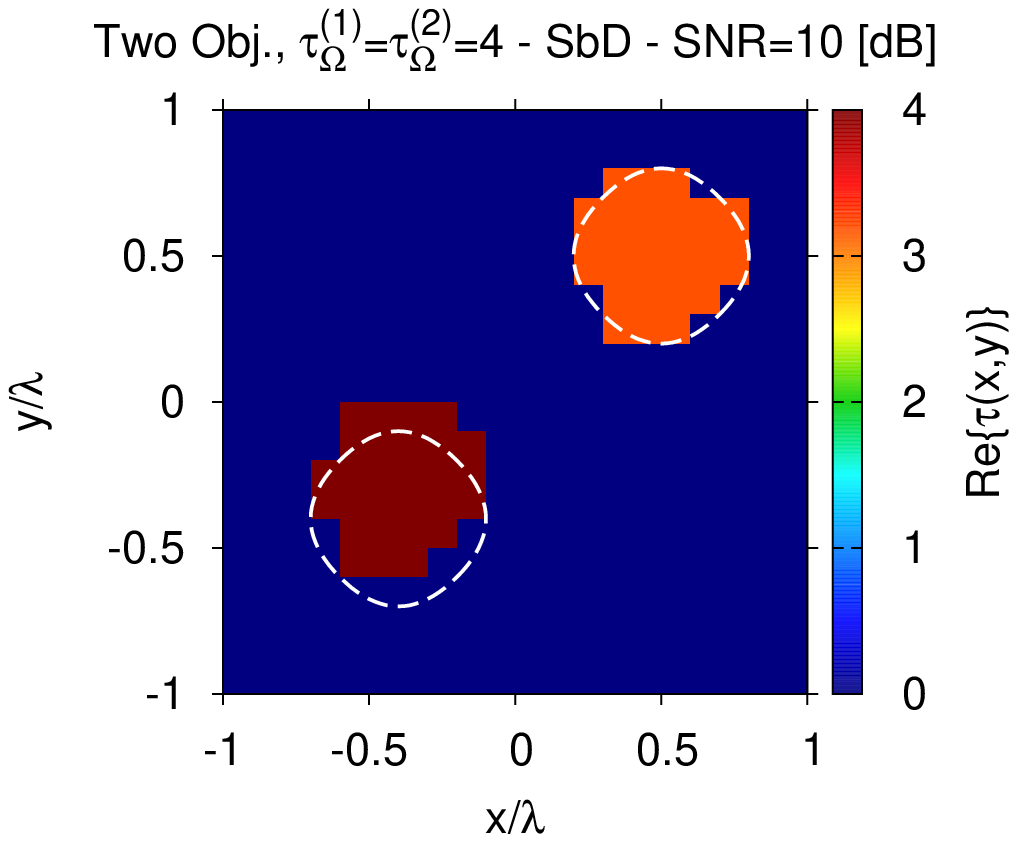}\tabularnewline
\multicolumn{1}{c}{(\emph{a})}&
(\emph{b})\tabularnewline
\multicolumn{1}{c}{}&
\tabularnewline
\multicolumn{1}{c}{\includegraphics[%
  width=0.50\columnwidth]{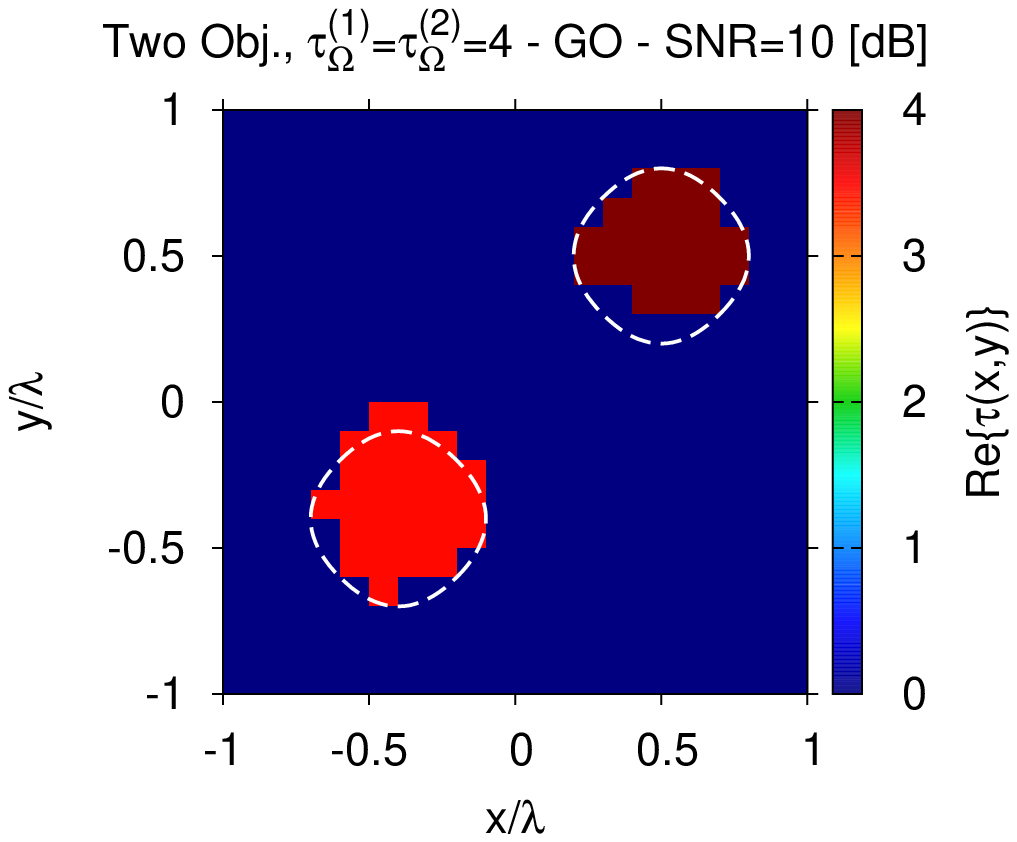}}&
\includegraphics[%
  width=0.50\columnwidth]{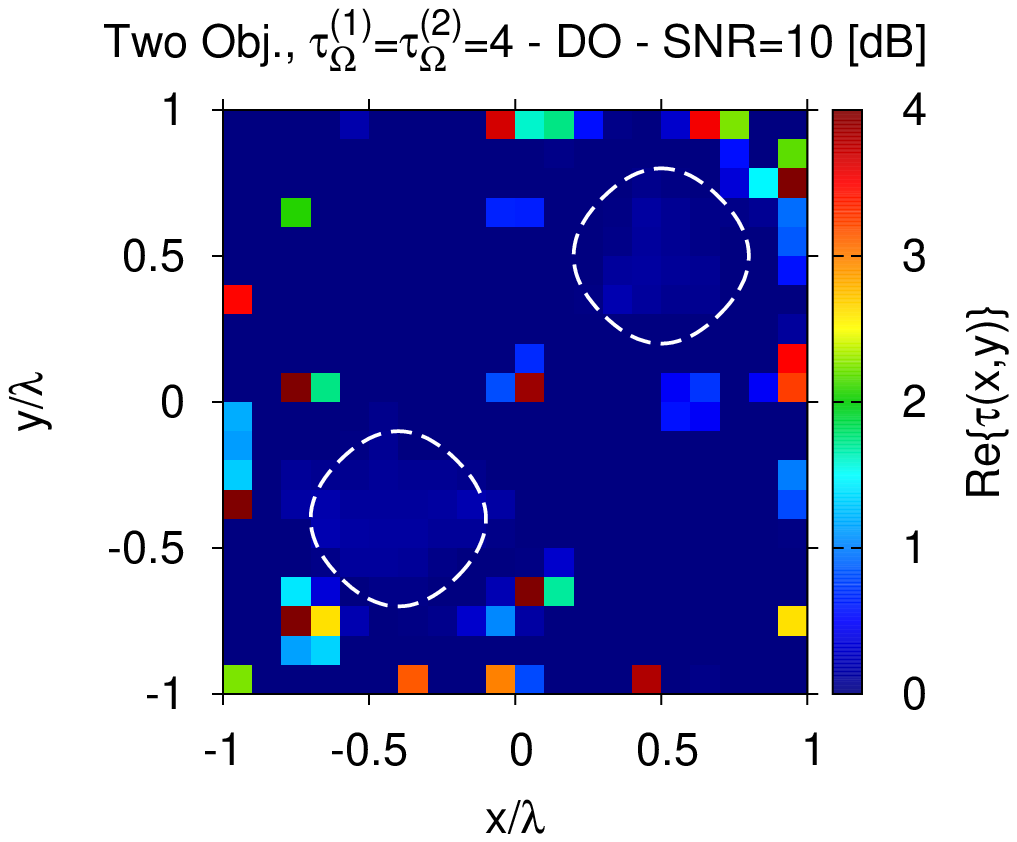}\tabularnewline
\multicolumn{1}{c}{(\emph{c})}&
(\emph{d})\tabularnewline
\end{tabular}\end{center}

\begin{center}~\vfill\end{center}

\begin{center}\textbf{Fig. 20 - M. Salucci} \textbf{\emph{et al.}}\textbf{,}
\textbf{\emph{{}``}}Learned Global Optimization ...''\end{center}

\newpage
\begin{center}~\vfill\end{center}

\begin{center}\begin{tabular}{cc}
\multicolumn{1}{c}{\includegraphics[%
  width=0.50\columnwidth]{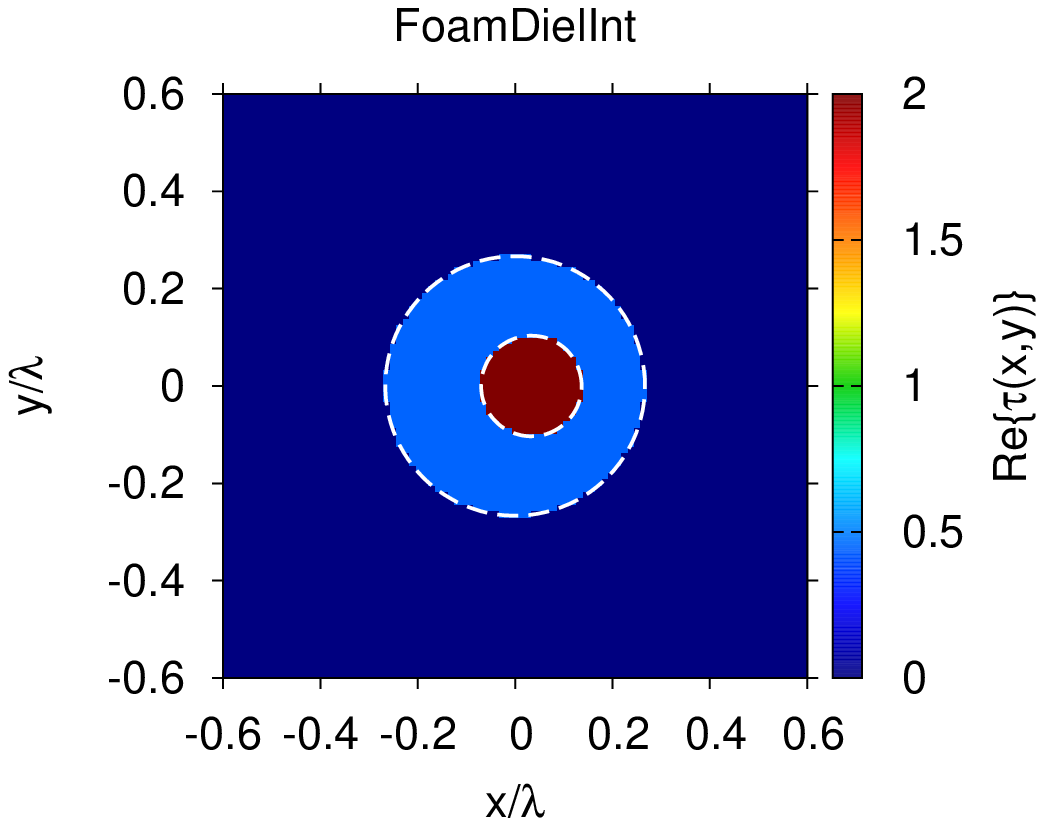}}&
\includegraphics[%
  width=0.50\columnwidth]{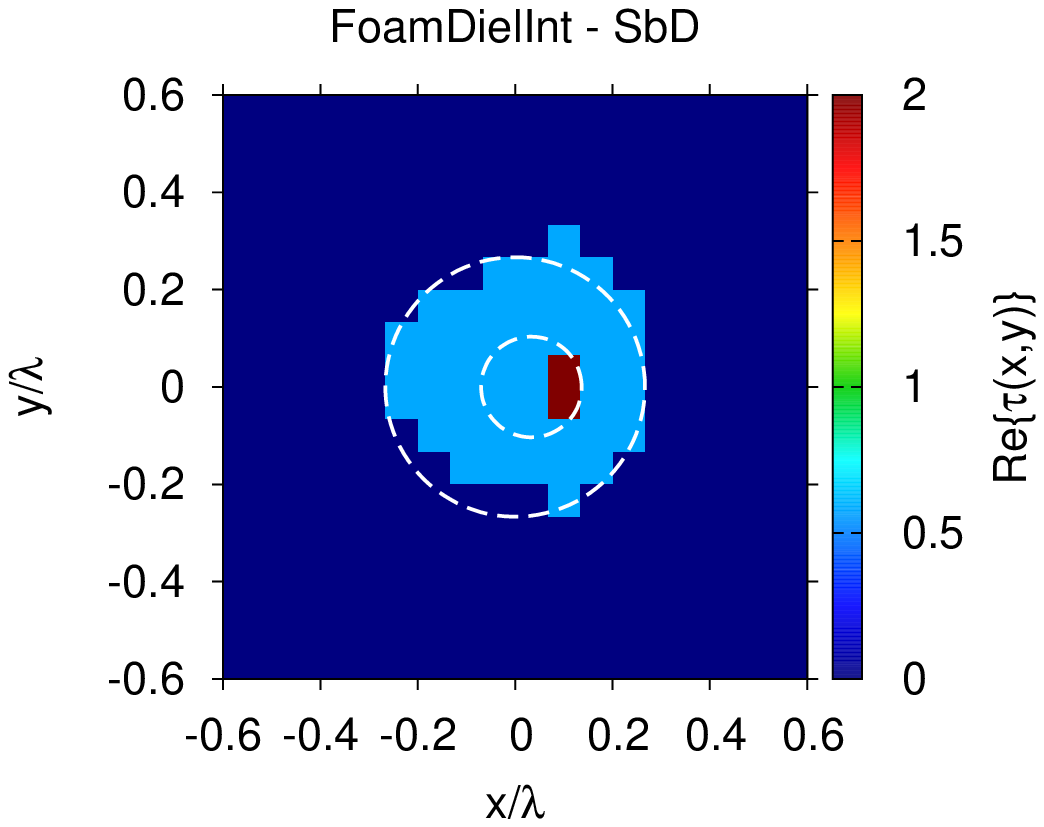}\tabularnewline
\multicolumn{1}{c}{(\emph{a})}&
(\emph{b})\tabularnewline
\multicolumn{1}{c}{}&
\tabularnewline
\multicolumn{1}{c}{\includegraphics[%
  width=0.50\columnwidth]{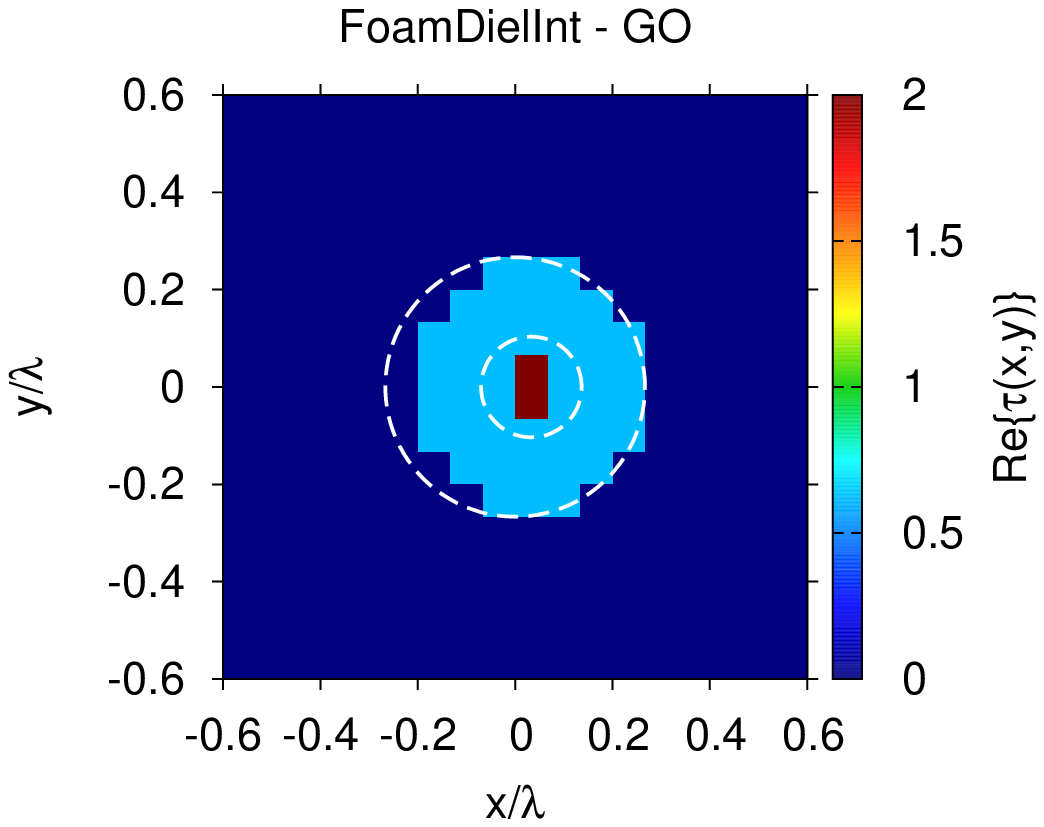}}&
\includegraphics[%
  width=0.50\columnwidth]{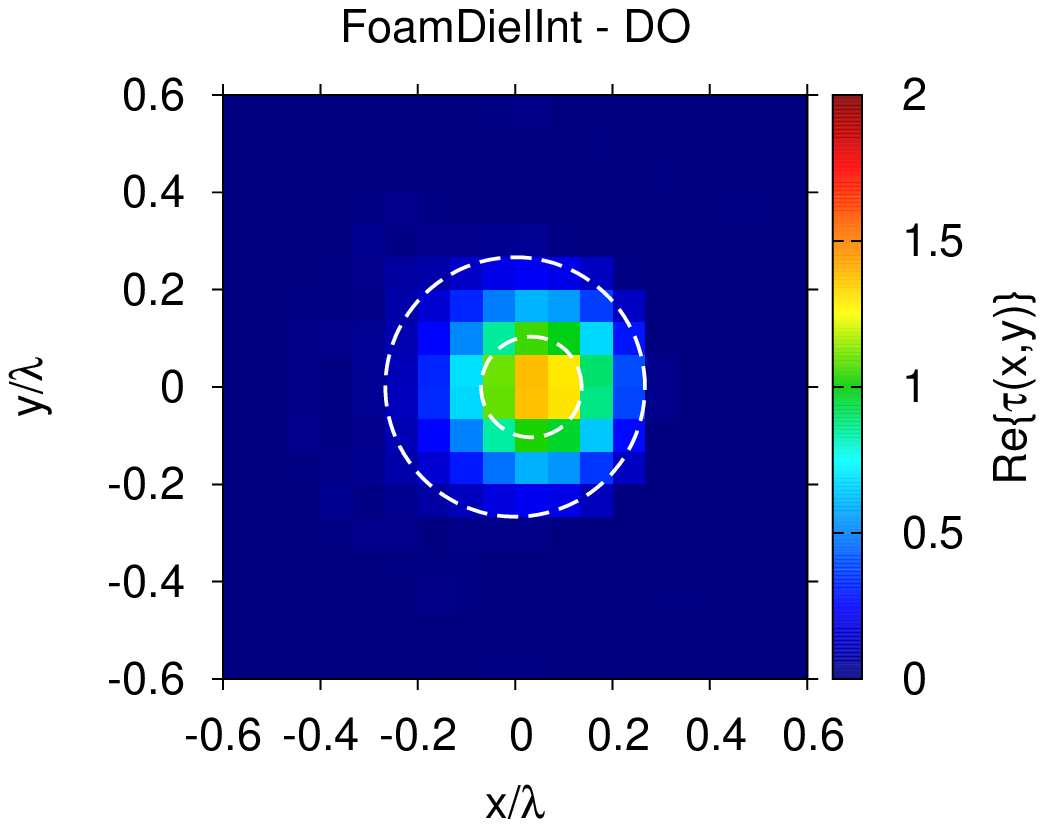}\tabularnewline
\multicolumn{1}{c}{(\emph{c})}&
(\emph{d})\tabularnewline
\end{tabular}\end{center}

\begin{center}~\vfill\end{center}

\begin{center}\textbf{Fig. 21 - M. Salucci} \textbf{\emph{et al.}}\textbf{,}
\textbf{\emph{{}``}}Learned Global Optimization ...''\end{center}

\newpage
\begin{center}~\vfill\end{center}

\begin{center}\begin{tabular}{|c|c|c|c|c|c|c|c|}
\hline 
Object&
$K$&
$S_{0}$&
$I_{SbD}$&
Profile&
$\left(x_{\Omega},\, y_{\Omega}\right)$ {[}$\lambda${]}&
$Q$&
$\underline{\rho}$ {[}$\lambda${]}\tabularnewline
\hline
\hline 
Fig. 4(\emph{a})&
$8$&
$40$&
$100$&
$\partial\Omega$&
$\left(0,\,0\right)$&
$4$&
$\left\{ 0.5,\,0.5,\,0.5,\,0.5\right\} $\tabularnewline
\hline 
Fig. 16(\emph{a})&
$12$&
$60$&
$80$&
$\partial\Omega$&
$\left(0.1,\,0.1\right)$&
$8$&
$\left\{ 0.6,\,0.2,\,0.2,\,0.2,\,0.4,\,0.6,\,0.6,\,0.1\right\} $\tabularnewline
\hline 
Fig. 18(\emph{a})&
$11$&
$55$&
$85$&
$\partial\Omega^{\left(out\right)}$&
$\left(0.2,\,0.2\right)$&
$4$&
$\left\{ 0.6,\,0.6,\,0.6,\,0.6\right\} $\tabularnewline
&
&
&
&
$\partial\Omega^{\left(int\right)}$&
$\left(0.2,\,0.2\right)$&
$4$&
$\left\{ 0.36,\,0.36,\,0.36,\,0.36\right\} $\tabularnewline
\hline
Fig. 18(\emph{b})&
$11$&
$55$&
$85$&
$\partial\Omega^{\left(out\right)}$&
$\left(-0.2,\,0.2\right)$&
$4$&
$\left\{ 0.6,\,0.6,\,0.6,\,0.6\right\} $\tabularnewline
&
&
&
&
$\partial\Omega^{\left(int\right)}$&
$\left(-0.2,\,0.2\right)$&
$4$&
$\left\{ 0.24,\,0.24,\,0.24,\,0.24\right\} $\tabularnewline
\hline
Fig. 20(\emph{a})&
$16$&
$80$&
$60$&
$\partial\Omega^{\left(1\right)}$&
$\left(-0.4,\,-0.4\right)$&
$4$&
$\left\{ 0.4,\,0.4,\,0.4,\,0.4\right\} $\tabularnewline
&
&
&
&
$\partial\Omega^{\left(2\right)}$&
$\left(0.5,\,0.5\right)$&
$4$&
$\left\{ 0.4,\,0.4,\,0.4,\,0.4\right\} $\tabularnewline
\hline
\end{tabular}\end{center}

\begin{center}~\vfill\end{center}

\begin{center}\textbf{Tab. I - M. Salucci} \textbf{\emph{et al.}}\textbf{,}
\textbf{\emph{{}``}}Learned Global Optimization ...''\end{center}

\begin{thebibliography}{10}
\bibitem{Chen 2018}X. Chen, \emph{Computational Methods for Electromagnetic Inverse Scattering.}
Hoboken, NJ, USA: Wiley, 2018.
\bibitem{Abubakar 2002}A. Abubakar, P. M. van den Berg, and J. Mallorqui, {}``Imaging of
biomedical data using a multiplicative regularized contrast source
inversion method,'' \emph{IEEE Trans. Microw. Theory Techn.}, vol.
50, no. 7, pp. 1761- 1771, Jul. 2002.
\bibitem{Mojabi 2009}P. Mojabi and J. LoVetri, {}``Microwave biomedical imaging using
the multiplicative regularized Gauss-Newton inversion,'' \emph{IEEE
Antennas Wireless Propag. Lett.}, vol. 8, pp. 645-648, Jul. 2009.
\bibitem{Gao 2017}Y. Gao and R. Zoughi, {}``Millimeter wave reflectometry and imaging
for noninvasive diagnosis of skin burn injuries,'' \emph{IEEE Trans.
Instrum. Meas.}, vol. 66, no. 1, pp. 77-84, Jan. 2017.
\bibitem{Afsari 2019}A. Afsari, A. M. Abbosh, and Y. Rahmat-Samii, {}``Modified Born iterative
method in medical electromagnetic tomography using magnetic field
fluctuation contrast source operator,'' \emph{IEEE Trans. Microw.
Theory Techn.}, vol. 67, no. 1, pp. 454-463, Jan. 2019.
\bibitem{Song 2019}X. Song, M. Li, F. Yang, S. Xu, and A. Abubakar, {}``Study on joint
inversion algorithm of acoustic and electromagnetic data in biomedical
imaging,'' \emph{IEEE J. Multiscale Multiphys. Comput. Techn.}, vol.
4, pp. 2-11, 2019.
\bibitem{Cui 2001}T. Cui, W. C. Chew, A. A. Aydiner, and S. Chen {}``Inverse scattering
of two-dimensional dielectric objects buried in a lossy earth using
the distorted Born iterative method,'' \emph{IEEE Trans. Geosci.
Remote Sens.}, vol. 39, no. 2, pp. 339-346, Feb. 2001.
\bibitem{Salucci 2015}M. Salucci, G. Oliveri, and A. Massa, {}``GPR prospecting through
an inverse scattering frequency-hopping multi-focusing approach,''
\emph{IEEE Trans. Geosci. Remote Sens.}, vol. 53, no. 12, pp. 6573-6592,
Dec. 2015.
\bibitem{Salucci 2017}M. Salucci, L. Poli, and A. Massa, {}``Advanced multi-frequency GPR
data processing for non-linear deterministic imaging,'' \emph{Signal
Proc}., vol. 132, pp. 306-318, Mar. 2017.
\bibitem{Salucci 2017b}M. Salucci, L. Poli, N. Anselmi and A. Massa, {}``Multifrequency
particle swarm optimization for enhanced multiresolution GPR microwave
imaging,'' \emph{IEEE Trans. Geosci. Remote Sens.}, vol. 55, no.
3, pp. 1305-1317, Mar. 2017.
\bibitem{Liu 2018}Z. Liu, C. Li, D. Lesselier, and Y. Zhong, {}``Fast full-wave analysis
of damaged periodic fiber-reinforced laminates,'' \emph{IEEE Trans.
Antennas Propag.}, vol. 66, no. 7, pp. 3540-3547, Jul. 2018.
\bibitem{Zoughi 2000}R. Zoughi, \emph{Microwave Nondestructive Testing and Evaluation}.
Amsterdam, The Netherlands: Kluwer, 2000.
\bibitem{Kharkovsky 2007}S. Kharkovsky and R. Zoughi, {}``Microwave and millimeter wave nondestructive
testing and evaluation - Overview and recent advances,'' \emph{IEEE
Instrum. Meas. Mag.}, vol. 10, no. 2, pp. 26-38, Apr. 2007.
\bibitem{Caorsi 2001}S. Caorsi, A. Massa, and M. Pastorino, {}``A crack identification
microwave procedure based on a genetic algorithm for nondestructive
testing,'' \emph{IEEE Trans. Antennas Propag.}, vol. 49, no. 12,
pp. 1812-1820, Dec. 2001.
\bibitem{Xu 2018b}K. Xu, Y. Zhong, X. Chen, and D. Lesselier, {}``A fast integral equation-based
method for solving electromagnetic inverse scattering problems with
inhomogeneous background,'' \emph{IEEE Trans. Antennas Propag.},
vol. 66, no. 8, pp. 4228-4239, Aug. 2018.
\bibitem{Chu 2019}Y. Chu, K. Xu, Y. Zhong, X. Ye, T. Zhou, X. Chen, and G. Wang, {}``Fast
microwave through wall imaging method with inhomogeneous background
based on Levenberg-Marquardt algorithm,'' \emph{IEEE Trans. Microw.
Theory Techn.}, vol. 67, no. 3, pp. 1138-1147, Mar. 2019.
\bibitem{Fallahpour 2015}M. Fallahpour and R. Zoughi, {}``Fast 3-D qualitative method for
through-wall imaging and structural health monitoring,'' \emph{IEEE
Geosci. Remote Sens. Lett.}, vol. 12, no. 12, pp. 2463-2467, Dec.
2015.
\bibitem{LoVetri 2020}J. LoVetri, M. Asefi, C. Gilmore, and I. Jeffrey, {}``Innovations
in electromagnetic imaging technology: The stored-grain-monitoring
case,'' \emph{IEEE Antennas Propag. Mag.}, vol. 62, no. 5, pp. 33-42,
Oct. 2020.
\bibitem{Tobon 2020}J. A. Tobon Vasquez, R. Scapaticci, G. Turvani, M. Ricci, L. Farina,
A. Litman, M. R. Casu, L. Crocco, and F. Vipiana, {}``Noninvasive
inline food inspection via microwave imaging technology: An application
example in the food industry,'' \emph{IEEE Antennas Propag. Mag.},
vol. 62, no. 5, pp. 18-32, Oct. 2020.
\bibitem{Occhiuzzi 2020}C. Occhiuzzi, N. D'Uva, S. Nappi, S. Amendola, C. Gialluca, V. Chiabrando,
L. Garavaglia, G. Giacalone, and G. Marrocco, {}``Radio-frequency-identification-based
intelligent packaging: Electromagnetic classification of tropical
fruit ripening,'' \emph{IEEE Antennas Propag. Mag.}, vol. 62, no.
5, pp. 64-75, Oct. 2020.
\bibitem{Chew 1990}W. Chew and Y. Wang, {}``Reconstruction of two-dimensional permittivity
distribution using the distorted Born iterative method,'' \emph{IEEE
Trans. Med. Imag.}, vol. 9, no. 2, pp. 218-225, Jun. 1990.
\bibitem{Zhang 2009}W. Zhang, L. Li, and F. Li, {}``Multifrequency imaging from intensity-only
data using the phaseless data distorted Rytov iterative method,''
\emph{IEEE Trans. Antennas Propag.}, vol. 57, no. 1, pp. 290-295,
Jan. 2009.
\bibitem{Zhong 2016}Y. Zhong, M. Lambert, D. Lesselier, and X. Chen, {}``A new integral
equation method to solve highly nonlinear inverse scattering problems,''
\emph{IEEE Trans. Antennas Propag.}, vol. 64, no. 5, pp. 1788-1799,
May 2016.
\bibitem{Kleinman 1997}R. Kleinman and P. van den Berg, {}``A contrast source inversion
method,'' \emph{Inverse Probl.}, vol. 13, no. 6, pp. 1607-1620, Jul.
1997.
\bibitem{vandenBerg 2003}P. M. van den Berg, A. Abubakar, and J. Fokkema, {}``Multiplicative
regularization for contrast profile inversion,'' \emph{Radio Sci.},
vol. 38, no. 2, Apr. 2003.
\bibitem{Xu 2015}K. Xu, Y. Zhong, R. Song, X. Chen, and L. Ran, {}``Multiplicative-
regularized FFT twofold subspace-based optimization method for inverse
scattering problems,'' \emph{IEEE Trans. Geosci. Remote Sens.}, vol.
53, no. 2, pp. 841-850, Jun. 2015.
\bibitem{Chen 2010}X. Chen, {}``Subspace-based optimization method for solving inverse-scattering
problems,'' \emph{IEEE Trans. Geosci. Remote Sens.}, vol. 48, no.
1, pp. 42-49, Jan. 2010.
\bibitem{Zhong 2009}Y. Zhong and X. Chen, {}``Twofold subspace-based optimization method
for solving inverse scattering problems,'' \emph{Inverse Probl.},
vol. 25, 085003, Jul. 2009.
\bibitem{Zhong 2011}Y. Zhong and X. Chen, {}``An FFT twofold subspace-based optimization
method for solving electromagnetic inverse scattering problems,''
\emph{IEEE Trans. Antennas Propag.}, vol. 59, no. 3, pp. 914-927,
Mar. 2011.
\bibitem{Harada 1995}H. Harada, D. J. N. Wall, T. Takenaka, and M. Tanaka, {}``Conjugate
gradient method applied to inverse scattering problem,'' \emph{IEEE
Trans. Antennas Propag.}, vol. 43, no. 8, pp. 784-792, Aug. 1995.
\bibitem{Salucci 2017c}M. Salucci, G. Oliveri, N. Anselmi, F. Viani, A. Fedeli, M. Pastorino,
and A. Randazzo, {}``Three-dimensional electromagnetic imaging of
dielectric targets by means of the multiscaling inexact-Newton method,''
\emph{J. Opt. Soc. Am. A}, vol. 34, no. 7, pp. 1119-1131, 2017.
\bibitem{Rocca 2009}P. Rocca, M. Benedetti, M. Donelli, D. Franceschini, and A. Massa,
{}``Evolutionary optimization as applied to inverse scattering problems,''
\emph{Inverse Probl.}, vol. 25, no. 12, pp. 123003, Dec. 2009.
\bibitem{Pastorino 2007}M. Pastorino, {}``Stochastic optimization methods applied to microwave
imaging: A review,'' \emph{IEEE Trans. Antennas Propag.}, vol. 55,
no. 3, pp. 538-548, Mar. 2007.
\bibitem{Goudos 2021}S. Goudos, \emph{Emerging Evolutionary Algorithms for Antennas and
Wireless Communications}. SciTech/IET, 2021 (ISBN-13: 978-1-78561-552-8).
\bibitem{Rocca 2011}P. Rocca, G. Oliveri, and A. Massa, {}``Differential Evolution as
applied to electromagnetics,'' \emph{IEEE Antennas Propag. Mag.},
vol. 53, no. 1, pp. 38-49, Feb. 2011.
\bibitem{Caorsi 2003}S. Caorsi, M. Donelli, D. Franceschini, and A. Massa, {}``A new methodology
based on an iterative multiscaling for microwave imaging,'' \emph{IEEE
Trans. Microw. Theory Tech.}, vol. 51, pp. 1162-1173, 2003.
\bibitem{Donelli 2009}M. Donelli, D. Franceschini, P. Rocca, and A. Massa, \char`\"{}Three-dimensional
microwave imaging problems solved through an efficient multi-scaling
particle swarm optimization,\char`\"{} \emph{IEEE Trans. Geosci. Remote
Sens.}, vol. 47, no. 5, pp. 1467-1481, May 2009.
\bibitem{Massa 2019}A. Massa, D. Marcantonio, X. Chen, M. Li, and M. Salucci, {}``DNNs
as applied to electromagnetics, antennas, and propagation - A review,''
\emph{IEEE Antennas Wireless Propag. Lett.}, vol. 18, no. 11, pp.
2225-2229, Nov. 2019.
\bibitem{Xu 2020}K. Xu, L. Wu, X. Ye, and X. Chen, {}``Deep learning-based inversion
methods for solving inverse scattering problems with phaseless data,''
\emph{IEEE Trans. Antennas Propag.}, vol. 68, no. 11, pp. 7457-7470,
Nov. 2020.
\bibitem{Li 2019}L. Li, L. G. Wang, F. L. Teixeira, C. Liu, A. Nehorai, and T. J. Cui,
{}``DeepNIS: Deep neural network for nonlinear electromagnetic inverse
scattering,\char`\"{} \emph{IEEE Trans. Antennas Propag.}, vol. 67,
no. 3, pp. 1819-1825, Mar. 2019.
\bibitem{Zhou 2021}Y. Zhou, Y. Zhong, Z. Wei, T. Yin, and X. Chen, \char`\"{}An improved
deep learning scheme for solving 2D and 3D inverse scattering problems,''
\emph{IEEE Trans. Antennas Propag.}, doi: 10.1109/TAP.2020.3027898.
\bibitem{Massa 2021}A. Massa and M. Salucci, {}``On the design of complex \emph{EM} devices
and systems through the System-by-Design paradigm - A framework for
dealing with the computational complexity,'' \emph{IEEE Trans. Antennas
Propag.} (\emph{under review}).
\bibitem{Massa 2017}A. Massa, G. Oliveri, M. Salucci, N. Anselmi, and P. Rocca, {}``Learning-by-examples
techniques as applied to electromagnetics,'' \emph{J. Electromagn.
Waves Appl.}, pp. 1-16, 2017.
\bibitem{Salucci 2019b}M. Salucci, N. Anselmi, S. Goudos, and A. Massa, {}``Fast design
of multiband fractal antennas through a system-by-design approach
for NB-IoT applications,'' \emph{EURASIP J. Wirel. Commun. Netw.},
vol. 2019, no. 1, pp. 68-83, Mar. 2019.
\bibitem{Oliveri 2017}G. Oliveri, M. Salucci, N. Anselmi and A. Massa, {}``Multiscale system-by-design
synthesis of printed WAIMs for waveguide array enhancement,'' \emph{IEEE
J. Multiscale Multiphysics Computat. Techn.}, vol. 2, pp. 84-96, 2017.
\bibitem{Oliveri 2020}G. Oliveri, A. Gelmini, A. Polo, N. Anselmi, and A. Massa, {}``System-by-design
multi-scale synthesis of task-oriented reflectarrays,'' \emph{IEEE
Trans. Antennas Propag.}, vol. 68, no. 4, pp. 2867-2882, Apr. 2020.
\bibitem{Salucci 2019}M. Salucci, L. Tenuti, G. Gottardi, A. Hannan, and A. Massa, {}``System-by-design
method for efficient linear array miniaturisation through low-complexity
isotropic lenses,'' \emph{Electron. Lett.}, vol. 55, no. 8, pp. 433-434,
May 2019.
\bibitem{Shimrat 1962}M. Shimrat, {}``Algorithm 112: Position of point relative to polygon,''
\emph{Communications of the ACM}, vol. 5, no. 8, p. 434, 1962.
\bibitem{Forrester 2008}A. I. J. Forrester, A. Sobester, and A. J. Keane, \emph{Engineering
Design via Surrogate Modelling: A Practical Guide}. Hoboken, N.J.:
John Wiley \& Sons, 2008.
\bibitem{Jones 1998}D. R. Jones, M. Schonlau, and W.J. Welch, {}``Efficient global optimization
of expensive black-box functions,'' \emph{J. Global Opt}., vol. 13,
pp. 455-492, 1998.
\bibitem{Garud 2017}S. S. Garud, I. A. Karimi, and M. Kraft, ''Design of computer experiments:
a review,'' \emph{Comput. Chem. Eng.}, vol\emph{.} 106, pp. 71-95,
May 2017.
\bibitem{Geffrin 2005}J. Geffrin, P. Sabouroux, and C. Eyraud, {}``Free space experimental
scattering database continuation: experimental set-up and measurement
precision,'' \emph{Inverse Probl.}, vol. 21, no. 6, pp. 117-130,
Nov. 2005.
\bibitem{Wolpert 1997}D. H. Wolpert and W. G. Macready, {}``No free lunch theorems for
optimization,'' \emph{IEEE Trans. Evol. Comput.}, vol. 1, no. 1,
pp. 67-82, Apr. 1997.
\end{thebibliography}
\end{document}